\documentclass[preprint2]{emulateapj}
\usepackage{subfigure}
\usepackage{amsmath}

\newcommand{\feh}{\hbox{$ [{\rm Fe}/{\rm H}]$ }} 
\newcommand{\mh}{\hbox{$ [{\rm M}/{\rm H}]$}} 
\def\kms{\,{\rm km\,s^{-1}}}
\def\SNR{{\rm SNR}}
\def\dex{\,{\rm dex}}
\def\mag{\,{\rm mag}}

\slugcomment{accepted by AJ, 15 Nov 2016}

\shorttitle{RAVE DR5}
\shortauthors{Kunder et al.}

\begin{document}
\title{The Radial Velocity Experiment (RAVE): Fifth Data Release}

\author{Andrea Kunder\altaffilmark{1},
Georges Kordopatis\altaffilmark{1},
Matthias Steinmetz\altaffilmark{1},
Toma\v{z} Zwitter\altaffilmark{2},
Paul~J.~McMillan\altaffilmark{3},
Luca Casagrande\altaffilmark{4},
Harry Enke\altaffilmark{1},
Jennifer Wojno\altaffilmark{1},
Marica Valentini\altaffilmark{1},
Cristina Chiappini\altaffilmark{1},
Gal Matijevi\v{c}\altaffilmark{1},
Alessandro Siviero\altaffilmark{5},
Patrick de Laverny\altaffilmark{6}, 
Alejandra Recio-Blanco\altaffilmark{6},
Albert Bijaoui\altaffilmark{6},
Rosemary~F.~G.~Wyse\altaffilmark{7}, 
James Binney\altaffilmark{8}, 
E.~K.~Grebel\altaffilmark{9},  
Amina Helmi\altaffilmark{10},  
Paula Jofre\altaffilmark{11,12},  
Teresa Antoja\altaffilmark{10}, 
Gerard Gilmore\altaffilmark{12},  
Arnaud Siebert\altaffilmark{13},  
Benoit Famaey\altaffilmark{13},  
Olivier Bienaym\'{e}\altaffilmark{13},  
Brad K. Gibson\altaffilmark{14}, 
Kenneth~C.~Freeman\altaffilmark{15}, 
Julio F. Navarro\altaffilmark{16},   
Ulisse Munari\altaffilmark{5},  
George Seabroke\altaffilmark{17}, 
Borja Anguiano\altaffilmark{18,19}, 
Maru\v sa \v Zerjal\altaffilmark{2}, 
Ivan Minchev\altaffilmark{1}, 
Warren Reid\altaffilmark{19,20},  
Joss Bland-Hawthorn\altaffilmark{21}, 
Janez Kos\altaffilmark{21}, 
Sanjib Sharma\altaffilmark{21}, 
Fred Watson\altaffilmark{22}, 
Quentin A. Parker\altaffilmark{23,24}, 
Ralf-Dieter Scholz\altaffilmark{1}, %
Donna Burton\altaffilmark{18},
Paul Cass\altaffilmark{18},
Malcolm Hartley\altaffilmark{18},
Kristin Fiegert\altaffilmark{18},
Milorad Stupar\altaffilmark{18,19},
Andreas Ritter\altaffilmark{25},
Keith Hawkins\altaffilmark{11,26}
Ortwin Gerhard\altaffilmark{27},
W. J. Chaplin\altaffilmark{28,29}
G. R. Davies\altaffilmark{28,29},
Y. P. Elsworth\altaffilmark{28,29},
M.~N. Lund\altaffilmark{27,29},
A. Miglio\altaffilmark{27,29},
B. Mosser\altaffilmark{30}
}
\altaffiltext{1}{Leibniz-Institut f\"{u}Ÿr Astrophysik Potsdam (AIP), An der Sternwarte 16, D-14482 Potsdam, Germany}
\affil{E-mail: akunder@aip.de}
\altaffiltext{2}{Faculty of Mathematics and Physics, University of Ljubljana, Jadranska 19, 1000 Ljubljana, Slovenia}
\altaffiltext{3}{Lund Observatory, Lund University, Department of Astronomy and Theoretical Physics, Box 43, SE-22100, Lund, Sweden}
\altaffiltext{4}{Research School of Astronomy \& Astrophysics, Mount Stromlo Observatory, The Australian National University, ACT 2611, Australia}
\altaffiltext{5}{Dipartimento di Fisica e Astronomia Galileo Galilei, Universita' di Padova, Vicolo dell'Osservatorio 3, I-35122 Padova, Italy}
\altaffiltext{6}{Laboratoire Lagrange, Universit\'{e} C\^{o}te d'Azur , Observatoire de la C\^ote d\'Azur, CNRS, Bd de l\'Observatoire, CS 34229, 06304 Nice cedex 4, France}
\altaffiltext{7}{Department of Physics and Astronomy, Johns Hopkins University, 3400 N. Charles St, Baltimore, MD 21218, USA}
\altaffiltext{8}{Rudolf Peierls Centre for Theoretical Physics, Keble Road, Oxford OX1 3NP, UK}
\altaffiltext{9}{Astronomisches Rechen-Institut, Zentrum f\"ur Astronomie der Universit\"at Heidelberg, M\"onchhofstr.\ 12--14, 69120 Heidelberg, Germany}
\altaffiltext{10}{Kapteyn Astronomical Institute, University of Groningen, P.O. Box 800, 9700 AV Groningen, The Netherlands 0000-0003-3937-7641}
\altaffiltext{11}{Institute of Astronomy, University of Cambridge, Madingley Road, CB3 0HA Cambridge, UK}
\altaffiltext{12}{N\'ucleo de Astronom\'ia, Facultad de Ingenier\'ia, Universidad Diego Portales,  Av. Ejercito 441, Santiago, Chile}
\altaffiltext{13}{Observatoire astronomique de Strasbourg, Universit\'{e} de Strasbourg, CNRS, UMR 7550, 11 rue de l'Universit\'{e}, F-67000 Strasbourg, France}
\altaffiltext{14}{E.A. Milne Centre for Astrophysics, University of Hull, Hull, HU6 7RX, United Kingdom}
\altaffiltext{15}{Research School of Astronomy \& Astrophysics, Australian National University, Cotter Rd., Weston, ACT 2611, Australia}
\altaffiltext{16}{CIfAR Senior Fellow, Department of Physics and Astronomy, University of Victoria, Victoria, BC, Canada V8P5C2}
\altaffiltext{17}{Mullard Space Science Laboratory, University College London, Holmbury St Mary, Dorking, RH5 6NT, UK}
\altaffiltext{18}{Australian Astronomical Observatory, P.O. Box 915, North Ryde, NSW 1670, Australia}
\altaffiltext{19}{Department of Physics and Astronomy, Macquarie University, Sydney, NSW 2109, Australia}
\altaffiltext{20}{University of Western Sydney, Penrith South DC, NSW 1797, Australia}
\altaffiltext{21}{Sydney Institute for Astronomy, School of Physics, A28, The University of Sydney, NSW 2006, Australia}
\altaffiltext{22}{Anglo-Australian Observatory, P.O. Box 296, Epping, NSW 1710, Australia}
\altaffiltext{23}{Department of Physics, CYM Building, The University of Hong Kong, Hong Kong, China}
\altaffiltext{24}{The Laboratory for Space Research, The University of Hong Kong, Hong Kong, China}
\altaffiltext{25}{Department of Astrophysical Sciences, Princeton University, 4 Ivy Ln, Princeton, NJ 08544, USA}
\altaffiltext{26}{Department of Astronomy, Columbia University, 550 W. 120 st., New York, New York, USA}
\altaffiltext{27}{Max-Planck-Institut fuer Ex. Physik, Giessenbachstrasse, D-85748 Garching b. Muenchen, Germany}
\altaffiltext{28}{School of Physics and Astronomy, University of Birmingham, Edgbaston, Birmingham B15 2TT, UK}
\altaffiltext{29}{Stellar Astrophysics Centre, Department of Physics and Astronomy, Aarhus University, DK-8000 Aarhus C, Denmark}
\altaffiltext{30}{Observatoire de Paris, PSL Research University, CNRS, Université Pierre et Marie Curie, Université Paris Diderot, 92195, Meudon, France}

\begin{abstract}
Data Release 5 (DR5) of the Radial Velocity Experiment (RAVE) is the fifth
data release from a magnitude-limited ($9< I < 12$) survey of stars randomly
selected in the southern hemisphere.  The RAVE medium-resolution spectra
($R\sim7500$) covering the Ca-triplet region (8410-8795\,\AA) span the
complete time frame from the start of RAVE observations in 2003 to their
completion in 2013.  Radial velocities from 520\,781 spectra of 457\,588
unique stars are presented, of which 255\,922 stellar observations have
parallaxes and proper motions from the Tycho-Gaia astrometric solution (TGAS)
in {\it Gaia} DR1.  For our main DR5 catalog, stellar parameters (effective temperature, 
surface gravity, and overall metallicity) are computed using the RAVE DR4 stellar pipeline, 
but calibrated using recent K2 Campaign 1 seismic gravities and Gaia benchmark stars,
as well as results obtained from high-resolution studies.  Also included are 
temperatures from the Infrared Flux Method, and we provide a catalogue of red 
giant stars in the dereddened color $(J-Ks)_0$ interval (0.50,0.85) for which the 
gravities were calibrated based only on seismology.  Further
data products for sub-samples of the RAVE stars include individual abundances
for Mg, Al, Si, Ca, Ti, Fe, and Ni, and distances found using isochrones.
Each RAVE spectrum is complemented by an error spectrum, which has been
used to determine uncertainties on the parameters.  The data can be
accessed via the RAVE Web site or the Vizier database. 
\end{abstract}


\section{Introduction}

The kinematics and spatial distributions of Milky Way stars help define the Galaxy we
live in, and allow us to trace parts of the formation of the Milky Way.  In
this regard, large spectroscopic surveys that provide measurements
of fundamental structural and dynamical parameters for a statistical sample
of Galactic stars, have been extremely successful in advancing the
understanding of our Galaxy.  Recent and ongoing spectroscopic surveys of the Milky Way
include the RAdial Velocity Experiment \citep[RAVE,][]{steinmetz06}, the Sloan 
Extension for Galactic Understanding and Exploration \citep[SEGUE,][]{yanny09}, 
the APO Galactic Evolution Experiment
\citep[APOGEE,][]{eisenstein11}, the LAMOST Experiment for Galactic
Understanding and Exploration \citep[LAMOST,][]{zhao12},
the Gaia--ESO Survey \citep[GES,][]{gilmore12} and the GALactic
Archaeology with HERMES \citep[GALAH,][]{desilva15}.  These surveys were made possible by the
emergence of wide field multi-object spectroscopy (MOS) fibre systems,
technology that especially took off in the 1990s.  Each survey has its own
unique aspect, and together form complementary samples in terms of
capabilities and sky coverage.

Of the above mentioned surveys, RAVE was the first, designed to provide
stellar parameters to complement missions that focus on astrometric
information.  The four previous data releases, DR1 \citep{steinmetz06}, DR2
\citep{zwitter08}, DR3 \citep{siebert11} and DR4 \citep{kordopatis13} have
been the foundation for a number of studies which have advanced our
understanding of especially the disk of the Milky Way \citep[see review
by][]{kordopatis14}.  For example, in recent years a
wave-like pattern in the stellar velocity distribution was uncovered
\citep{williams13} and the total mass of the Milky Way was measured using the RAVE
extreme-velocity stars \citep{piffl14a}, as was the local dark
matter density \citep{bienayme14, piffl14b}.  Moreover, chemo-kinematic 
signatures of the dynamical effect of mergers on the Galactic 
disk \citep{minchev14}, and
signatures of radial migration were detected \citep{kordopatis13b, wojno16}.
Stars tidally stripped from globular clusters were also identified
\citep{kunder14, anguiano15, anguiano16}. RAVE further allowed for the
creation of pseudo-3D maps of the diffuse interstellar band at 
8620~\AA~\citep{kos14} and high-velocity stars to be studied
\citep{hawkins15}.  

RAVE DR5 includes not only the final RAVE observations taken in 2013, but
also earlier discarded observations recovered from previous years, resulting in an additional
$\sim30\,000$ RAVE spectra.  This is the first RAVE data release in which 
error spectrum was generated for each
RAVE observation, so we can provide realistic uncertainties and probability
distribution functions for the derived radial velocities and stellar
parameters.  We have performed a recalibration of stellar metallicities,
especially improving stars of super-solar metallicity.  Using the Gaia
benchmark stars \citep{jofre14, heiter15} as well as 72 RAVE stars with Kepler-2
asteroseismic $\log g$ parameters (Valentini et~al. submitted, hereafter V16), 
the RAVE $\log g$ values have been recalibrated, resulting in more accurate gravities
especially for the giant stars in RAVE.  The distance pipeline
\citep{binney14} has been improved and extended to process more accurately
stars with low metallicities (${\rm [M/H]} < -0.9\dex$).  Finally, by
combining optical photometry from APASS \citep{munari14} with 2MASS
\citep{skrutskie06} we have derived temperatures from the Infrared Flux
Method \citep{c10}.

Possibly the most unique aspect of DR5 is the extent to which it complements
the first significant data release from {\it Gaia}. 
The successful completion of the Hipparcos Mission and publication
of the catalogue \citep{esa97} demonstrated that
space astrometry is a powerful technique to measure accurate distances to 
astronomical objects.
Already in RAVE-DR1\citep{steinmetz06}, we looked forward 
to the results from the ESA cornerstone mission {\it Gaia}, as this space-based mission's astrometry
of Milky Way stars will have $\sim$100 times better astrometric accuracies than its
predecessor, Hipparcos.  Although {\it Gaia} has been launched and data collection
is ongoing, a long enough time baseline has to have elapsed for
sufficient accuracy of a global reduction of observations \citep[e.g., five years 
for {\it Gaia} to yield positions, parallaxes and annual proper motions at an accuracy 
level of 5 -- 25 $\rm \mu$as,][]{michalik15}.  To expedite the use of the first {\it Gaia}
astrometry results, the approximate positions at the 
earlier epoch (around 1991) provided by the Tycho-2 Calalogue \citep{hog00}
can be used to disentangle the 
ambiguity between parallax and proper motion in a shorter stretch of
Gaia observations.  These TGAS stars therefore
contain positions, parallaxes, and proper motions earlier than the
global astrometry from Gaia can be released.  
There are $215\,590$ unique RAVE stars in TGAS, so for these stars
we now have space-based parallaxes and proper motions from {\it Gaia} DR1 
in addition to stellar parameters, radial velocities, and in many cases chemical abundances.
The Tycho-2 stars observed by RAVE in a homogeneous and 
well-defined manner can be combined with the released TGAS stars to exploit the 
larger volume of stars for which milliarcsecond accuracy astrometry
exists, for an extraordinary return in scientific results.  We note that in a companion paper, 
a data-driven re-analysis of the RAVE spectra using {\it The Cannon} model has been
carried out (Casey et~al. 2016, in prep, hereafter C16), which presents the derivation of 
$T_{\rm eff}$, surface gravity $\log{g}$ and $\rm [Fe/H]$, as well as
chemical abundances of giants of up to seven elements (O, Mg, Al, Si, Ca, Fe, Ni). 

In \S\ref{sec:SF}, the selection function of the RAVE DR5 stars is presented --
further details can be found in Wojno et~al. submitted, hereafter W16.  The RAVE observations and 
reductions are summarised in \S\ref{sec:S}.  An explanation of how the error
spectra were obtained is found in \S\ref{sec:ES}, and \S\ref{sec:RV} summarises 
the derivation of radial velocities from the spectra.  In \S\ref{sec:P}, the procedure used to
extract atmospheric parameters from the spectrum is described and the
external verification of the DR5 $T_{\rm eff}$, $\log g$ and [M/H] values
is discussed in \S\ref{sec:EV}. The dedicated
pipelines to extract elemental abundances, and distances are described in
\S\S\ref{chemicalpipeline} and \ref{sec:D}, respectively -- DR5 gives radial
velocities for all RAVE stars but elemental abundances and distances are
given for sub-samples of RAVE stars that have SNR $>$ 20 and the most well-defined 
stellar parameters.  Temperatures from the
Infrared Flux Method are presented in \S\ref{sec:IRFM}.  
In \S\ref{sec:AC} we present for the red giants gravities based on
asteroseismology by the method of V16.  A comparison of the
stellar parameters in the RAVE DR5 main catalog to
other stellar parameters for RAVE stars (e.g., those from C16) 
is provided in \S\ref{sec:BP}. The final 
sections, \S\ref{sec:diff} and \S\ref{sec:conclude} provide a
summary of the difference between DR4 and DR5, and 
an overview of DR5, respectively.   

\section{Survey Selection Function}\label{sec:SF}

Rigorous exploitation of DR5 requires knowledge of RAVE's selection function,
which was recently described by W16. Here we provide only a
summary.  

The stars for the RAVE input catalogue were selected from their $I$-band
magnitudes, focusing on bright stars ($9<I <12$) in the southern hemisphere,
but the catalogue does contain some stars that are either brighter or
fainter, in part because stars were selected by extrapolating data from other
sources, such as Tycho-2 and SuperCOSMOS before DENIS was available 
in 2006 \citep[see DR4 paper,][\S2 for details]{kordopatis13}.  As the
survey progressed, the targets in the input catalog were grouped into four
$I$-band magnitude bins: 9.0 -- 10.0, 10.0 -- 10.75, 10.75 -- 11.5, and 11.5
-- 12.0, which helped mitigate fibre cross-talk problems.  This led to a
segmented distribution of RAVE stars in $I$-band magnitudes, but the
distributions in other passbands are closely matched by Gaussians \citep[see
e.g., Fig.~11 in][]{munari14}.  For example, in the $B$-band, the stars
observed by RAVE have a nicely Gaussian distribution, peaking at $B=12.62$
with $\sigma=1.11\mag$.

The initial target selection was based only on
the apparent $I$-band magnitude, but  a colour
criterion ($J-K_s \ge 0.5$) was later imposed in regions close to
the Galactic plane (Galactic latitude $|b|<25^\circ$) to bias the survey
towards giants. Therefore, the
probability, $S$, of a star being observed by the RAVE survey is
\begin{equation} 
\label{eq:selection_function}
 S \propto S_{\rm select}(l,b,I,J-K_s),
\end{equation}
where $l$ is Galactic longitude.
W16 determine the function $S_{\rm select}$ both on a
field-by-field basis, so  time-dependent effects can be captured, and
with Hierarchical Equal-Area iso-Latitude Pixelisation (HEALPix)
\citep[e.g.,][]{gorski05}, which divides the sky into equal-area
pixels, as regularly distributed as possible.  The sky is divided into 12\,288
pixels ($N_{\rm side} = 32$) which results in a pixel area of $\simeq3.36\deg^2$,
and we only consider the selection function evaluated with HEALPix, 
because RAVE fields overlap on the sky for quality control and variability tests.  

The parent RAVE sample is constructed by first discarding all repeat
observations, keeping only the observation with the highest SNR.  Then
observations which were not conducted as part of the typical observing
strategy (e.g., calibration fields) were removed.  Finally, all stars with
$|b|<25^\circ$ that were observed despite violating the colour criterion
$J-K_s \ge 0.5$ were dismissed.  After applying these cuts, we are left with
448\,948 stars, or 98\% of all stars targeted by RAVE.  These define the RAVE 
DR5 core sample (survey footprint).  The core sample is complemented by 
targeted observations (e.g., open clusters), mainly for calibration and testing.

The number of RAVE stars ($N_{\rm RAVE}$) in each HEALPix pixel is then
counted as a function of $I_{\rm 2MASS}$.  We apply the same criteria to two
photometric all-sky surveys, 2MASS and Tycho-2 to discover how many stars
could, in principle, have been observed.  After these catalogues were
purged of spurious measurements, we obtain $N_{\rm 2MASS}$ and $N_{\rm
TYCHO2}$ and can compute the completeness of RAVE as a function of magnitude
for both 2MASS and TYCHO2 as $N_{\rm RAVE}/N_{\rm 2MASS}$ and $N_{\rm
RAVE}/N_{\rm TYCHO2}$. 

\begin{figure}[htb]  
\includegraphics[width=9cm]{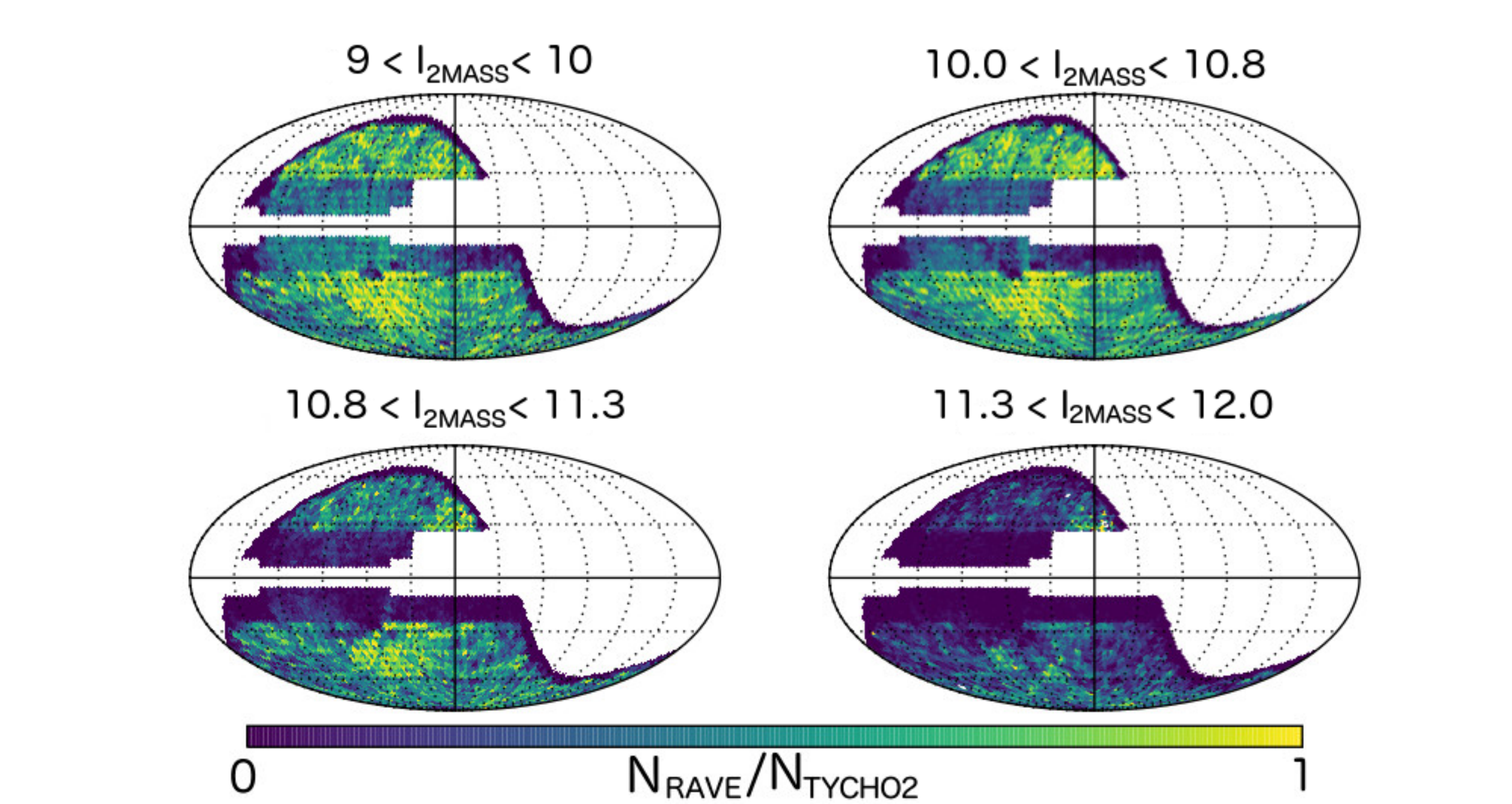}
\caption{Mollweide projection of Galactic coordinates of the completeness of the stars in 
Tycho-2 for which RAVE DR5 radial velocity measurements are available for the core sample. 
Each panel shows the completeness over a different magnitude bin, where the HEALPix pixels are colour-coded 
by the fractional completeness ($N_{\rm RAVE}$/$N_{\rm TYCHO2}$).  
\label{skycompleteness_tycho}}
\end{figure}

Figure~\ref{skycompleteness_tycho} shows the DR5 completeness with respect to
Tycho-2 as a function of magnitude.  It is evident that RAVE avoids the Galactic
plane, and we find that the coverage on the sky is highly anisotropic, with a
significant drop-off in completeness at the fainter magnitudes.
A similar result is seen for $N_{\rm RAVE}/N_{\rm 2MASS}$ (W16).
However, in $N_{\rm RAVE}/N_{\rm 2MASS}$, there is a significantly higher
completeness at low Galactic latitudes ($|b|<25^\circ$) for the fainter
magnitude bins.

Because stars that passed the photometric cuts were randomly selected for
observation, RAVE DR5 is free of kinematic bias.  Hence, the contents of DR5
are representative of the Milky Way for the specific magnitude interval.  A
number of peculiar and rare objects are included.  The morphological flags of
\citet{matijevic12} allow one to identify the normal single stars (90 -
95\%), and those that are unusual -- the peculiar stars include various types
of spectroscopic binary and chromospherically active stars.  The stars
falling within the RAVE selection function footprint described in W16 are 
provided in \doi{10.17876/rave/dr.5/005}.

%\begin{figure}[htb]  
%\includegraphics[width=9cm]{coverage_r_t.pdf}
%\caption{Same as Figure~\ref{skycompleteness}, but for $N_{\rm RAVE}/N_{\rm TYCHO2}$.
%\label{skycompleteness_tycho}}
%\end{figure}

\section{Spectra and their Reduction}\label{sec:S}
The RAVE spectra were taken using the multi-object spectrograph 6dF (6 degree
field) on the $1.2\,$m UK Schmidt Telescope of the Australian Astronomical
Observatory (AAO).  A total of 150 fibres could be allocated in one pointing,
and the covered spectral region (8410--8795\,\AA) at an effective resolution
of $R = \lambda/\Delta \lambda \sim7\,500$ was chosen as analogous to
the wavelength range of Gaia's Radial Velocity Spectrometer
\citep[see DR1 paper,][\S2 and \S3 for details]{steinmetz06}.

The RAVE reductions are described in detail in DR1 \S4 and upgrades to the
process are outlined in DR3 \S2.  In DR5 further improvements have been made
to the Spectral Parameters And Radial Velocity (SPARV) pipeline, the DR3
pipeline that carries out the continuum normalisation, masks bad pixels, and
provides RAVE radial velocities.  The most significant is that instead of the
reductions being carried out on a field-by-field basis, single fibre
processing was implemented.  Therefore, if there were spectra within a RAVE
field which simply could not be processed, instead of the whole field failing
and being omitted in the final RAVE catalogue, only the problematic spectra are
removed.  This is one reason DR5 has more stars than the previous RAVE data
releases.

The DR5 reduction pipeline is able to processes the problematic DR1 spectra,
and it produces error spectra.  An overhaul of bookkeeping and process
control lead to identification of multiple copies of the same observation and
of spectra with corrupted FITS headers.  Some RAVE IDs
have changed from DR4, and some stars released in DR4 could not be processed
by the DR5 pipeline.  The vast majority of these stars have low
signal-to-noise ratios ($\SNR<10$).  Details are provided in Appendix~A; less
than 0.1\% of RAVE spectra were affected by bookkeeping inconsistencies.

\begin{table*}
\begin{scriptsize}
\begin{center}
\caption{Contents of RAVE DR5 } 
\label{ravegcs}
\begin{tabular}{ p{6.5cm}p{2.5cm}}
 & in DR5  \\ 
\hline
\hline
RAVE stellar spectra & 520,781 \\
Unique stars observed & 457,588 \\
Stars with $\geq 3$ visits  & 8000 \\
Spectra / unique stars with  $\SNR> 20$ & 478,161 / 423,283 \\
Spectra / unique stars with  $\SNR> 80$ & 66,888 / 60,880 \\
Stars with  ${\tt AlgoConv}\neq1$$^a$  & 428,952 \\
Stars with elemental abundances & 339,750 \\  
Stars with morphological flags n,d,g,h,o & 394,612 \\
\hline
Tycho2 + RAVE stellar spectra/unique stars & 309,596 / 264,276 \\
TGAS + RAVE stellar spectra/unique stars & 255\,922 / 215,590 \\
\hline
\hline
$^a$For a discussion of {\tt AlgoConv} see \S\ref{subsec:P1}
%TGAS Stars with $\geq 3$ visits  & --- \\
%TGAS Stars with  $\SNR> 20$ & --- \\
%TGAS Stars with  $\SNR> 80$ & --- \\
%TGAS Stars with  ${\tt AlgoConv}\neq1$ & --- \\
%TGAS Stars with elemental abundances & ---- \\
%TGAS Stars with morphological flags n,d,g,h,n,o & ---- \\

\end{tabular}
\end{center}
\end{scriptsize}
\end{table*}

\section{Error Spectra}\label{sec:ES}
\label{errspec}
The wavelength range of the RAVE spectra is dominated by strong spectral
lines: for a majority of stars, the dominant absorption features are due to the
infra-red calcium triplet (CaT), which in hot stars gives way to Paschen
series of hydrogen. Also present are weaker metallic lines for the Solar type
stars and molecular bands for the coolest stars. Within an absorption trough
the flux is small, so shot noise is more significant in the middle of a line
than in the adjacent continuum. Error levels increase also at wavelengths of
airglow sky emission lines, which have to be subtracted during reductions. As
a consequence, a single number, usually reported as a SNR ratio, is not an
adequate quantification of the observational errors associated with a given
spectrum. 

For this reason, DR5 provides error spectra which comprise uncertainties
(``errors") for each pixel of the spectrum. RAVE spectra generally have a
high SNR in the continuum (its median value is SNR~$\sim 40$), and there shot
noise dominates the errors.  Denoting number of counts accumulated in the
spectrum before sky subtraction by $N_{\rm u}$, the corresponding number
after sky subtraction by $N_{\rm s}$, and the effective gain by $g$, the
shot noise is $N = \sqrt{g N_{\rm u}}$ and the signal is $S ={g N_{\rm s}}$.
The appearance of $N_{\rm u}$ rather than $N_{\rm s}$ in the relation for $N$
reflects the fact that noise is enhanced near night-sky emission lines. As a
consequence the SNR ratio is decreased both within profiles of strong stellar
absorption lines (where $N_{\rm s}$ is small) and near sky emission lines.
The gain $g$ is determined using the count vs.\ magnitude relation \citep[see
eq.\ 1 from ][]{zwitter08}. Its value ($g = 0.416 \mathrm{e}^- /
\mathrm{ADU}$) reflects systematic effects on a pixel-to-pixel scale that
lower the effective gain to this level.  

Telluric absorptions are negligible in the RAVE wavelength range
\citep{munari99}. RAVE observations from Siding Spring generally show a sky
signal with a low continuum level, even when observed close to the Moon. The
main contributors to the sky spectrum are therefore airglow emission lines, which
belong to three series: OH transitions 6-2 at $\lambda < 8651\,$\AA, OH
transitions 7-3 at $\lambda > 8758\,$\AA, and O$_2$ bands at $8610\,\AA<
\lambda< 8710\,$\AA.  Wavelengths of OH lines are listed in the file {\it
linelists\$skylines.dat} which is part of the IRAF\footnote{IRAF is
distributed by the National Optical Astronomy Observatory, which is operated
by the Association of Universities for Research in Astronomy (AURA) under a
cooperative agreement with the National Science Foundation.} reduction
package, while the physics of their origin is nicely summarised at {\tt
http://www.iafe.uba.ar/aeronomia/airglow.html}.  One needs to be careful when
analysing stellar lines with superimposed airglow lines. Apart from
increasing the noise levels, these lines may not be perfectly
subtracted, as they can be variable on angular scales of degrees and on
timescales of minutes, whereas the telescope's field of view is $6.7^\circ$
and the  exposure time was typically 50 minutes.

Evaluation of individual reduction steps \citep[see][]{zwitter08} shows that
fibre cross-talk and scattered light have only a small influence on error
levels. In particular, a typical level of fibre-cross talk residuals is
$0.0014 f$, where $f$ is the ratio between flux of an object in an adjacent
fibre and flux of the object in question. Fibre cross-talk suffers from
moderate systematic effects (variable point spread function profiles across
the wavelength range), but even at the edges of the spectral range these
effects do not exceed a 1\%\ level. Scattered light typically contributes
$\sim 5$\%\ of the flux level of the spectral tracing. So its effect on noise
estimation is not important, and we were not able to identify any
systematics. Finally, RAVE observes in the near IR and uses a thinned CCD
chip, so an accurate subtraction of interference fringes is needed. Tests
show that fringe patterns for the same
night and for the same focal plate typically stay constant to within 1\%\ of
the flat-field flux level.  As a result scattered
light and fringing only moderately increase the final noise levels. Together,
scattered light and fringing are estimated to contribute a
relative error of $\sim 0.8$\%, which is added in quadrature to the prevailing
contribution of shot noise discussed above. 

Finally we note that fluxes and therefore noise levels for individual pixels
of a given spectrum are not independent of each other, but are correlated
because of a limited resolving power of RAVE spectra. So the final noise
spectrum was smoothed with a window with a width of 3 pixels in the
wavelength direction, which corresponds to the FWHM for a resolving power of
RAVE spectra. 

For each pixel in a RAVE spectrum, we invoke a Gaussian with a mean
and standard deviation as measured from the same pixel of the corresponding
error spectra.  A new spectrum is therefore generated which can be roughly interpreted as
an alternative measurement of the star (although note the error spectrum does not take
every possible measurement uncertainty into account as discussed above).  We then 
can redetermine our radial velocity for these resampled data which will
differ slightly from that obtained from the actual observed spectrum.   Repeating this resampling 
process and monitoring the resulting radial velocity estimates, we get a distribution of the
radial velocity from which we can then infer an uncertainty.

\begin{figure}[htb]  
\includegraphics[width=9cm]{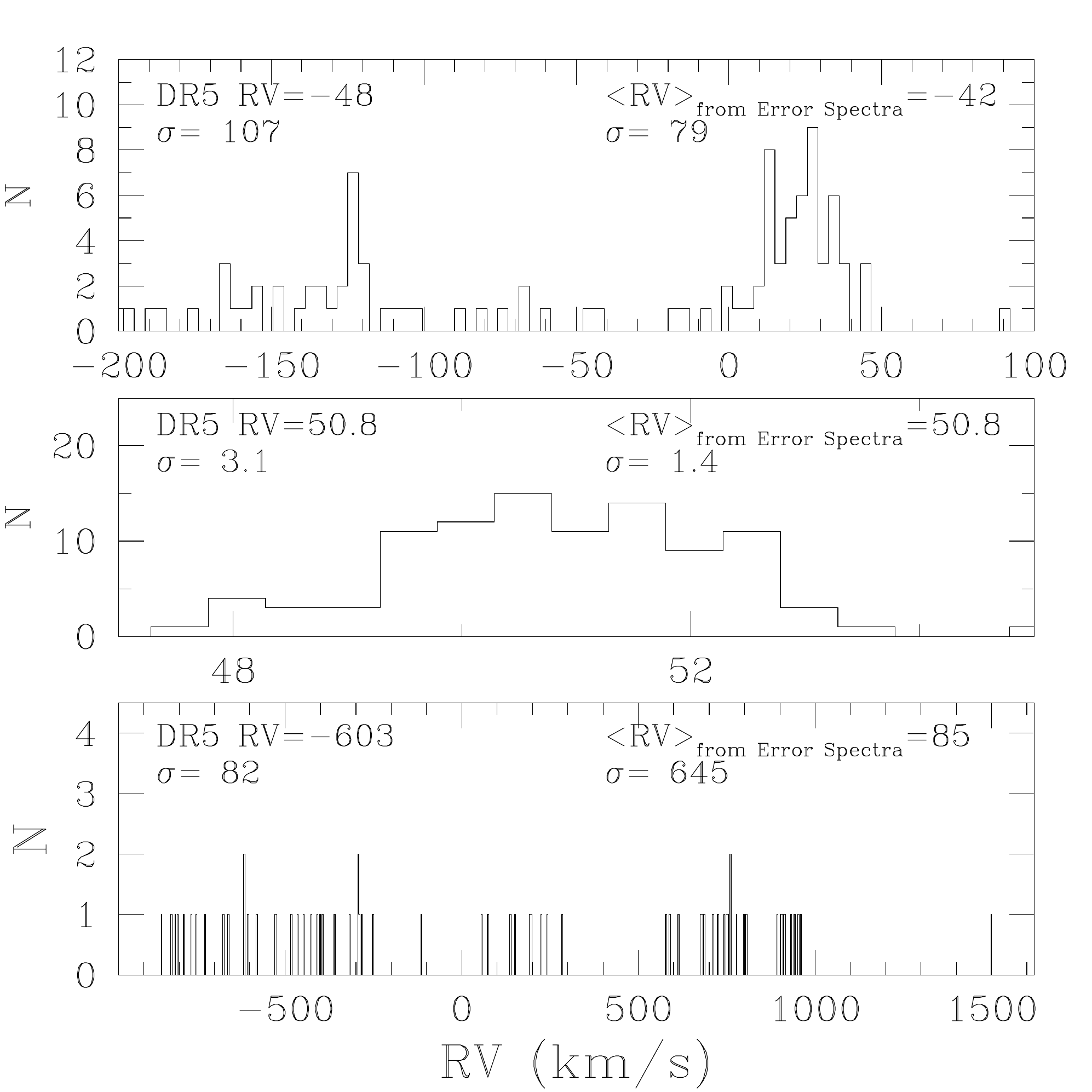}
\caption{The derived radial velocities and dispersion from resampling the RAVE spectra 100 times
using the error spectra.
The top panel shows the radial velocity distribution from a SNR=5 star with a $T_{\rm eff}$=3620~K, the middle panel
shows the radial velocity distribution from a SNR=13 star with a $T_{\rm eff}$=5050~K, and the bottom panel shows
the radial velocity distribution from a SNR=8 star with a $T_{\rm eff}$=7250~K. The standard deviation
of the radial velocity as derived from the error spectrum leads to more realistic uncertainty estimates
for especially the hot stars.
\label{err_spec_ex}}
\end{figure}

The raw errors as derived in the error spectra are propagated both into the radial velocities
and stellar parameters presented here.  This process allows a better assessment of the 
uncertainties, especially of stars with low SNR or hot stars, where the CaT is not as prominent.
Figure~\ref{err_spec_ex} shows the mean radial velocity from the resulting radial velocity 
estimates of 100 resampled spectra for low SNR stars.  
For most RAVE stars, the radial velocity errors are consistent with 
a Gaussian (see middle panel), but for the more problematic hot stars, or those with low SNR, 
this is clearly not the case.

Each RAVE spectrum was resampled from its error spectrum ten times.  Whereas
our tests indicate that a larger number of resamplings ($\sim60$) would be
ideal for the more problematic spectra, ten resamplings were chosen as a
compromise between computing time and the relatively small number of RAVE
spectra with low SNR and hot stars that would benefit from additional
resamplings.  For $\sim97.5$\% of the RAVE sample, there is one-sigma or less
difference in the radial velocity and radial velocity dispersions when
resampling the spectrum 10 or 100 times.  In DR5, we provide both the formal
error in radial velocity, which is a measure of how well the
cross-correlation of the RAVE spectrum against a template spectrum was
matched, and the standard deviation and the median absolute deviation
(MAD) in heliocentric radial velocity from a spectrum resampled ten times.

%The error distribution of the measured data can be derived from these 
%error spectrum.  Error estimation by resampling the measured data is a particularly accurate
%way to asses uncertainties.   

%% clarify

Each RAVE spectrum was resampled from its error spectrum ten times.  Whereas
our tests indicate that a larger number of resamplings ($\sim60$) would be
ideal for the more problematic spectra, ten resamplings were chosen as a
compromise between computing time and the relatively small number of RAVE
spectra with low SNR and hot stars that would benefit from additional
resamplings.  For $\sim97.5$\% of the RAVE sample, there is one-sigma or less
difference in the radial velocity and radial velocity dispersions when
resampling the spectrum 10 or 100 times.  
%In DR5, we provide both the formal
%error in radial velocity, which is a measure of how well the
%cross-correlation of the RAVE spectrum against a template spectrum was
%matched, and the standard deviation and the median absolute deviation
%(MAD) in heliocentric radial velocity from a spectrum resampled ten times.

\section{Radial Velocities}\label{sec:RV}

The DR5 radial velocities are derived in an identical manner to 
in those in DR4. The process of velocity determination  is explained by \citet{siebert11}.  Templates 
are used to measure the 
radial velocities in a two-step process.  First, using a subset of 10 template spectra,
a preliminary estimate of the RV is obtained, which has a typical accuracy
better than 5 km s$^{-1}$.  A new template is then constructed using the
full template database described in \citet{zwitter08}, from which
the final, more precise RV is obtained.  This has a typical accuracy
better than 2 km s$^{-1}$.  

\begin{figure}
\includegraphics[width=\linewidth]{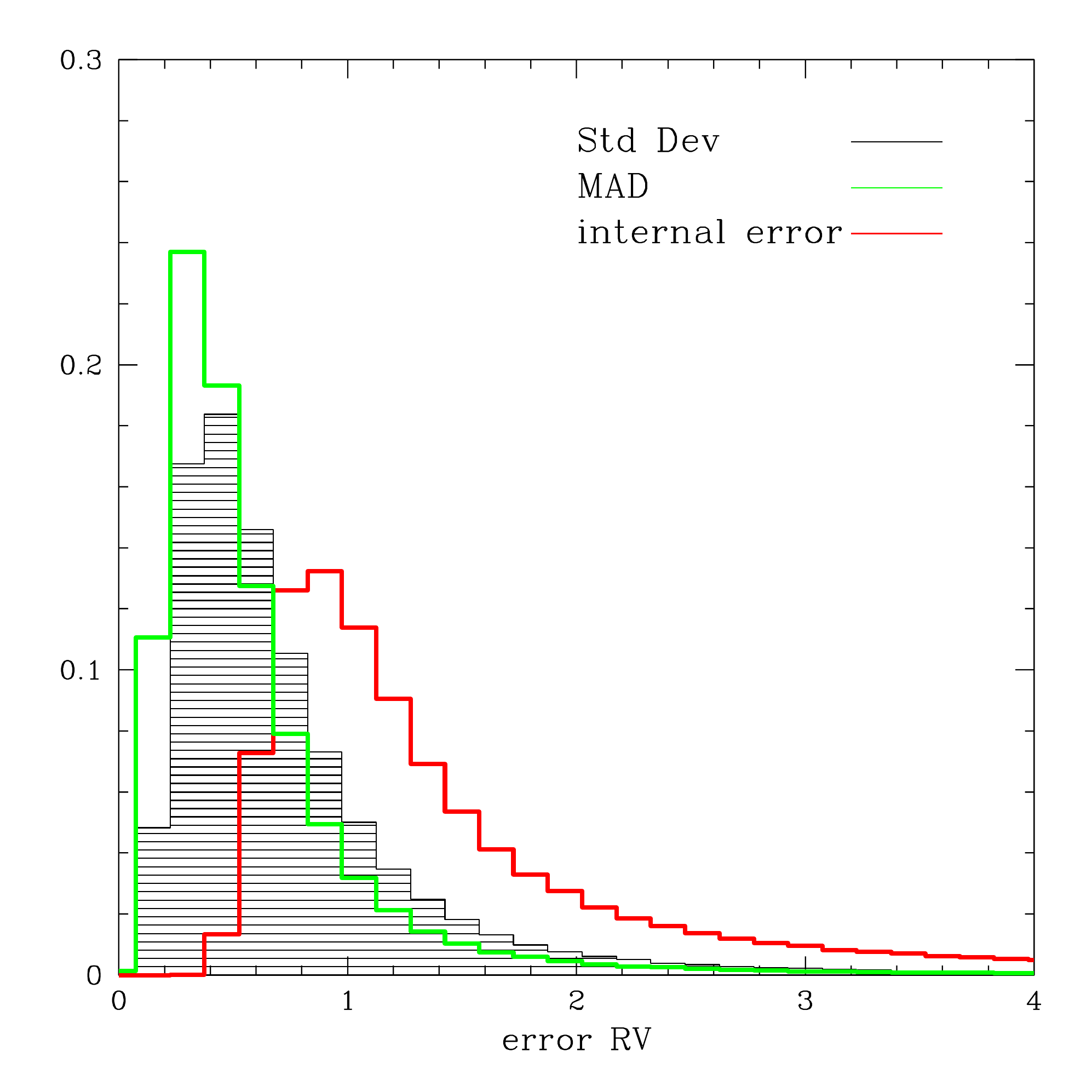}\\
\caption{Histograms of the errors on the radial velocities
of the DR5 stars, derived from the resampling of the DR5 spectra
ten times using their associated error spectra. 
The filled black histogram shows the standard deviation distributions 
and the green histogram shows the MAD estimator distribution. 
The red histogram shows the internal radial velocity error obtained from
cross correlating the RAVE spectra with a template.
\label{rvuncertainties}}
\end{figure}

The internal error in RV, $\rm \sigma(RV)$, comes from the {\tt xcsao} task
within IRAF, and therefore describes the error on the determination of the
maximum of the correlation function.  It was noticed that for some stars,
particularly those with $\rm \sigma(RV) >10\kms$, $\rm \sigma(RV)$ was
underestimated.  The inclusion of error spectra in DR5 largely remedies this
problem, and the standard deviation and MAD provide independent
measures  of the RV uncertainties (see Fig.~\ref{err_spec_ex}). Uncertainties
derived from the error spectra are especially useful for stars
that have low SNR or high temperatures.
Figure~\ref{rvuncertainties} shows the errors from the resampled spectra
compared to the internal errors.  For the majority of RAVE stars, the
uncertainty in RV is dominated by the cross-correlation between the RAVE
spectrum and the RV template, and not by the array of uncertainties
(``errors") for each pixel of the RAVE spectrum.

Repeated RV measurements have been used to characterise the uncertainty in 
the RVs.  There are 43\,918 stars that have been observed more than once; 
the majority (82\%) of these stars have two measurements, and six RAVE stars were 
observed 13 times.
The histogram of the RV scatter between the
repeat measurements peaks at 0.5 km~s$^{-1}$, and has a long tail at larger scatter.  
This extended scatter is due both to variability from stellar binaries and 
problematic measurements.  If stars are selected that have radial velocities derived
with high confidence, e.g.,  stars with $|{\tt correctionRV}| < 10\kms$, 
$\sigma(RV) <8\kms$, and  ${\tt correlationCoeff}>10$ \citep[see][]{kordopatis13},
the scatter of the repeat measurements peaks at $0.17\kms$ and the tail
is reduced by 90\%.

The zero-point in RV has already been evaluated in the previous data
releases.  The exercise is repeated here, with the inclusion of a comparison
to APOGEE and Gaia-ESO, and the summary of the comparisons to different
samples is given in Table~\ref{externalrvs}.  Our comparison sample comprises
of the GCS \citep{nordstrom04} data as well as high-resolution echelle
follow-up observations of RAVE targets at the ANU 2.3 m telescope, the Asiago
Observatory, the Apache Point Observatory \citep{ruchti11}, and Observatoire
de Haute Provence using the Elodie and Sophie instruments.  Sigma-clipping is
used to remove contamination by spectroscopic binaries or problematic
measurements, and the mean $\rm \Delta (RV)$ given is $\rm \Delta (RV) =
RV_{DR5} - RV_{ref}$.  As seen previously, the agreement in zero-point
between RAVE and the external sources is better than 1 km~s$^{-1}$.
\begin{table*}
\begin{center}
\caption{External RV samples Compared to RAVE DR5 } 
\begin{tabular}{ p{2.2cm}p{1.15cm}p{1.45cm}p{3.2cm}}
Sample & $N_{obs}$ & $<\Delta RV>$ & $\sigma_{\Delta RV}$~($\sigma_{clip}$, $n_{rej}$ \\ 
\hline
GCS & 1020 & 0.31 & 1.76 (3,113) \\
Chubak & 97 & $-$0.07 & 1.28 (3,2) \\
Ruchti & 443 & 0.79 & 1.79 (3,34) \\
Asiago & 47 & $-$0.22 & 2.98 (3,0) \\
ANU 2.3m & 197 & $-$0.58 & 3.13 (3,16) \\
OHP Elodie & 13 & $-$0.49 & 2.45 (3,2) \\
OHP Sophie & 43 & 0.83 & 1.58 (3,4) \\
APOGEE & 1121 & $-$0.11 & 1.87 (3,144) \\
Gaia-Eso & 106 & $-$0.14 & 1.68 (3,15) \\
%Schlaufman & 84 & 0.55 & 1.68 & (3,15) \\
\hline
\hline
\end{tabular}
\end{center}
\label{externalrvs}
\end{table*}

\section{Stellar Parameters and Abundances}\label{sec:P}
\subsection{Atmospheric parameter determinations} 
\label{subsec:P1}
RAVE DR5 stellar atmospheric parameters, $T_{\rm eff}$, $\log g$ and $\rm [M/H]$ 
have been determined using the same
stellar parameter pipeline as in DR4. The details can be found in
\citet{kordopatis11} and the DR4 paper \citep{kordopatis13}, but a summary is
provided here. 

The pipeline is based on the combination of a decision tree, DEGAS \citep{bijaoui12}, to renormalise 
iteratively the spectra and obtain stellar parameter estimations for the low SNR spectra, and a projection 
algorithm MATISSE \citep{recio-blanco06} to derive the parameters for stars having high SNR. 
The threshold above which MATISSE is preferred to DEGAS is based on tests performed with 
synthetic spectra \citep[see][]{kordopatis11} and has been set to SNR=30~pixel$^{-1}$. 

The learning phase of the pipeline is carried out using synthetic spectra computed with 
the Turbospectrum code \citep{alvarez98} combined with MARCS model atmospheres \citep{gustafsson08} 
assuming local thermodynamic equilibrium (LTE) and hydrostatic equilibrium.  
The cores of the CaT lines are masked in order to avoid issues such as non-LTE 
effects in the observed spectra, which could affect our parameter determination. 

The stellar parameters covered by the grid are between 3000\,K and 8000\,K for $T_{\rm eff}$, 
0 and 5.5 for $\log g$ and $-$5 to +1 dex in metallicity.  Varying $\alpha$-abundances ($\rm [\alpha/Fe]$) 
as a function of metallicity are also included in the learning grid, but are not a free parameter.  
The line-list was calibrated on the Sun and Arcturus \citep{kordopatis11}.  

The pipeline is run on the continuum normalised, radially-velocity corrected 
RAVE spectra using a soft conditional constraint based on the 2MASS $J-K_s$ colours of each 
star.  This restricted the solution space and minimised the spectral degeneracies that exist in the 
wavelength range of the CaT \citep{kordopatis11}. Once a first set of parameters is obtained 
for a given observation, we select pseudo-contrinuum windows to re-normalize
the input spectrum based on the pseudo-continuum shape of the
synthetic spectrum having the parameters determined by the code, and the pipeline is run again on 
the modified input.  This step is repeated ten times, which is usually enough for
convergence of the continuum shape to be reached and hence to obtain a final set of 
parameters (see, however, next paragraph).  

Once the spectra have been parameterised, the pipeline provides one of the four quality flags for 
each spectrum\footnote{The flags are unchanged as compared to DR4.}:  
\begin{itemize}
\item
`0': The analysis was carried out as desired. The re-normalisation process converged, 
as did MATISSE (for high SNR spectra) or DEGAS (for low SNR spectra). 

\item
`1': Although the spectrum has a sufficiently high SNR to use the projection algorithm, the MATISSE 
algorithm did not converge.  Stellar parameters for stars with this flag are not reliable.  Approximately
6\% of stars are affected by this.

\item
`2':  The spectrum has a sufficiently high SNR to use the projection algorithm, but MATISSE 
oscillates between two solutions. The reported parameters are the mean of these two solutions. 
In general the oscillation happens for a set of parameters that are nearby in parameter space and 
computing the mean is a sensible thing to do.  However, this is not always the case,
for example, if the spectrum contains artefacts. Then the mean may not provide accurate
stellar parameters. Spectra with a flag of `2' could be used for analyses, but with caution.  

\item `3': MATISSE gives a solution that is extrapolated from the parameter
range of the learning grid, and the solution is forced to be the one from
DEGAS.  For spectra having artefacts but high-SNR overall, this is a sensible
thing to do, as DEGAS is less sensitive to such discrepancies.  However, for
the few hot stars that have been observed by RAVE, adopting this approach is
not correct.  A flag of `3' and a $T_{\rm eff} >$ 7750\,K is very likely to
indicate that this is a hot star with $T_{\rm eff} >$ 8000\,K and hence that
the parameters associated with that spectrum are not reliable. 
 
 \item
  `4': This flag will only appear for low SNR stars.  For metal-poor giants, the spectral 
  lines available are neither strong enough nor numerous enough to have DEGAS successfully parametrise the star.
Tests on synthetic spectra have shown that to derive reliable parameters the settings used to explore the
branches of the decision tree need to be changed compared to the parameters
adopted for the rest of the parameter space.  A flag `4' therefore marks this change in the setting for
book-keeping purposes, and the spectra associated with this flag should be safe
for any analysis.  
  
\end{itemize}

The several tests performed for DR4 as well as the sub-sequent science
papers, have indicated that the stellar parameter pipeline is globally robust
and reliable. However, being based on synthetic spectra that may not match
the real stellar spectra over the entire parameter range, the direct outputs
of the pipeline need to be calibrated on reference stars in order to minimise
possible offsets. 
\subsection{Metallicity calibrations}

\begin{figure}[htb]  
\includegraphics[width=\linewidth]{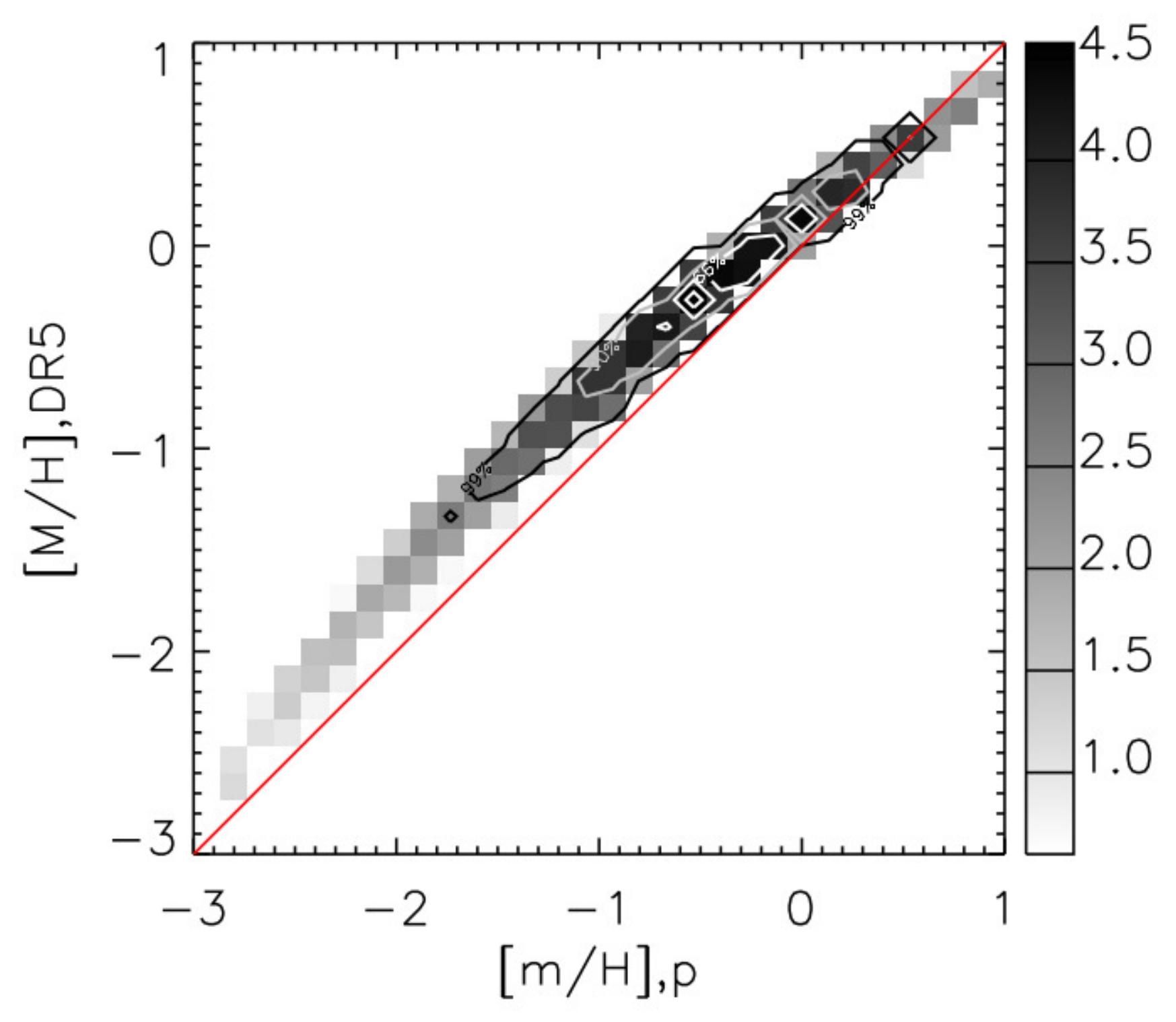}\\
\includegraphics[width=\linewidth]{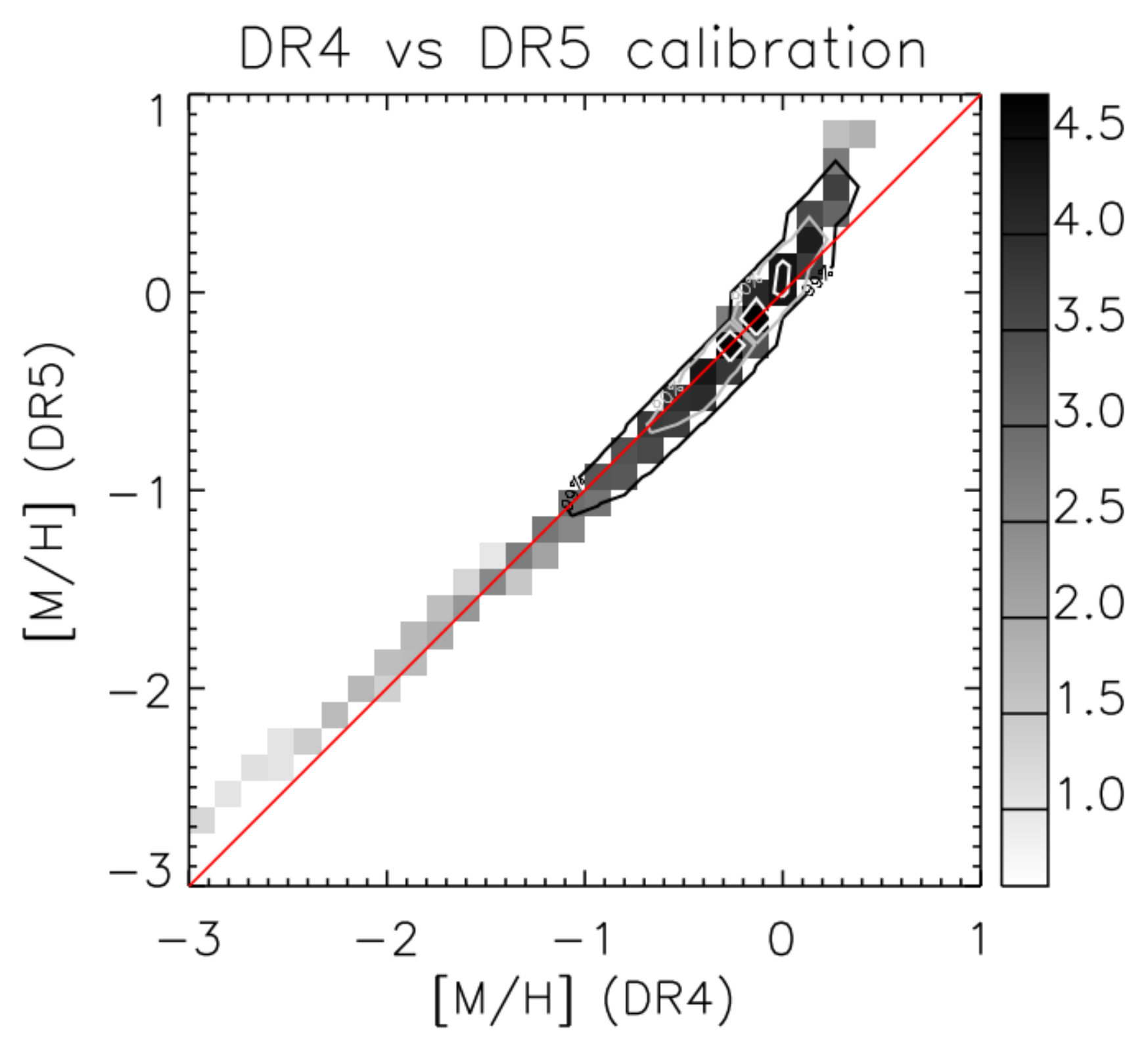}
\caption{Top: The calibrated DR5 $\rm [M/H]$ is compared to the uncalibrated
DR5 $\rm [M/H]_{\rm p}$.  Bottom: A comparison of $\rm [M/H]$ from DR5 with
$\rm [M/H]$ from DR4. The changes occur mostly in the metal-rich end, as our
reference sample now contains more high-metallicity stars.  The grey scale
bar indicates the $\log_{10}(N)$ of stars in a bin, and the contour lines
contain 33, 66, 90 and 99 per cent of the sample.
\label{fig:calibration_metallicity}}
\end{figure}

In DR4, metallicity calibration proved to be the most critical and important
one.  Using a set of reference stars for which metallicity determinations
were available in the literature (usually derived from high-resolution
spectra), a second order polynomial correction, based on  surface gravity
and raw metallicity, was applied in DR4.  This corrected the
metallicity offsets with the external datasets of \citet{pasquini04,
pancino10, cayrel04, ruchti11} and the PASTEL database \citep{soubiran10}.
For DR5, we relied on the same approach.  However, we added reference stars
to the set used in DR4, with the focus on expanding our calibrating sample
towards the high metallicity end to better calibrate the tails of the
distribution function.  This calibration is based on the crossmatch of RAVE
targets with the \citet{worley12} and \citet{adibekyan13} catalogues, as well as
the Gaia benchmark stellar spectra.  The
metallicity of the Gaia benchmark stars is taken from \citet{jofre14}, where a 
library of Gaia benchmark stellar spectra was specially prepared to match 
RAVE data in terms of wavelength coverage, resolution and spectral spacing. 
This was done following the procedure described in \citet{blancocuaresma14}. 
Our calibration has already been successfully used in
\citet{kordopatis15, wojno16} and Antoja et al. (submitted).  The calibration
relation for DR5 is: 

\begin{equation}
\begin{split}
\rm [M/H]= [M/H]_{\rm p} - (-0.276 + 0.044~{log~{\it g_{\rm p}}}\\
 -0.002~{ \rm log~{\it g_{\rm p}}^2} +0.248~{\rm [M/H]}_{\rm p}\\
 -0.007~{ \rm [M/H]_{\rm p}}~{ \rm log~{\it g_{\rm p}}}+ 0.078~{
 \rm [M/H]_{\rm p}^2)},
\end{split}
\end{equation}
where $\rm [M/H]$ is the calibrated metallicity, $\rm [M/H]_{\rm p}$ and $\log g_{\rm p}$
are, respectively, the uncalibrated (raw output from the pipeline)
metallicity and surface gravity.  The effect of the calibration on the raw
output can be seen in the top panel of
Fig.~\ref{fig:calibration_metallicity}. The bottom panel shows that in the
range $(-2,0)$ the DR5 and DR4 values are very similar.  Above $\rm
[M/H]\sim$ 0, the DR5 metallicities are higher than the DR4 ones and are in
better agreement with the chemical abundance pipeline presented below
(\S\ref{chemicalpipeline}).  We note that after metallicity calibration we do 
not re-run the pipeline to see if other stellar parameters change with this new metallicity.

\subsection{Surface gravity calibrations}

Measuring the surface gravity spectroscopically, and in particular from
medium resolution spectra around the IR calcium triplet, is challenging.
Nevertheless, the DR4 pipeline proved to perform in a relatively reliable
manner, so no calibrations was performed on $\log g_{\rm p}$.  The
uncertainties in the DR4 $\log g_{\rm p}$ values are of the order of $\sim
0.2-0.3\,$dex, with any offsets being mainly confined to the
giant stars.  In particular, an offset in $\log g_{\rm p}$ of $\sim 0.15$ was
detected for the red clump stars. 

For the main DR5 catalogue, the surface gravities are calibrated using both
the asteroseismic $\log g$ values of 72 giants from V16 and
the Gaia benchmark dwarfs and giants \citep{heiter15}.
Although the calibration presented in V16 focuses only on
giant stars and should therefore perform better for these stars (see
\S\ref{aseis}), the global DR5 $\log g$ calibration is valid for all stars
for which the stellar parameter pipeline provides $T_{\rm eff}$, $\log g$ and
$\rm [M/H]$. 

Biases in $\log g_{\rm p}$ depended mostly on $\log g_{\rm p}$, so for the surface
gravity calibration, we computed the offset between the pipeline output and
the reference values, as a function of the pipeline output, and a low-order 
polynomial fitted to the residuals (for a more quantitative assessment, see V16). 
This quadratic expression defines our
surface gravity calibration:
\begin{equation}
\rm log~{\it g_{\rm DR5}}=\rm log~{\it g_{\rm p}} - (-0.515+0.026~{log~{\it g_{\rm p}}} +0.023~{log~{\it g_{\rm p}}^2}).
\end{equation}

The calibration above affects mostly the giants but also allows a smooth
transition of the calibration for the dwarfs.  The red clump is now at $\log
g\sim2.5\,$dex, consistent with isochrones for thin disk stars of metallicity
$\rm [M/H]=-$0.1 and age of 7.5 Gyr (see
Sect.~\ref{sec:summary_calibrations}).  This calibration has the effect of
increasing the minimum published $\log g$ from 0 (as set by the learning
grid) to $\sim0.5$.  The maximum reachable $\log g$ is $\sim5.2$ (instead of
5.5, as in DR4).  Tests carried out with the Galaxia model \citep{sharma11}, where the RAVE
selection function has been applied (W16) show that the
calibration improves $\log g$ even at these boundaries.  We do caution,
however, that special care should be taken for stars with $\log
g\lesssim0.75$ or $\log g\gtrsim5$. 

\begin{figure}
\includegraphics[width=\linewidth]{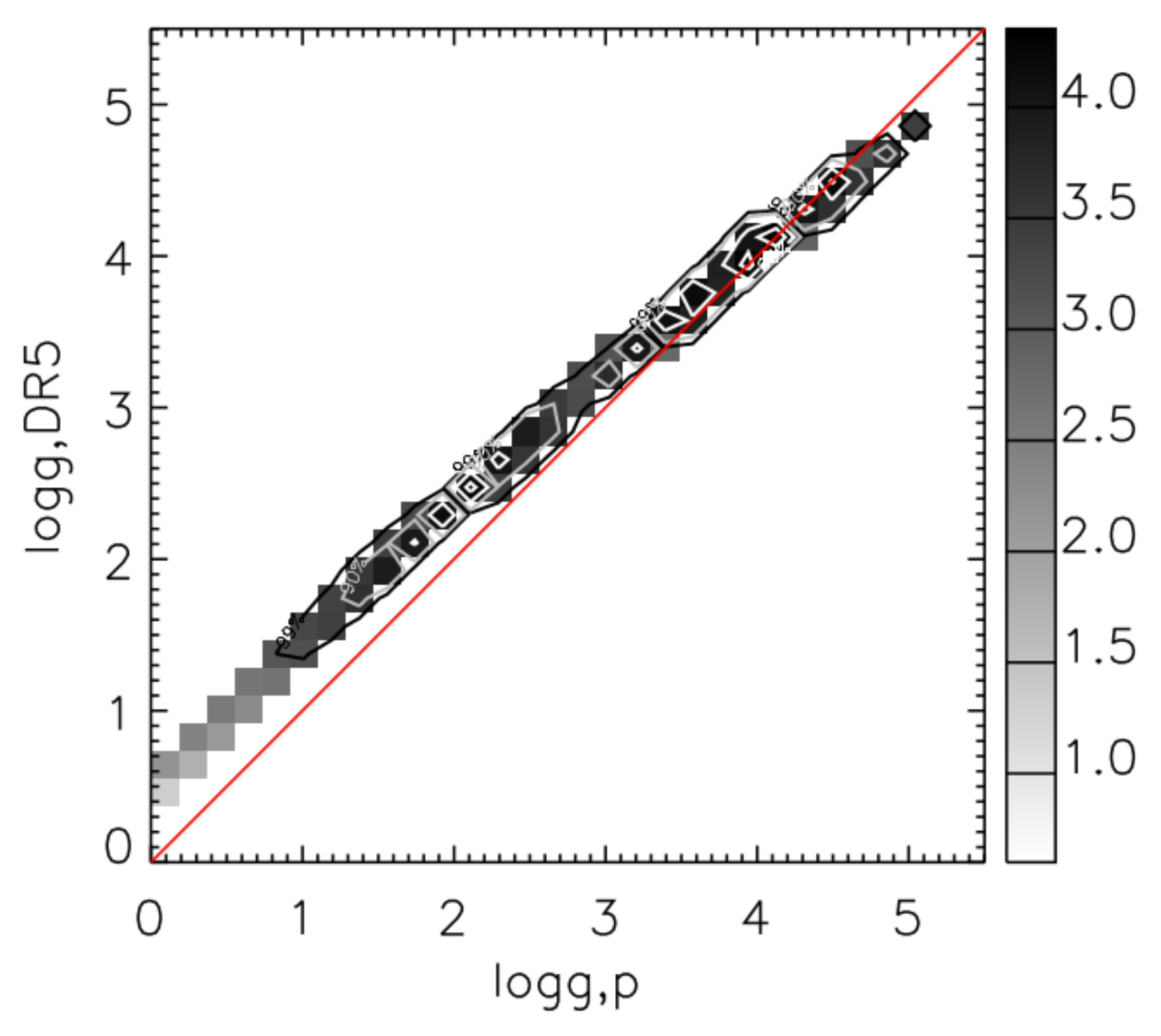} 
\caption{As
Figure~\ref{fig:calibration_metallicity} except it compares the calibrated
DR5 $\log g$ with the uncalibrated DR5 $\log g_{\rm p}$.
Contours as in Figure~\ref{fig:calibration_metallicity}.
\label{fig:calibration_gravity}}
\end{figure}

\subsection{Effective temperature calibrations} 

\begin{figure}[htb]  
\includegraphics[width=\linewidth]{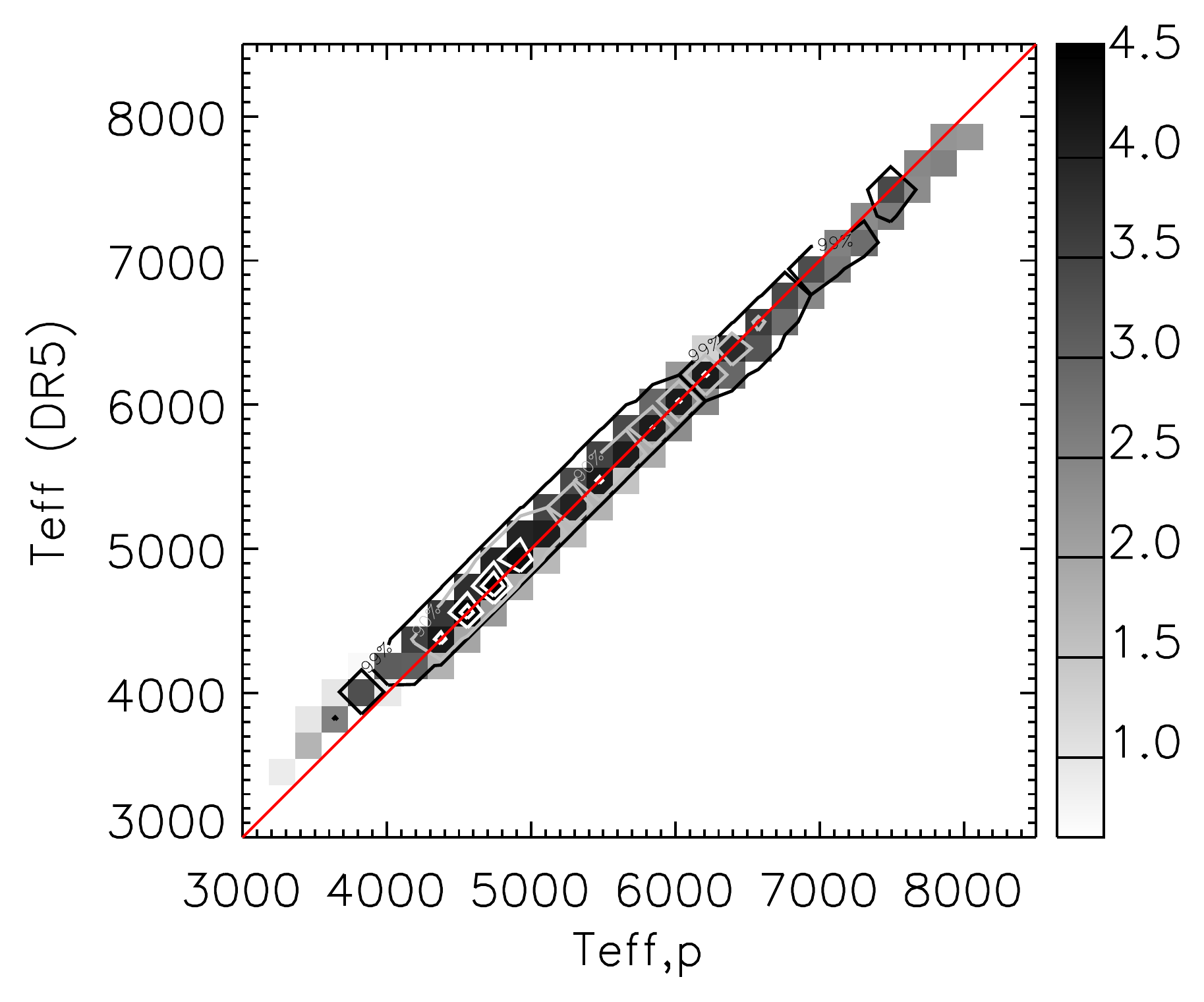}
\caption{As Figure~\ref{fig:calibration_metallicity} except it compares 
the calibrated DR5 $T_{\rm eff}$ with the uncalibrated DR5 $T_{\rm eff,p}$. 
Contours as in Figure~\ref{fig:calibration_metallicity}.
\label{fig:calibration_Teff}}
\end{figure}

\citet{munari14} showed that the DR4 effective temperatures for warm stars
($T_{\rm eff}\gtrsim 6000$\,K) are under-estimated by $\sim 250$\,K.  This
offset is evident when plotting the residuals against the reference
(photometric) $T_{\rm eff}$, but is barely discernible when plotting them
against the pipeline $T_{\rm eff}$. Consequently, it is difficult to
correct for this effect.
The calibration that we carry out changes $T_{\rm eff,p}$
only modestly, and does not fully
compensate for the (fortunately small) offsets (see
Fig.~\ref{fig:calibration_Teff}).  The adopted calibration for effective
temperatures is
\begin{equation}
\rm T_{\rm eff, DR5}=T_{\rm eff,p} + (285-0.073{T_{\rm eff,p}}+40{\log{g_{\rm p}}}).
\end{equation}

%%%
%%%
\subsection{Summary of the calibrations}
\label{sec:summary_calibrations}

\begin{figure*}[htb]  
\includegraphics[width=\linewidth]{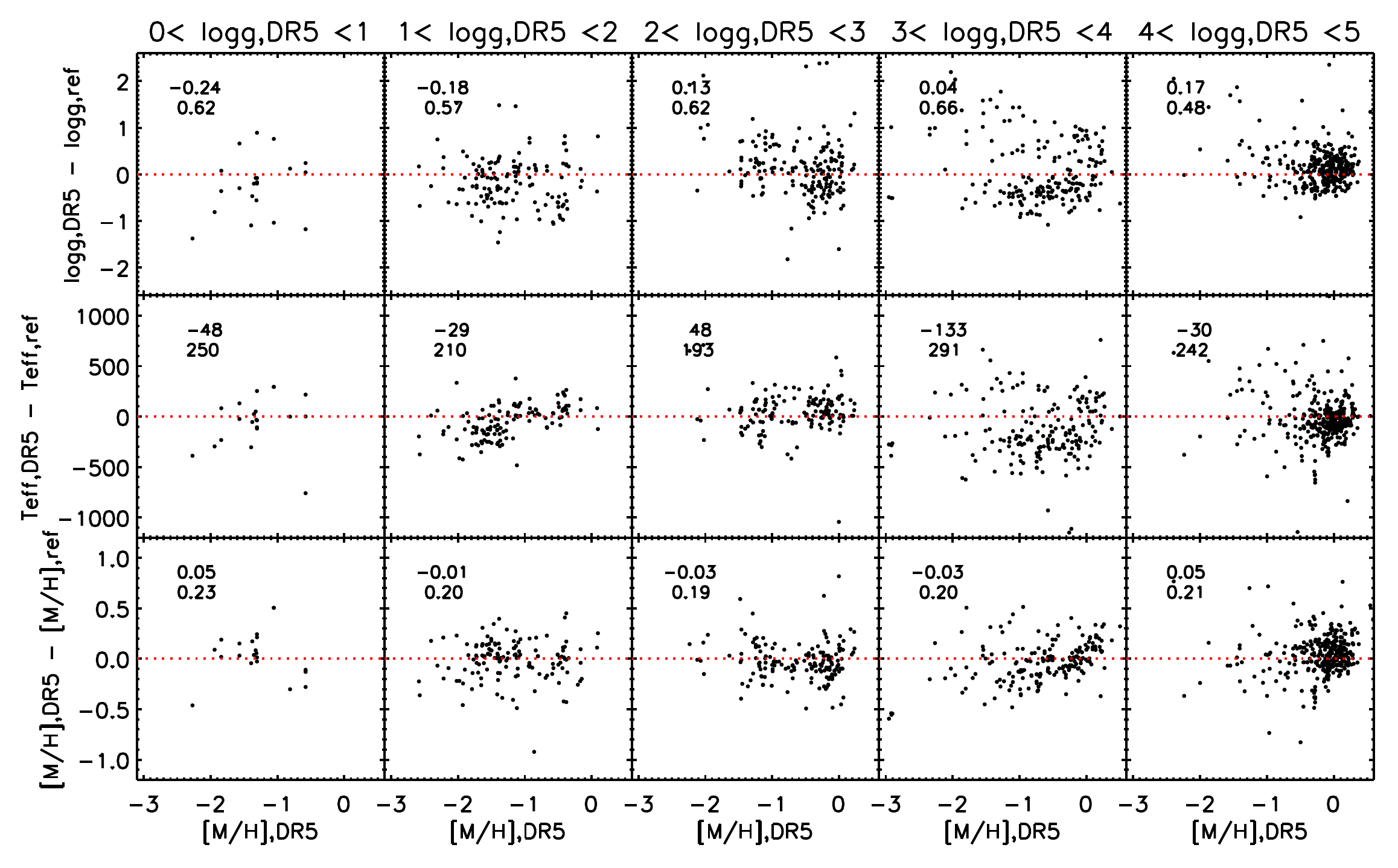}
\caption{Residuals between the calibrated DR5 parameters and the reference values, as a function of the calibrated DR5 metallicity, for different calibrated DR5 logg bins. The numbers inside each panel indicate the mean difference (first line) and the dispersion (second line) for each considered subsample.
\label{fig:calibration_reference_logg}}
\end{figure*}

\begin{figure*}[htb]  
\includegraphics[width=\linewidth]{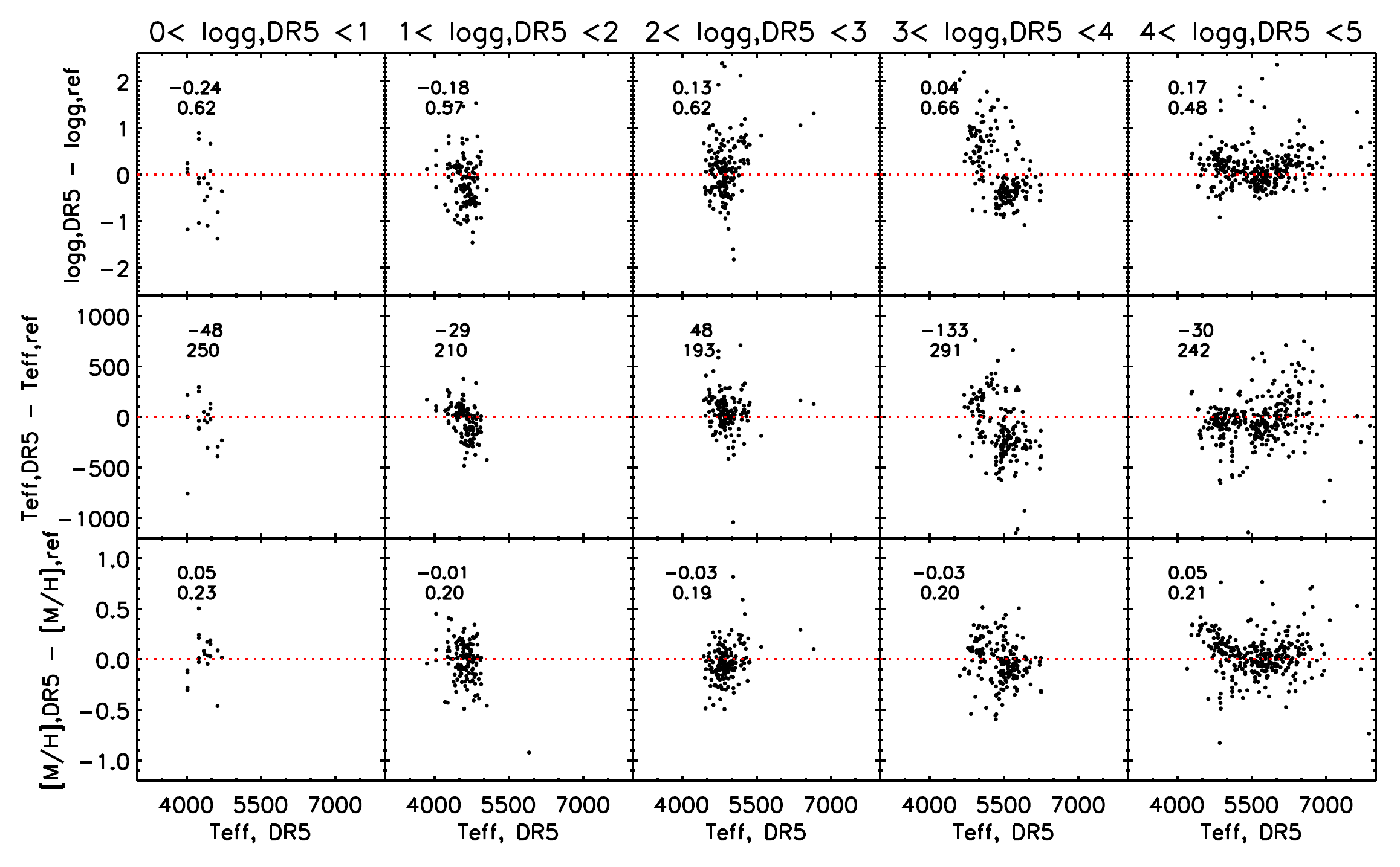}
\caption{Same as Fig.~\ref{fig:calibration_reference_logg}, but showing on the $x$-axis the calibrated DR5 $T_{\rm eff}$.
\label{fig:calibration_reference_teff}}
\end{figure*}

Figures~\ref{fig:calibration_reference_logg} and
\ref{fig:calibration_reference_teff} show, as functions of metallicity and
effective temperature, respectively, the residuals between the calibrated
values and the set of reference stars that have been used. We show the $\log
g$ comparison (first rows of Figs.~\ref{fig:calibration_reference_logg},
\ref{fig:calibration_reference_teff}), for all sets of stars, and not only
the  stars in V16 and \citet{jofre14}, which in the end were the
only samples used to define the calibration.  Although the V16 and \citet{jofre14}
derivations of $\log g$ are independent of each other, the shifts in $\log~g$ between the two
samples are small, so there is no concern that we could end up with non-physical 
combinations of parameters. 

Overall there are no obvious trends as a function of any stellar parameter, except the already mentioned 
mild trend in $T_{\rm eff}$ for the stars having $4<\log g<$5 (seen at the middle row, 
last column of Fig.~\ref{fig:calibration_reference_teff}).  The absence of any strong bias in the 
parameters is also confirmed in the next sections, with additional comparisons with 
APOGEE, Gaia-ESO and LAMOST stars (\S~\ref{sec:EV}).

\begin{figure}
\includegraphics[width=\linewidth]{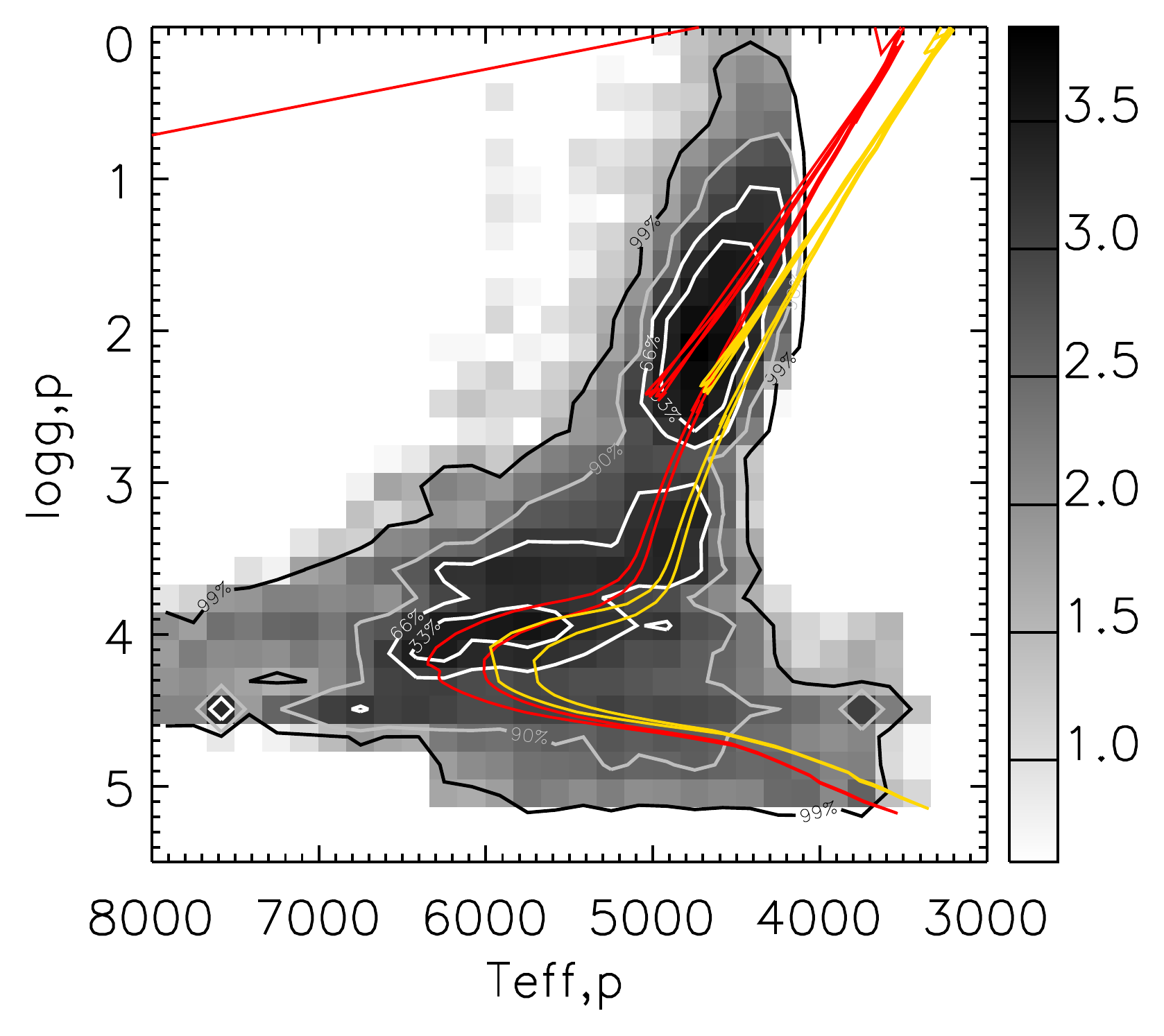}\\
\includegraphics[width=\linewidth]{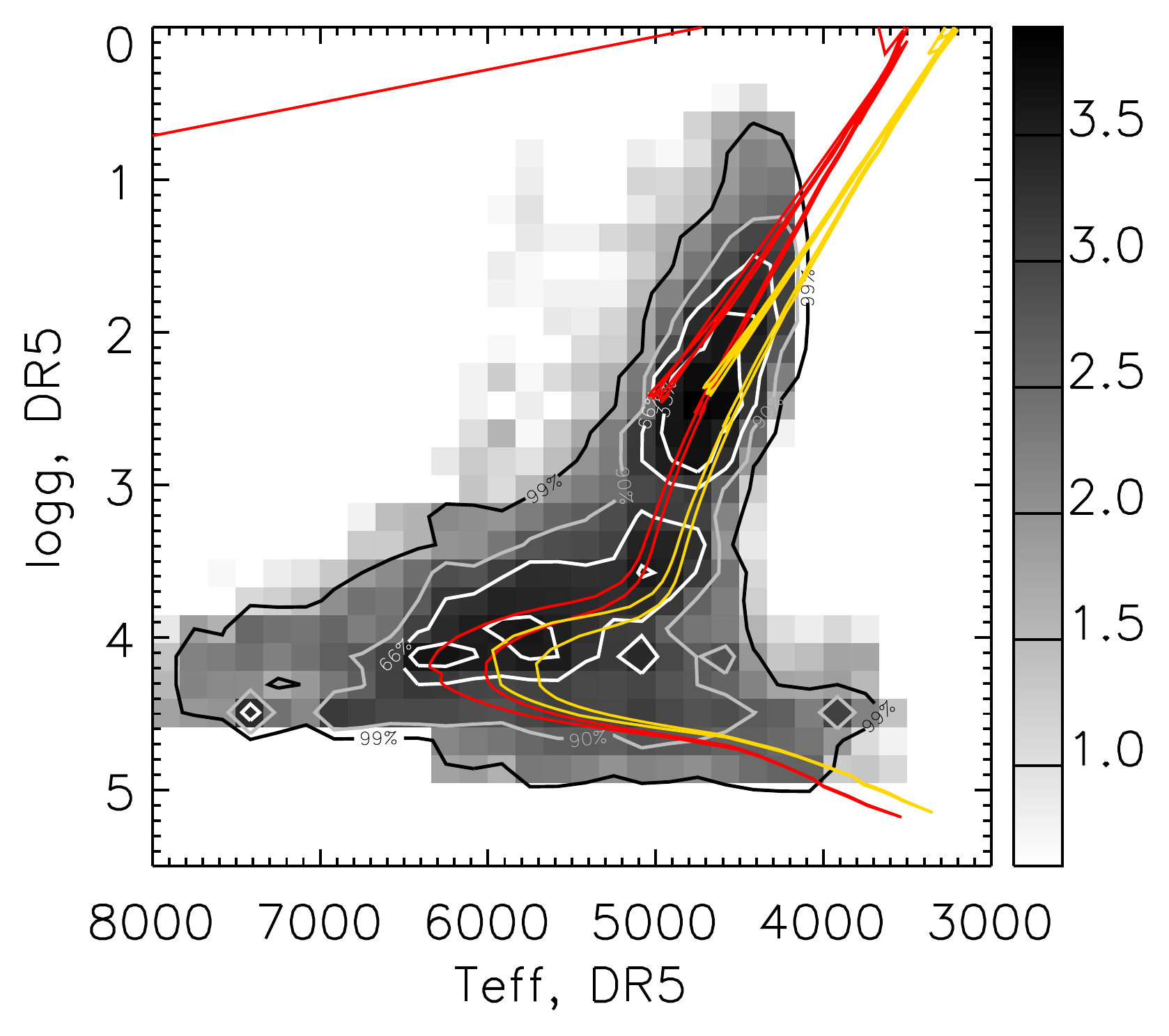}
\caption{Top: $T_{\rm eff}$-$\log g$ diagram for the raw output of the pipeline, 
$i.e.$, before calibration. Bottom: $T_{\rm eff}$-$\log g$ diagram for the calibrated DR5 parameters. 
Both plots show in red two Padova isochrones at metallicity $-$0.5 and ages 7.5 and 12.5 Gyr,
and in yellow two Padova isochrones at metallicity $-$0.1 and ages 7.5 and 12.5 Gyr.  
For the new calibration, the locus of the red-clump agrees better with 
stellar evolution models, as does the position of the turn-off.
\label{fig:calibration_HRD}}
\end{figure}

The effect of the calibrations on the $(T_{\rm eff},\log g)$ diagram is shown
in Fig.~\ref{fig:calibration_HRD}.  The calibrations
bring the distribution of stars into better agreement with the predictions of
isochrones for the old thin disk and thick disk (yellow and red, respectively).
%%
%%%
\subsection{Estimation of the atmospheric parameter errors and robustness of the pipeline}
Using the error spectrum of each observation, 10 re-sampled spectra were
computed for the entire database (see also \S\ref{errspec}). The SPARV algorithm was run on these spectra, 
the radial velocity estimated and the spectra shifted to  the rest-frame. 
Subsequently, the \citet{kordopatis13} pipeline was run on these radial velocity-corrected spectra.
 
The dispersion of the derived parameters among the re-sampled spectra of each
observation give us an indication of the individual errors on $T_{\rm eff}$,
$\log g$ and $\rm [M/H]$ and of the robustness of the pipeline. That said,
because the noise is being introduced twice (once during the initial
observation and once when re-sampling), the results should be considered as
an over-estimation of the errors (since we are dealing with an overall lower
SNR). 

\begin{figure}
\includegraphics[width=\linewidth]{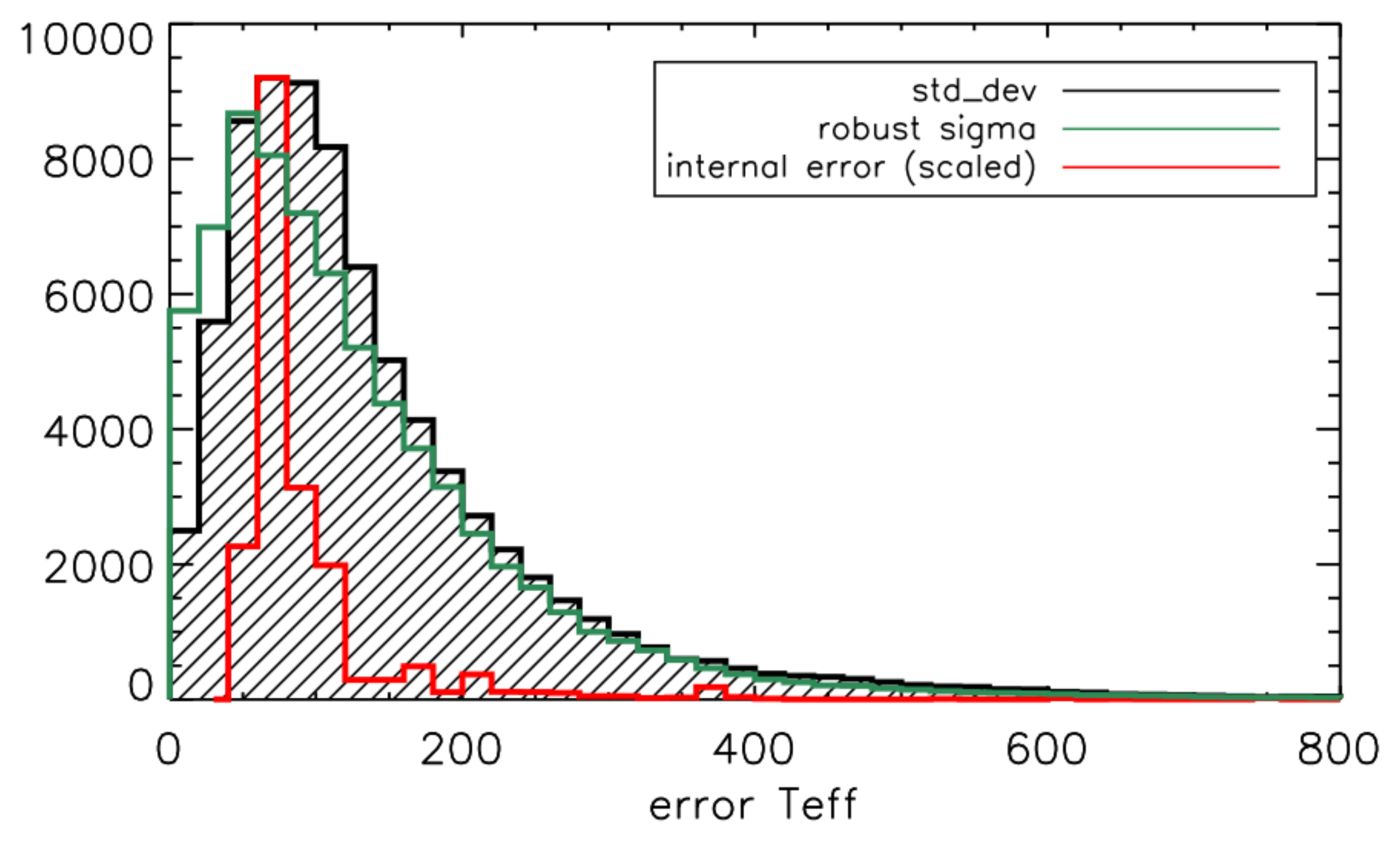}\\
\includegraphics[width=\linewidth]{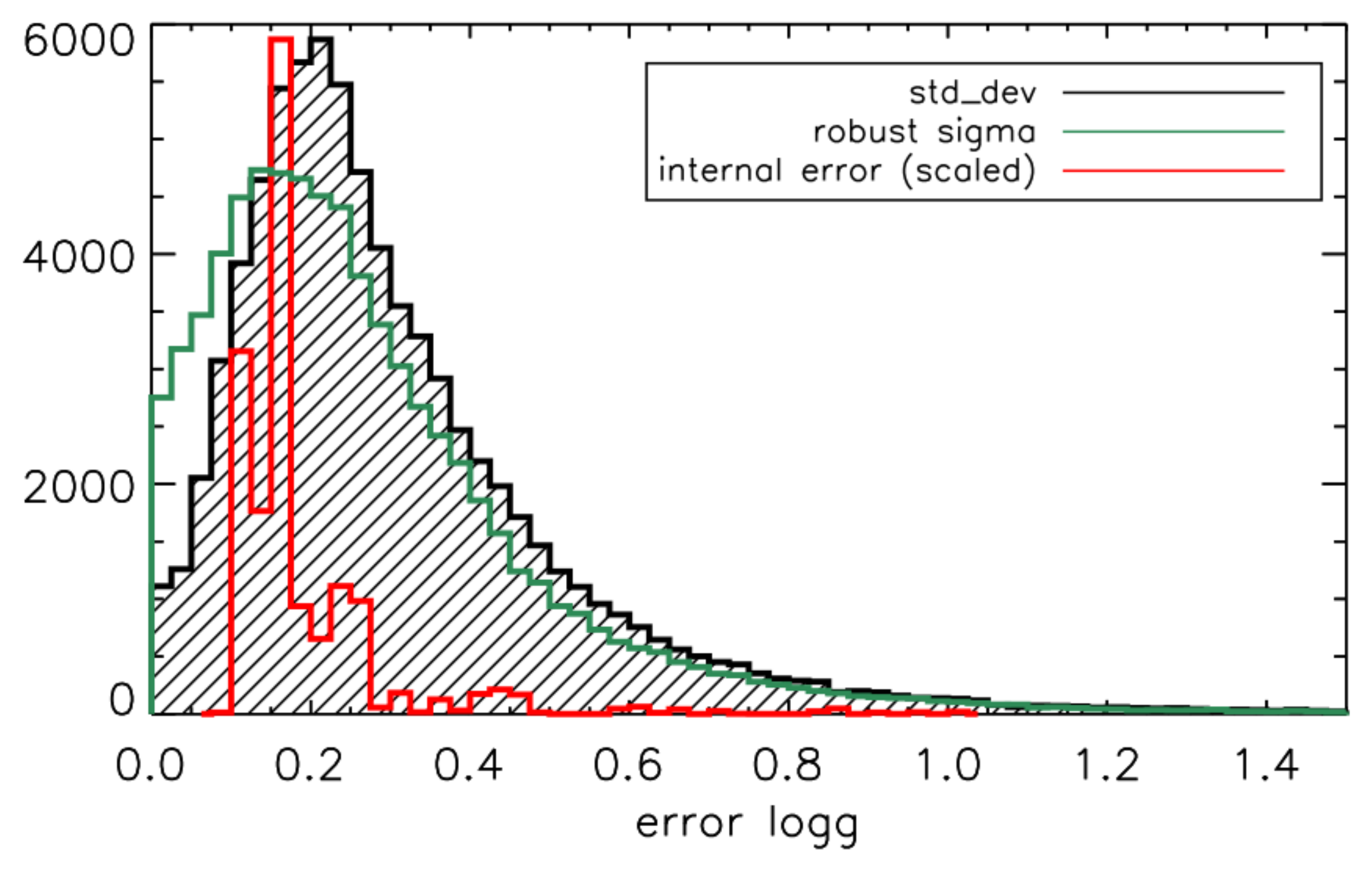}\\
\includegraphics[width=\linewidth]{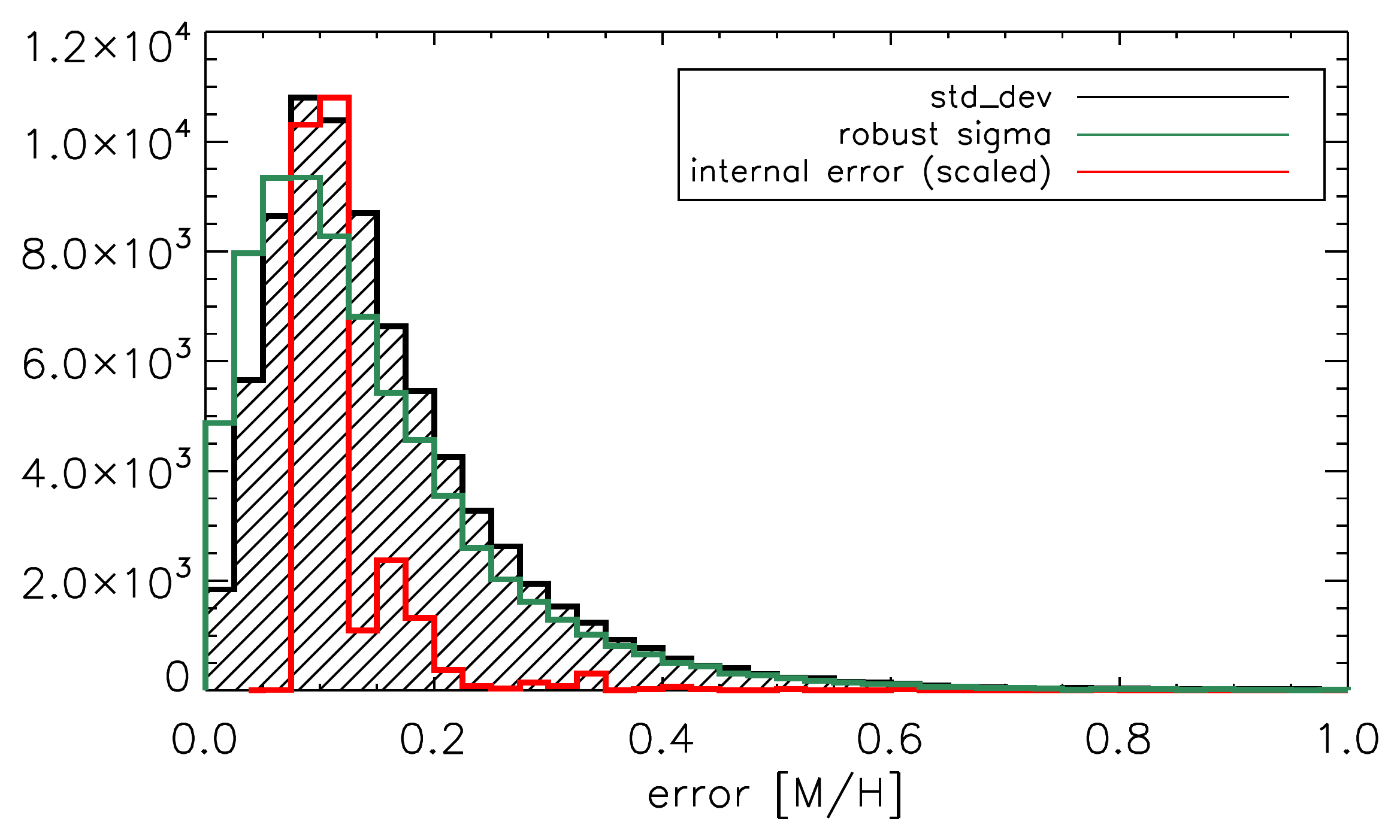}
\caption{Histograms of the errors in the uncalibrated 
parameters (top: $T_{\rm eff}$, middle: $\log g$, bottom: $\rm [M/H]_{\rm p}$), 
obtained from the analysis of all the spectra gathered in 2006, 
resampled ten times using their associated error spectra. 
The filled black histograms show the standard deviation distributions 
whereas green histograms show the MAD estimator distribution. 
The red histograms are normalised to peak of the standard deviation distribution, 
and show the distributions of the internal errors as estimated by the 
stellar parameter pipeline. 
\label{fig:uncalibrated_parameter_dispersion}}
\end{figure}

Figure~\ref{fig:uncalibrated_parameter_dispersion} shows the dispersion of each
parameter determined from the spectra collected in 2006. We show both the
simple standard deviation, and the Median Absolute Deviation (MAD) estimator,
which is more robust to outliers.  The distribution of the internal errors
(normalised to the peak of the black histogram) as given in Tables~1 and 2 of
\citet{kordopatis13} is also plotted.
Figure~\ref{fig:uncalibrated_parameter_dispersion} shows that the internal
errors are consistent with the parameter dispersion we obtain from the
re-sampled spectra, though the uncertainties calculated from the error
spectra have a tail extending to larger error values.  Therefore, for some
stars, the true errors are considerably larger than those produced by the
pipeline.  This is not unexpected, as it reflects the degeneracies that
hamper the IR CaT region, and also the fact that the resampled spectra have 
a lower SNR than the true observations, since the noise is introduced a second time.  

\begin{figure}  
\includegraphics[width=\linewidth]{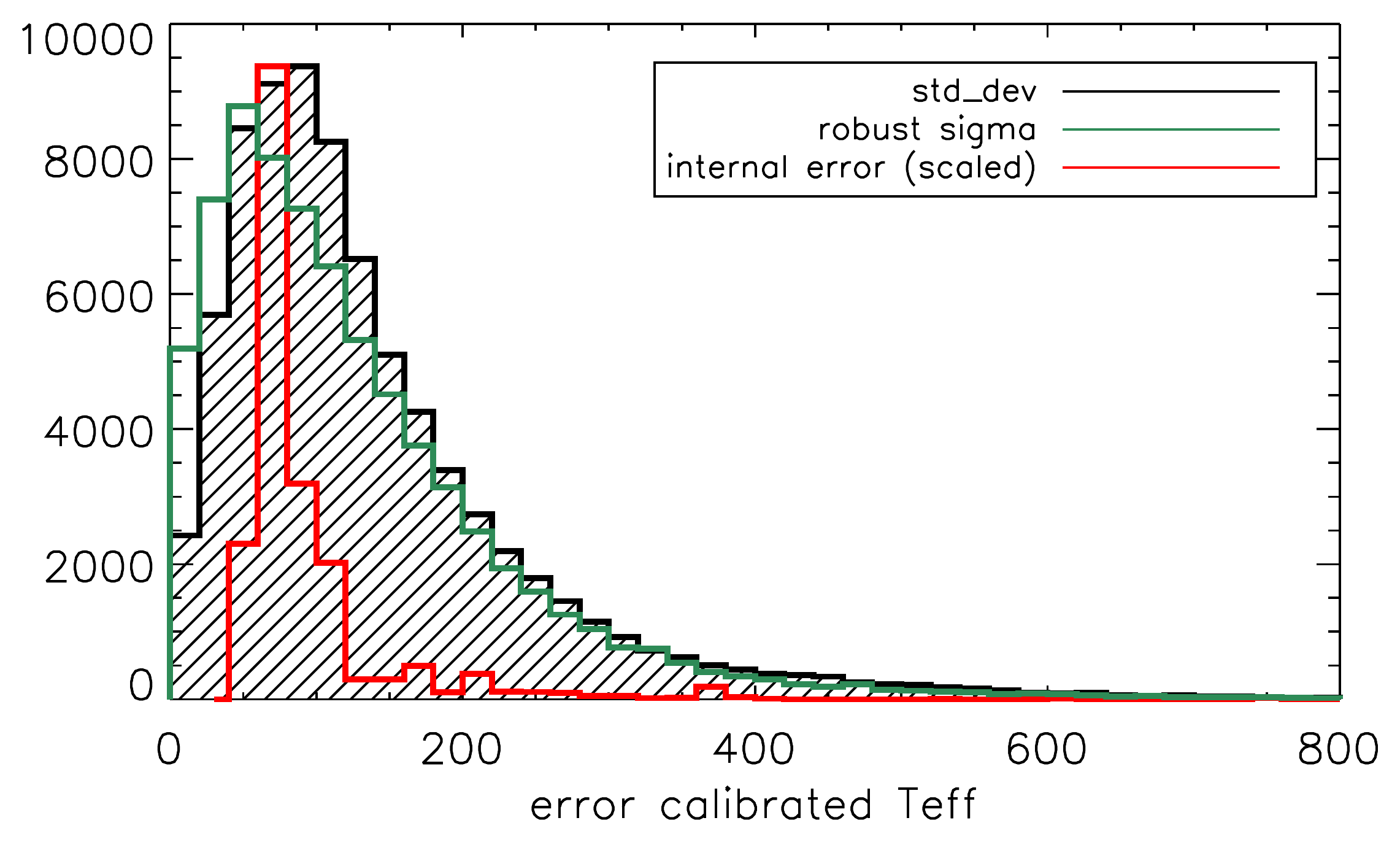}\\
\includegraphics[width=\linewidth]{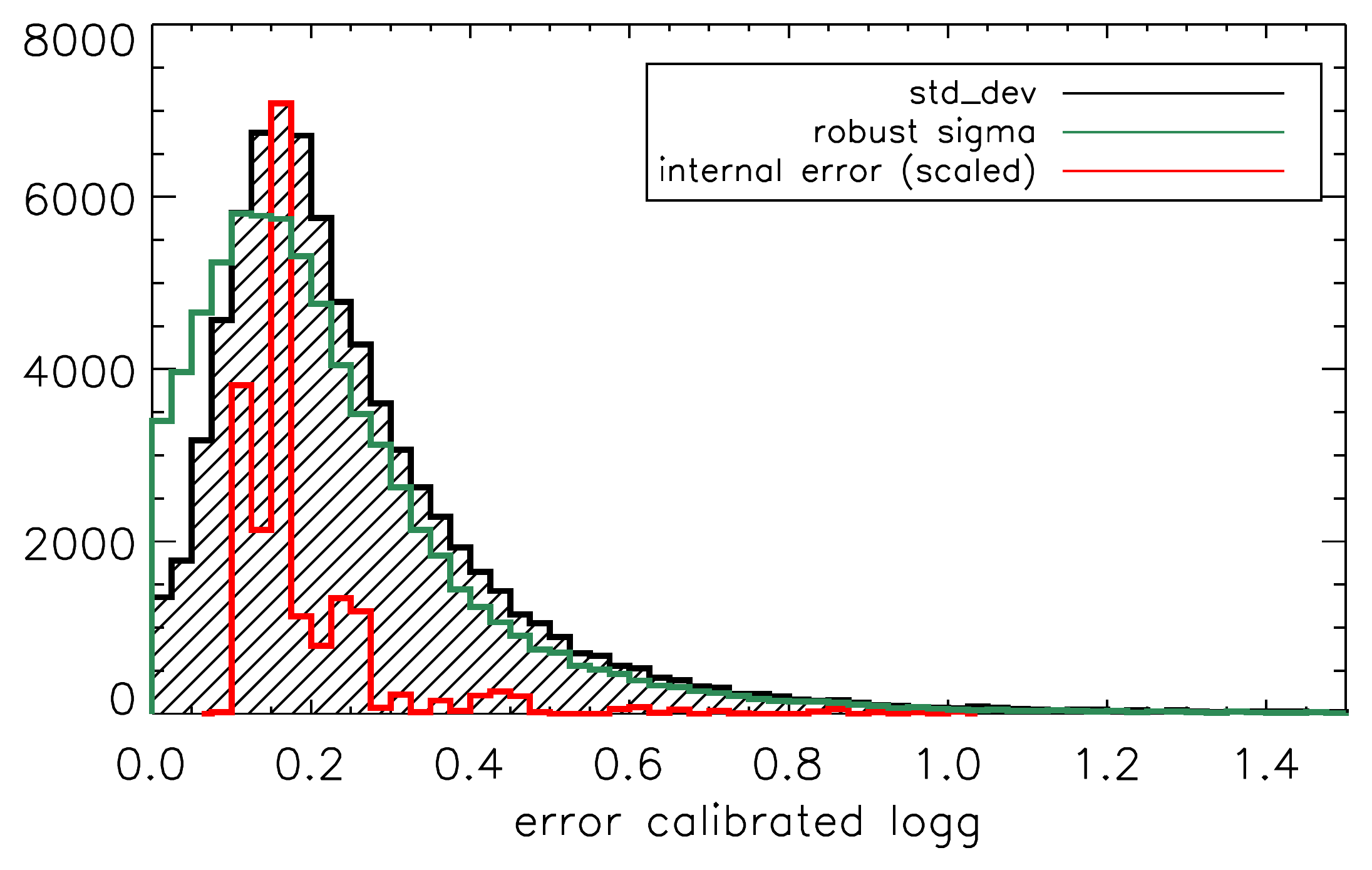}\\
\includegraphics[width=\linewidth]{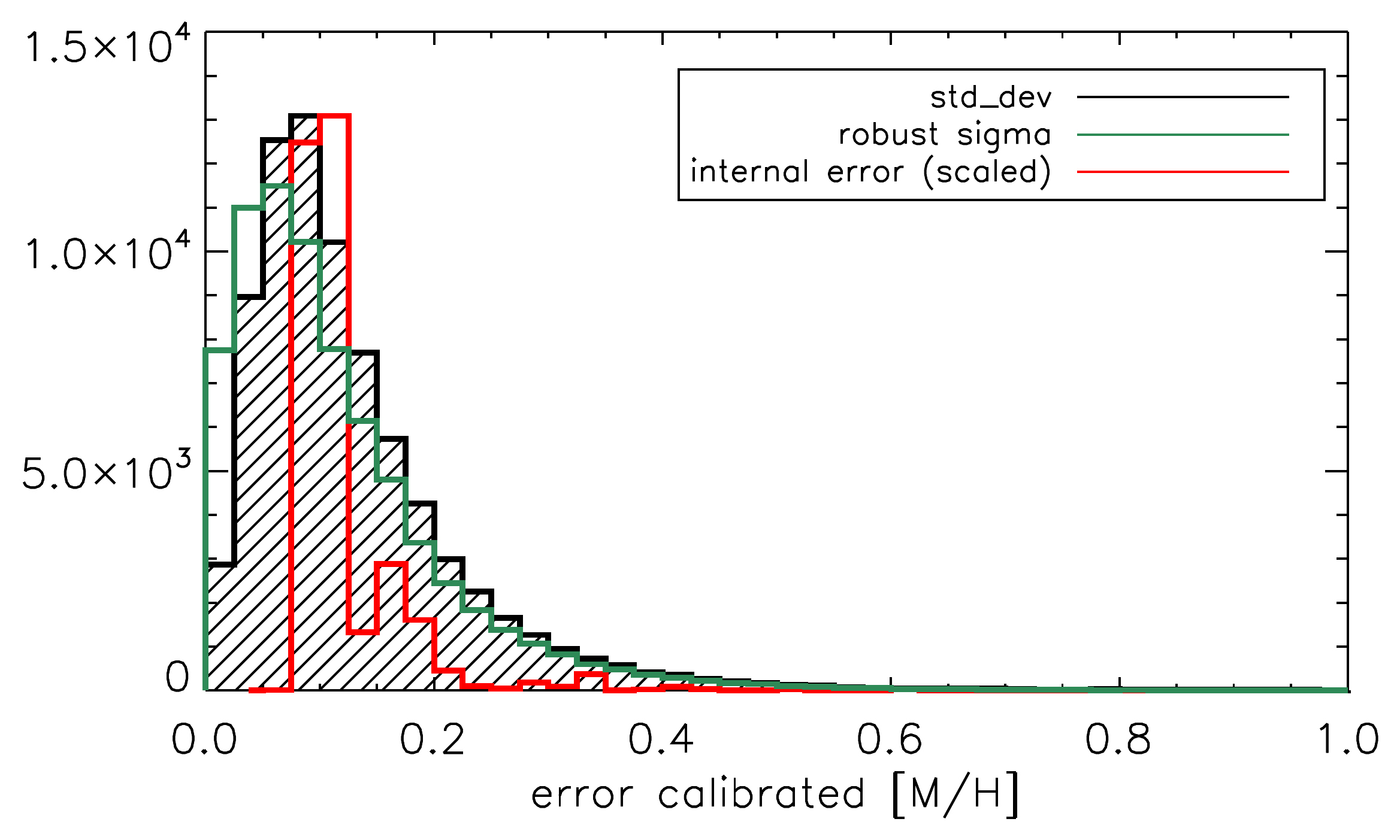}
\caption{Same as Fig.~\ref{fig:uncalibrated_parameter_dispersion} but showing the 
error histograms for the calibrated DR5 parameters. 
\label{fig:calibrated_parameter_dispersion}}
\end{figure}

The published DR5 parameters, however, are not the raw output of the pipeline, but are calibrated values. 
Since this calibration takes into 
account the output $T_{\rm eff}$, $\log g$ and $\rm [M/H]$, it is also valuable to test the dispersion of 
the calibrated values.  This is shown in Fig.~\ref{fig:calibrated_parameter_dispersion} for the 
same set of stars.  As before, no large differences are introduced, indicative again 
of a valid calibration and reliable stellar parameter pipeline. 

\subsection{Completeness of Stellar Parameters} 
\begin{figure}  
\includegraphics[width=\linewidth]{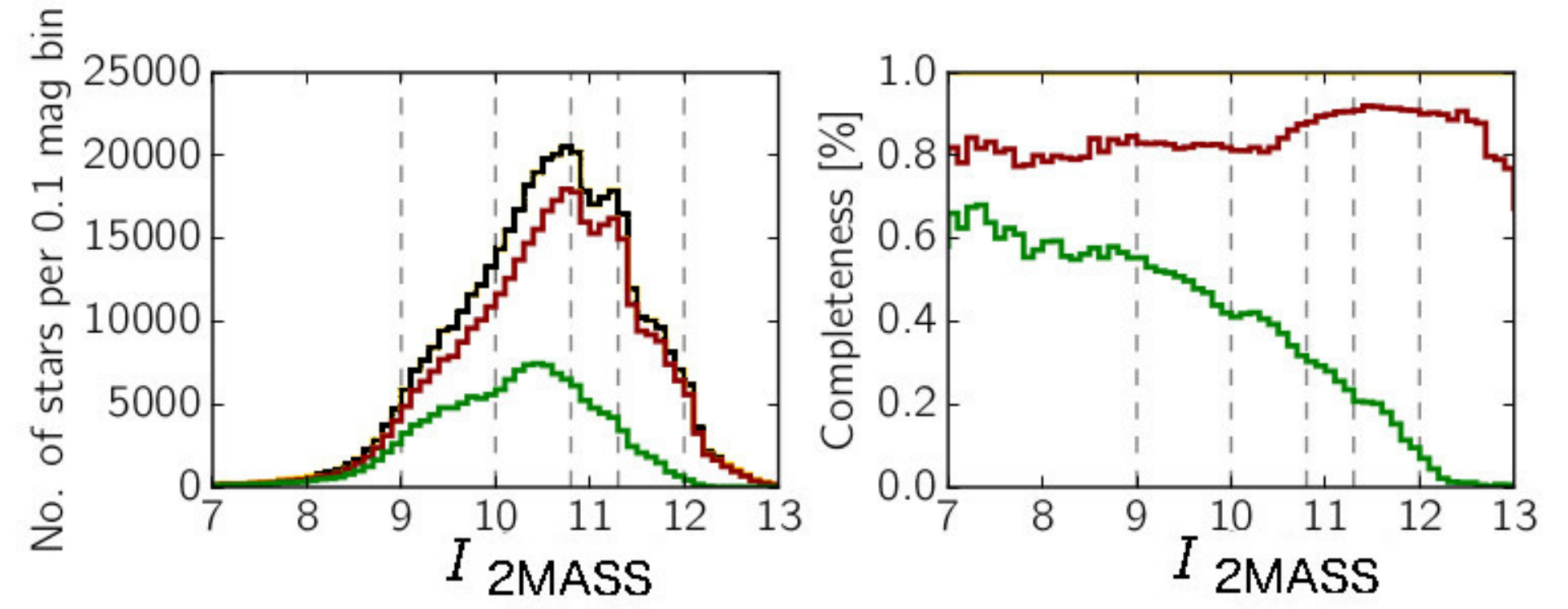}\\
\includegraphics[width=\linewidth]{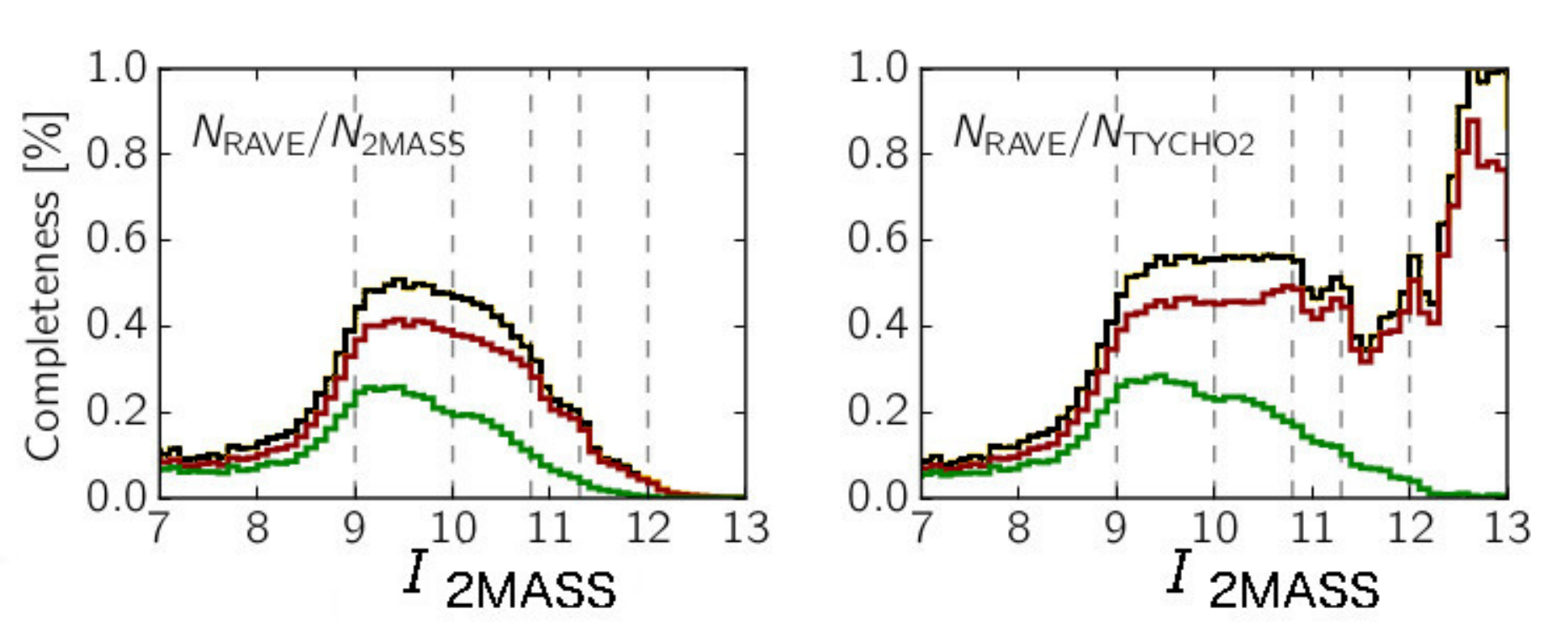}
\caption{{\it Top Left panel:}  The number of RAVE stars with 
spectral parameters (black), distances (red) and chemical abundances (green)
as a function of magnitude.  {\it Top Right panel:}  The completeness of the RAVE DR5 
sample is shown as a function of magnitude for stars with spectral parameters, 
distances and chemical abundances.  {\it Bottom left panel:}  The completeness 
of the RAVE DR5 sample with respect to the completeness of 2MASS is shown 
as a function of magnitude for stars with spectral parameters, distances and 
chemical abundances. {\it Bottom right panel:}  The same as the third panel, but for TYCHO2. 
\label{magcompleteness}}
\end{figure}

It is of value to consider
the completeness of DR5 with respect to derived stellar parameters.  To
evaluate this, the stars that satisfy the following criteria are selected:
SNR $\ge 20$, $|${\tt correctionRV}$| < 10\kms$, $\sigma(RV) <8\kms$ , and
{\tt correlationCoeff} $>10$ \citep[see][]{kordopatis13}.  The resulting
distributions are shown in Figure~\ref{magcompleteness}.  Whereas the $10.0 <
I_{\rm 2MASS} < 10.8$ magnitude bin has the highest number of stars with
spectral parameters, distances, and chemical abundances, the fractional
completeness compared to 2MASS (panel 3) peaks in the $9.0 < I_{\rm 2MASS} <
10.0$ magnitude bin. In this bin, we find that we determine stellar
parameters for approximately 50\% of 2MASS stars in the RAVE fields.  We further
estimate distances for 40\% of stars, and chemical abundances for $\sim
20\%$.  This fraction drops off significantly at fainter magnitudes.

Similarly, for the brighter bins we obtain stellar parameters for $\sim 55
\%$ of Tycho-2 stars, distances for $\sim 45 \%$ of stars and similar trends
in the completeness fraction of chemical abundances.

\section{External Verification}\label{sec:EV}
%\label{external}
Stars observed specifically for understanding the stellar parameters of RAVE, as well as stars 
observed that fortuitously overlap with high-resolution studies are compiled to further asses
the validity of the RAVE stellar parameter pipeline.  As discussed above, calibrating the RAVE 
stellar parameter pipeline is not straight-forward, and although a global calibration over the 
diverse RAVE stellar sample has been applied, the accuracy of the atmospheric parameters 
depends also on the stellar population probed.  Therefore, for the specific samples 
investigated in this section, Table~\ref{tab:externalcomps} summarises the results of the 
external comparisons split into hot, metal-poor dwarfs,
hot, metal-rich dwarfs, cool, metal-poor dwarfs, cool, metal-rich dwarfs, 
cool, metal-poor giants and cool, metal-rich giants.  The boundary between ``metal-poor"
and ``metal-rich" occurs at $\rm [M/H]$ = $-$0.5, and between ``hot" and ``cool" lies at
$T_{\rm eff}=5500$\,K.  The giants and dwarfs are divided at $\log g=3.5\,$dex.
From here on, only the calibrated RAVE stellar parameters are used.

\subsection{Cluster Stars}

In the 2011B, 2012 and 2013 RAVE observing semesters, stars in various open
and globular clusters were targeted with the goal of using the cluster stars
as independent checks on the reliability of RAVE stellar parameters and their
errors.  RAVE stars observed within the targeted clusters that have also
been studied externally from high-resolution spectroscopy are compiled, so
a quantitative comparison of the RAVE stellar parameters can be made.  
   
Table~\ref{raveclusters} lists clusters and their properties for which RAVE
observations could be matched to high-resolution studies.  The open cluster
properties come from the Milky Way global survey of star clusters
\citep{kharchenko13} and the globular cluster properties come from the Harris
catalog \citep[][2010 update]{harris96}.  The number of RAVE stars that were cross-matched and the literature
sources are also listed.  

\begin{figure}[htb]  
\includegraphics[width=\linewidth]{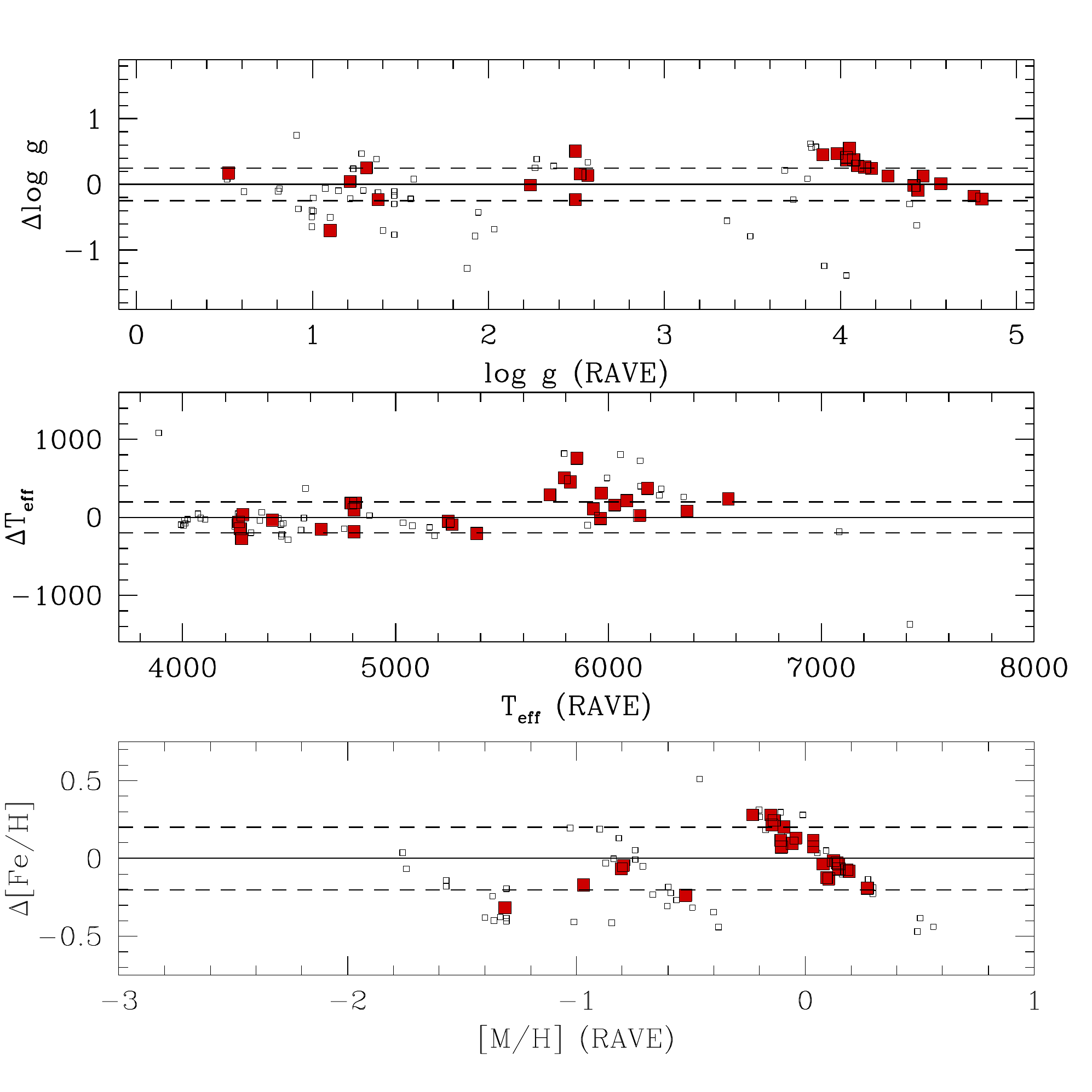}
\caption{A comparison between the stellar parameters presented here with those from cluster stars 
studied in the literature from various different sources (see Table~\ref{raveclusters}).  
The filled squares indicate the stars with ${\tt AlgoConv}=0$.
\label{clusters_comp}}
\end{figure}

Figure~\ref{clusters_comp} shows a comparison between the high-resolution
cluster studies and the RAVE cluster stars.  From this inhomogeneous sample
of 75 overlap RAVE cluster stars with an  ${\tt AlgoConv}\neq1$, the formal
uncertainties in $T_{\rm eff}$, $\log g$, and $\rm [M/H]$ are 300\,K,
0.6\,dex, and 0.04\,dex, respectively, but decrease by a factor of almost two
when only stars with $\SNR>50$ are considered (see
Table~\ref{tab:externalcomps_1}).  This is a $\sim15$\% improvement on the same
RAVE cluster stars in DR4.

\begin{table*}
\begin{scriptsize}
\begin{center}
\caption{RAVE Targeted Clusters } 
\label{raveclusters}
\begin{tabular}{ p{1.85cm}p{1.2cm}p{1.2cm}p{1.2cm}p{0.8cm}p{0.6cm}p{0.6cm}p{0.6cm}p{0.6cm}p{1.0cm}p{1cm}p{3.5cm}}
Cluster ID & Alternate Name & RA & Dec & Ang Rad (deg) & $\rm RV_{helio}$ &
$\rm [Fe/H]$ & Dist (kpc) & Age (Gyr) & Semester Targeted & Total \# RAVE
(${\tt AlgoConv}=0$)& Comments \\ 
\hline
\hline
Pleiades & Melotte 22, M45 & 03 47 00 & 24 07 00 & 6.2 & 5.5 & $-$0.036 & 0.130 & 0.14 & 2011B & 11 (8) & \citet{funayama09} \\ 
\hline
Hyades & Melotte 25 & 04 26 54 & 15 52 00 & 20 & 39.4 & 0.13 & 0.046 & 0.63 & 2011B & 5 (5) & \citet{takeda13} \\ %%
\hline
IC~4651 & -- & 17 24 49 & $-$49 56 00 & 0.24 & $-$31.0 & $-$0.102 & 0.888 & 1.8 & 2011B & 10 (4) & \citet{carretta04, pasquini04} \\ %%
\hline
47~Tuc $^{\rm GC}$ & NGC~104 & 00 24 05 & $-$72 04 53 & 0.42 & $-$18.0 & $-$0.72 & 4.5 & 13 & 2012B & 23 (12) & \citet{cordero14, koch08, carretta09} \\ %% 5 flds 
\hline
NGC~2477 & M93 & 07 52 10 & $-$38 31 48 & 0.45 & 7.3 & $-$0.192 & 1.450 & 0.82 & 2012B & 9 (4) & \citet{bragaglia08, mishenina15} \\ %% 8 flds 
\hline
M67 & NGC~2682 & 08 51 18 & 11 48 00 & 1.03 & 33.6 & $-$0.128 & 0.890 & 3.4 & 2012A + 2013 & 1 (1) & \citet{onehag14}  \\ %%
\hline
Blanco~1 & -- & 00 04 07 & $-$29 50 00 & 2.35 & 5.5 & 0.012 & 0.250 & 0.06 & 2013 & 1 (1) & \citet{ford05} \\ %%
\hline
Omega~Cen $^{\rm GC}$ & NGC~5139 & 09 12 03.10 & $-$64 51 48.6  & 0.12 & 101.6 & $-$1.14 & 9.6 & 10 & 2013 & 15 (2) & \citet{johnson10} \\ %%
\hline
NGC 2632 & Praesepe &  08 40 24.0 & +19 40 00 & 3.1 & 33.4 & 0.094 & 0.187 & 0.83 & 2012 & 1 (0) & \citet{yang15} \\
\hline
\end{tabular}
\end{center}
\end{scriptsize}
\end{table*}
\subsection{Field star surveys}

\begin{figure}[htb]  
\includegraphics[width=9cm]{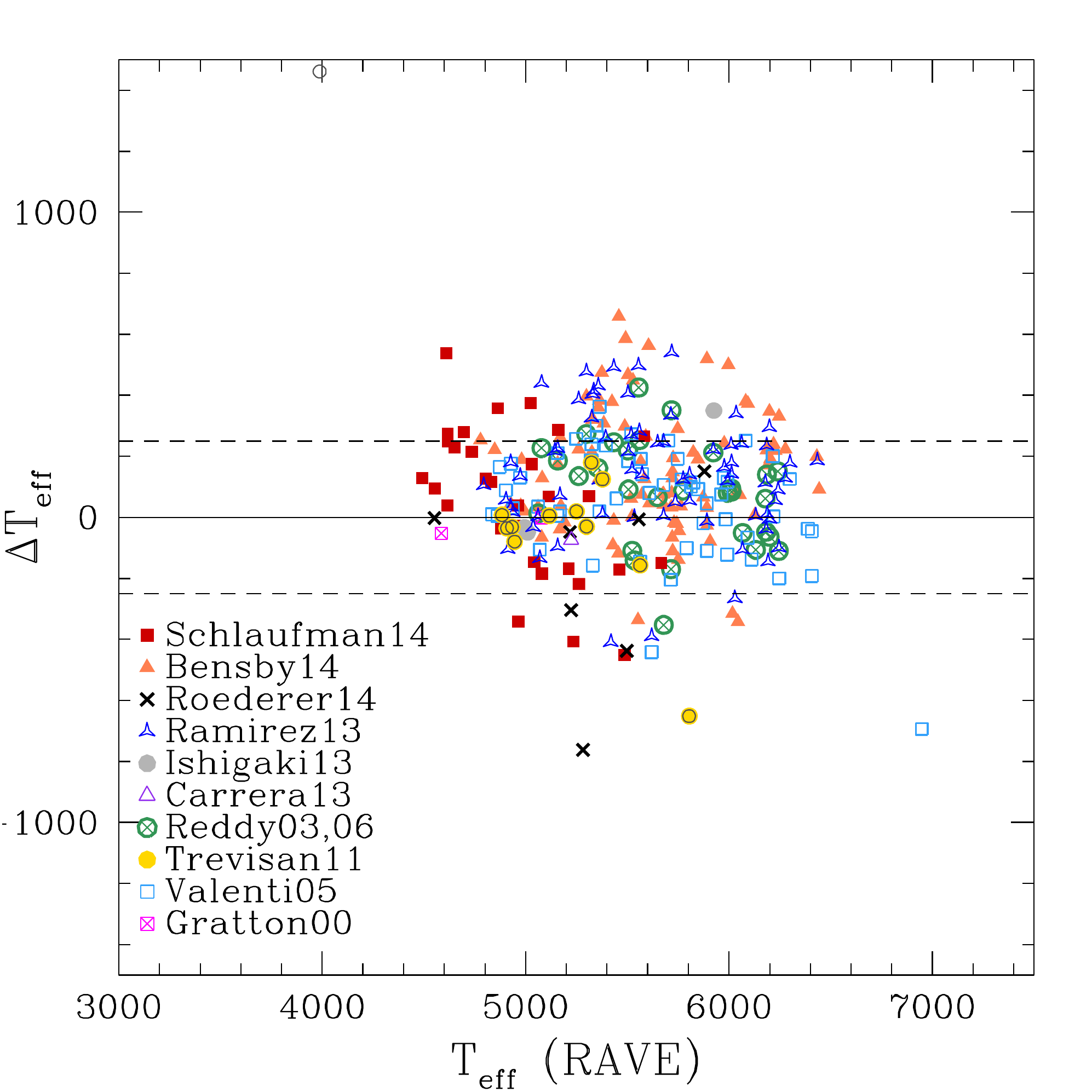}
\caption{A comparison between the $T_{\rm eff}$ presented here with those from field stars 
studied using high-resolution studies in the literature from various different sources.  
Only shown are stars with ${\tt AlgoConv} = 0$ and $T_{\rm eff}$ between 4000 - 8000~K. 
\label{field_teff1}}
\end{figure}

\begin{figure}[htb]  
\includegraphics[width=9cm]{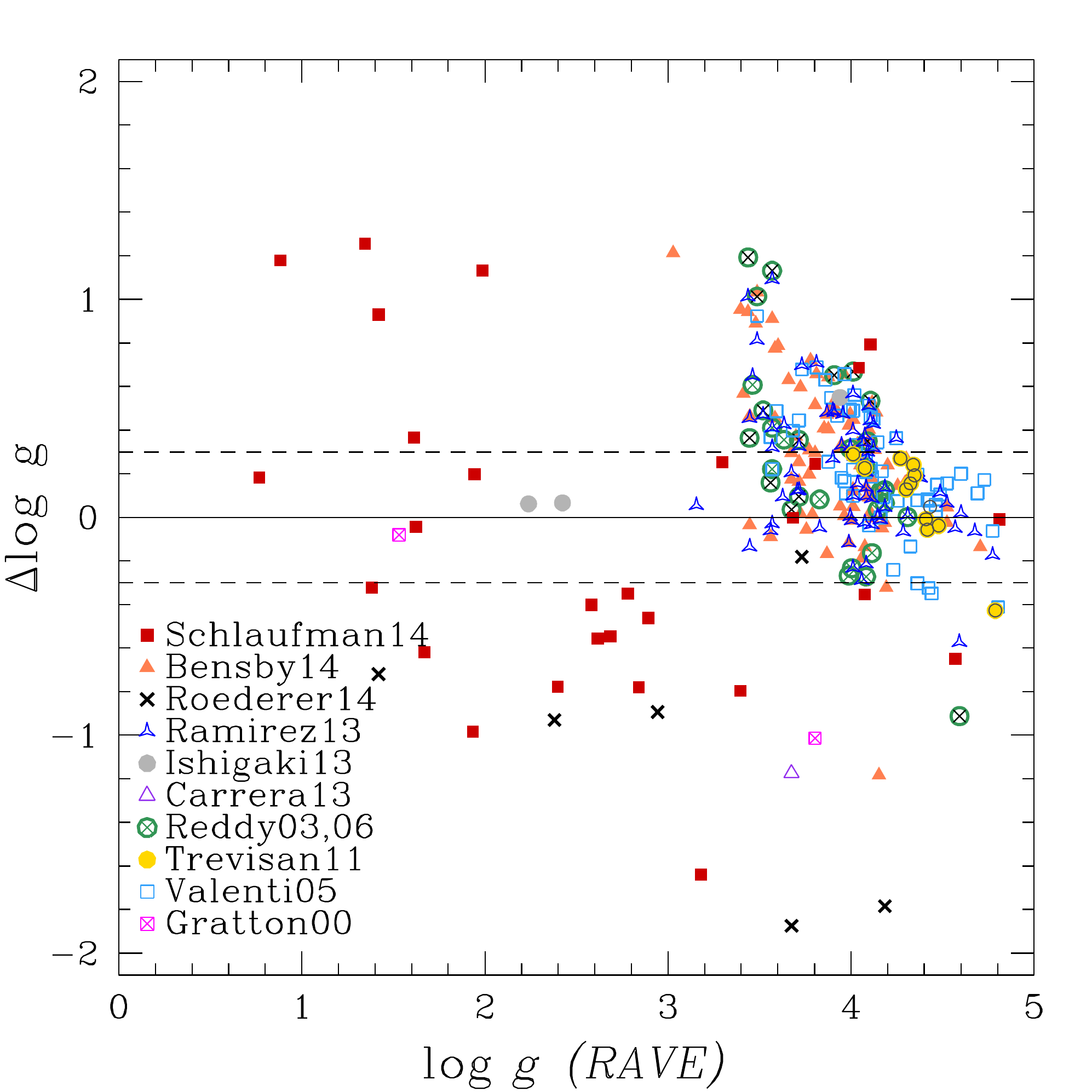}
\caption{A comparison between the $\log g$ presented here with those from field stars 
studied using high-resolution studies in the literature from various different sources.  
Only shown are stars with ${\tt AlgoConv} = 0$ and $T_{\rm eff}$ between 4000 - 8000~K. 
\label{field_logg1}}
\end{figure}

\begin{figure}[htb]  
\includegraphics[width=9cm]{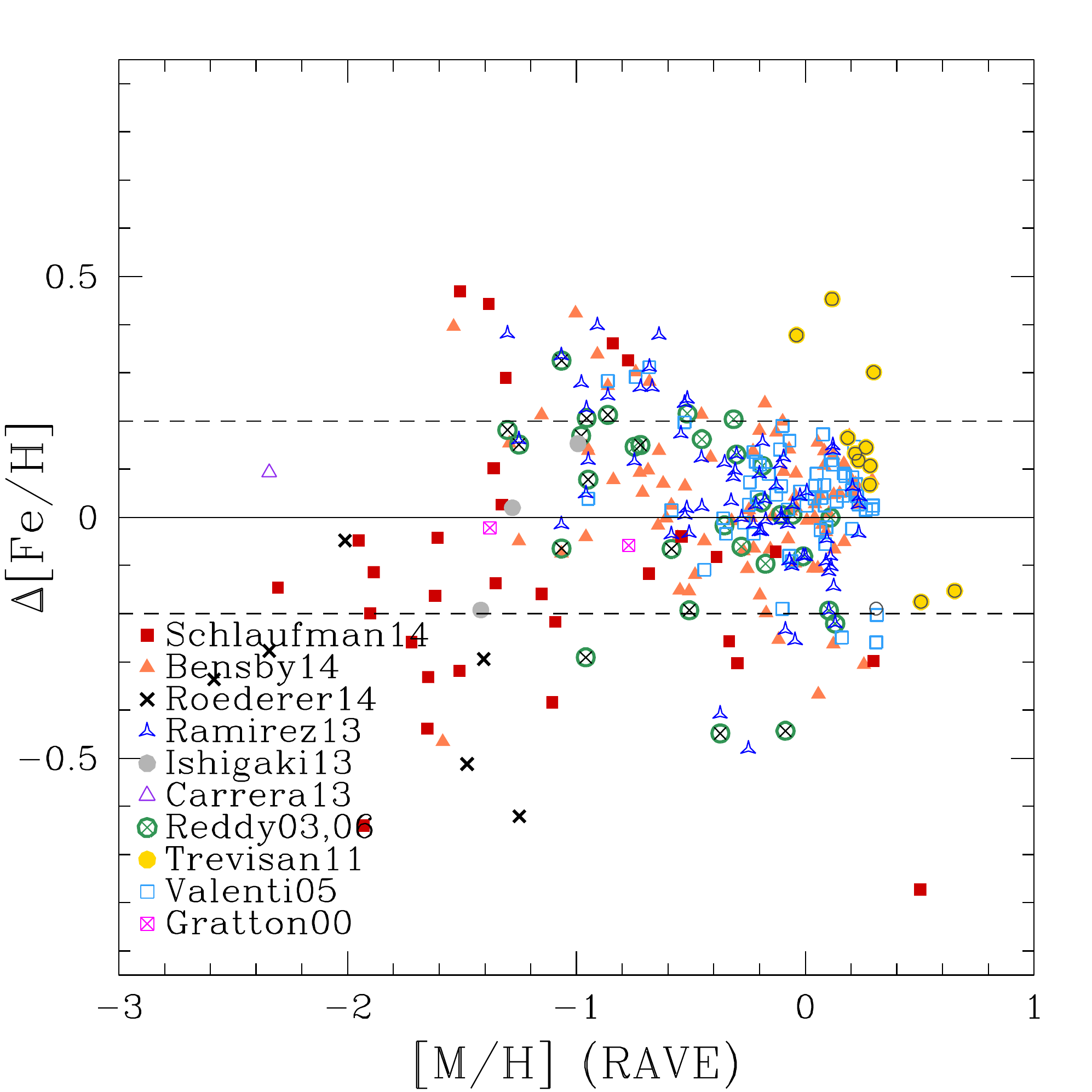}
\caption{A comparison between the $\rm [Fe/H]$ presented here with those from field stars 
studied using high-resolution studies in the literature from various different sources.  
Only shown are stars with ${\tt AlgoConv} = 0$ and $T_{\rm eff}$ between 4000 - 8000~K. 
\label{field_feh1}}
\end{figure}

We have matched RAVE stars with the high-resolution studies of
%\citet{gratton00, aoki02, hollek11, carrera13, ishigaki13, roederer14} and
\citet{gratton00, carrera13, ishigaki13, roederer14} and
\citet{schlaufman14}, which concentrate on bright metal-poor stars, the study
of \citet{trevisan11}, which concentrates on old, metal-rich stars, and the
studies of \citet{ramirez13, reddy03, reddy06, valenti05, bensby14}, which target FGK stars in
the solar neighbourhood.  Figures~\ref{field_teff1}, \ref{field_logg1}, and
\ref{field_feh1} compare stellar parameters from these studies with the DR5
values.  Trends are detectable in $\log g$ for both giants and dwarfs. For
the giants the same tendency for $\log g$ to be over-estimated when $\log g$
is small was evident in V16.  In Figure~\ref{field_logg1} a
similar, but less pronounced, tendency is evident in the $\log g$ values for
dwarfs.

\subsection{APOGEE}
The Apache Point Observatory Galactic Evolution Experiment,
part of the Sloan Digital Sky Survey and covering mainly the northern hemisphere, has made public
near-IR spectra with a resolution of R$\sim$22,500 for over 150,000 stars \citep[DR12,][]{holtzman15}. 
Stellar parameters are only provided for APOGEE giant stars, and temperatures, gravities, 
$\rm [Fe/H]$ metallicities and radial velocities are reported 
to be accurate to $\sim$100\,K (internal), $\sim$0.11~dex (internal), $\le$0.1~dex (internal) and 
$\sim$100 m~s$^{-1}$, respectively \citep{holtzman15, nidever12}.
Despite the different hemispheres targeted by RAVE and APOGEE, there are $\sim$1100 APOGEE
stars that overlap with RAVE DR5 stars, two thirds of these having valid APOGEE stellar parameters.

\begin{figure}[htb]  
\includegraphics[width=9cm]{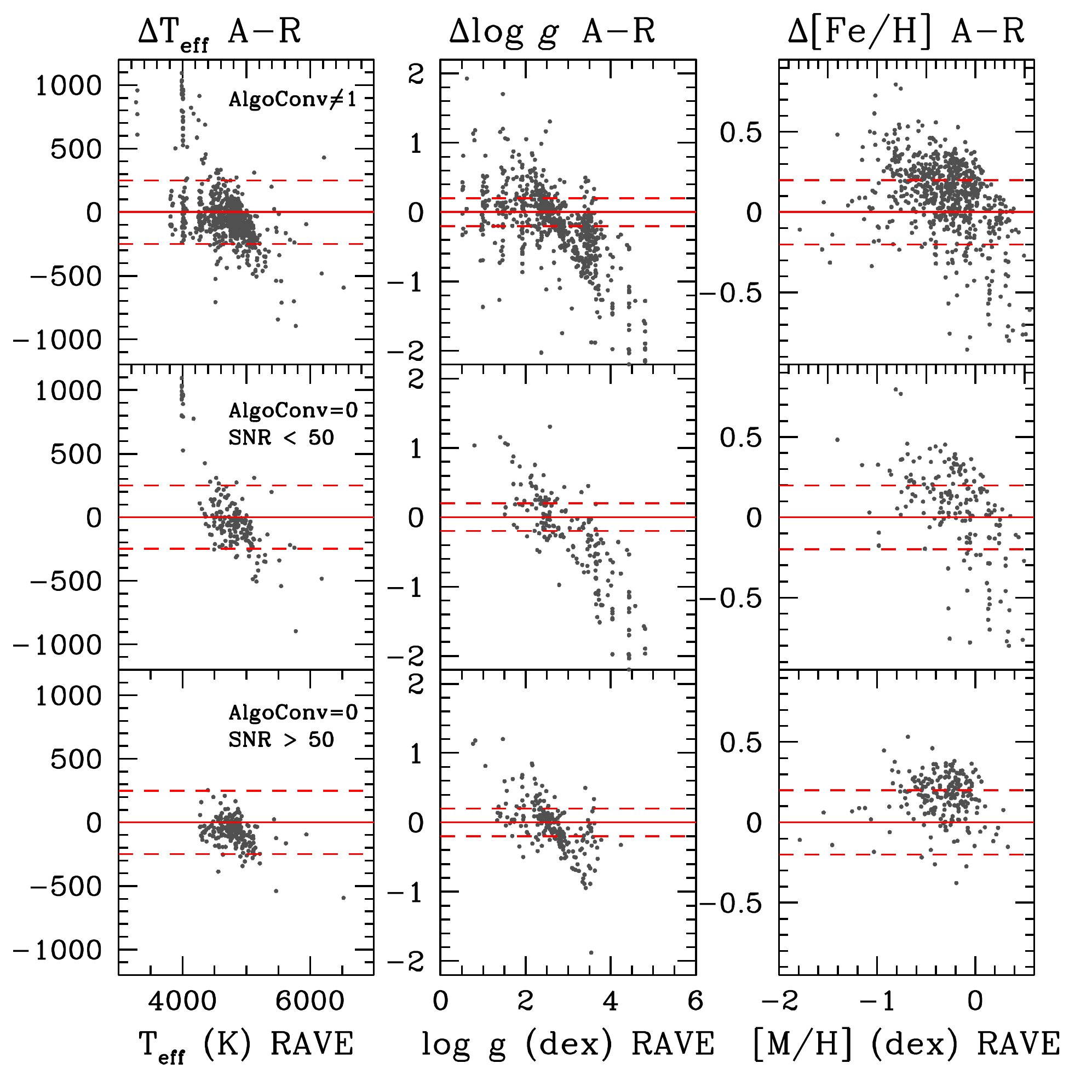}
\caption{A comparison between the stellar parameters of the RAVE stars that overlap with
APOGEE.  Different subsets of SNR and {\tt AlgoConv} cuts are shown.
\label{stelparam_apogee}}
\end{figure}

A comparison between the APOGEE and RAVE stellar parameters is shown in Figure~\ref{stelparam_apogee}. 
The zero-point and standard deviation for different subsets of SNR and
{\tt AlgoConv} are provided in Table~\ref{tab:externalcomps_1}.
There appears to be a $\sim$0.15~dex zero-point offset in $\rm [Fe/H]$ between APOGEE and RAVE,
as seen most clearly in the high SNR sample, and there is a noticeable break in $\log g$ where
the cool main-sequence stars and stars along the giant branch begin to overlap.  This is the
consequence of degeneracies in the CaT region that affect the determination of 
$\log g$ (see Tables 1 and 2 in DR4).

\subsection{LAMOST}

The Large sky Area Multi-Object Spectroscopic Telescope is an ongoing optical
spectroscopic survey with a resolution of $R\sim1\,800$, which has gathered
spectra for more than 4.2 million objects.  About 2.2 million stellar
sources, mainly with  $\SNR> 10$, have stellar
parameters.  Typical uncertainties are 150\,K, 0.25\,dex, 0.15\,dex,
$5\kms$ for $T_{\rm eff}$, $\log g$, metallicity and radial
velocity, respectively \citep{xiang15}.

\begin{figure}[htb]  
\includegraphics[width=9cm]{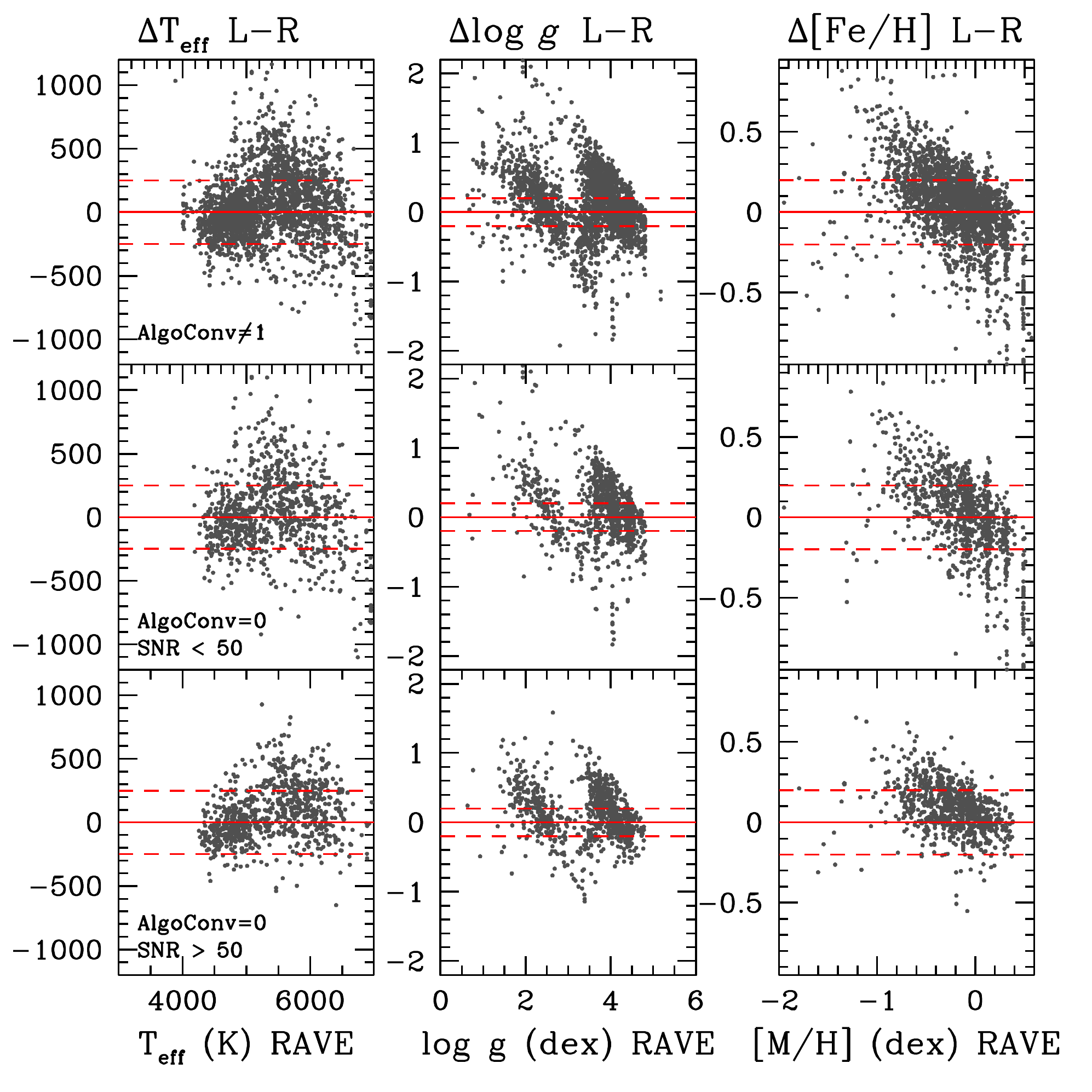}
\caption{A comparison between the stellar parameters of the stars presented here with those
from LAMOST.  There are 2700, 1026 and 987 stars in the top, middle and bottom panels,
respectively.
\label{stelparam_lamo}}
\end{figure}

The overlap between LAMOST and RAVE comprises almost 3000 stars, including
both giants and dwarfs.  Figure~\ref{stelparam_lamo} shows the comparison
between the stellar parameters of RAVE and LAMOST. The giants (stars with
$\log g<$3) and dwarfs (stars with $\log g>3$) exhibit different trends in
$\log g$, and the largest uncertainties in $\log g$ occur where these
populations overlap in $\log g$.  The zero-point and standard deviation for
the comparisons between RAVE and LAMOST stellar parameters are provided in
Table~\ref{tab:externalcomps}.

\subsection{GALAH} The Galactic Archaeology with HERMES (GALAH) Survey
is a high-resolution (R$\sim$28,000) spectroscopic survey using the HERMES spectrograph and 
Two Degree Field (2dF) fibre positioner on the 3.9m Anglo-Australian telescope.
The first data release provides  $T_{\rm eff}$, $\log g$, [$\alpha$/Fe], radial  velocity, 
distance modulus and reddening  for 9860 Tycho-2 stars \citep{martell16}.  
There are $\sim1800$ RAVE stars that overlap with a star observed in GALAH, spanning 
the complete range in temperature, gravity and metallicity.

\begin{figure}[htb]  
\includegraphics[width=9cm]{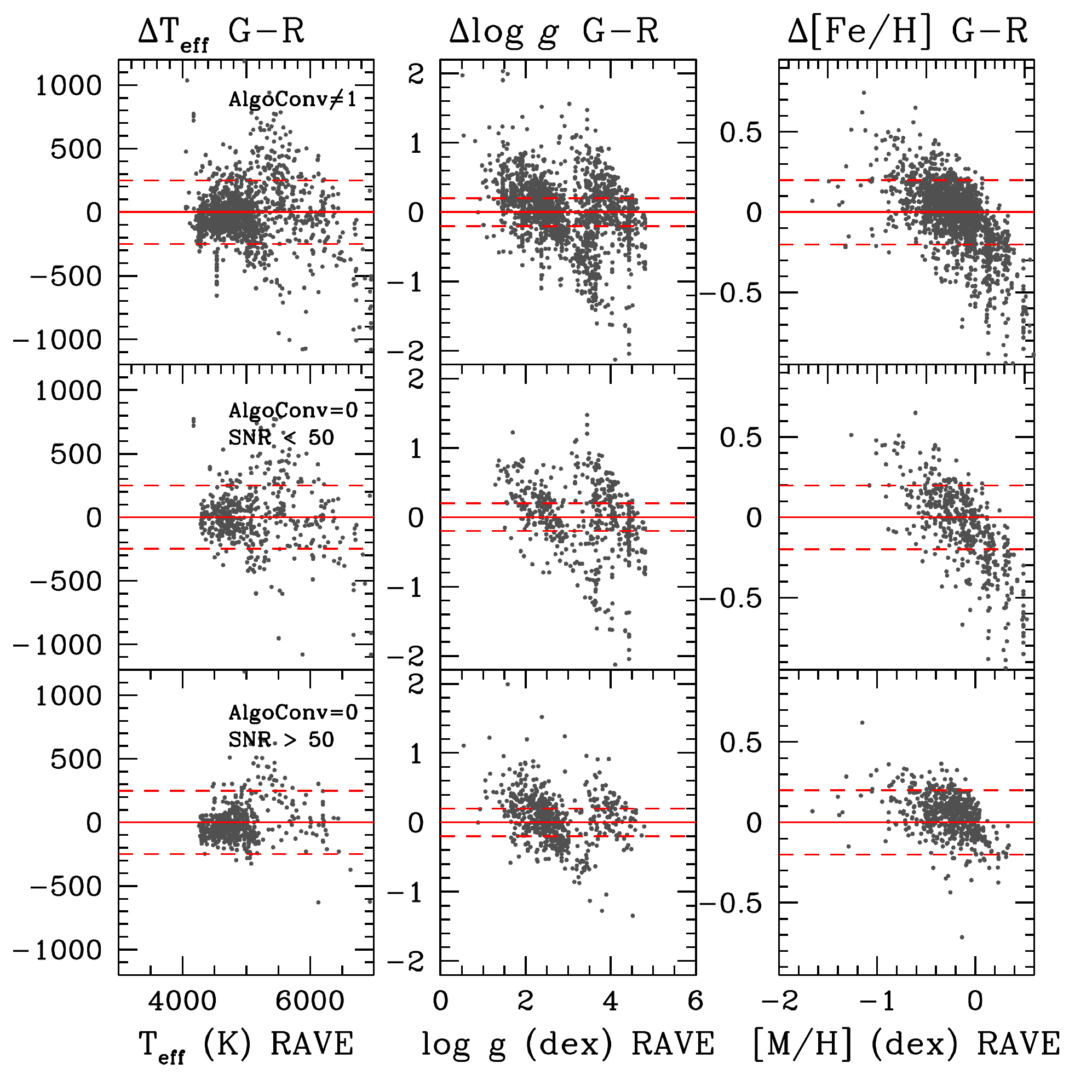}
\caption{A comparison between the stellar parameters of the stars presented here with those
from GALAH DR1.  %There are 53, 11 and 28 stars in the top, middle and bottom panels,
%respectively.
\label{stelparam_galah}}
\end{figure}

Figure~\ref{stelparam_galah} shows the comparison of stellar parameters between the
RAVE and Galah overlap stars, and Table~\ref{tab:externalcomps}
quantifies the agreement between these two surveys.

\subsection{GAIA-ESO} Gaia-ESO, a public spectroscopic survey observing stars
in all major components of the Milky Way using the Very Large Telescope
(VLT), provides 14\,947 unique targets in DR2.  The resolution of the stellar
spectra ranges from $R\sim17\,000$ to $R\sim47\,000$.  There are $\sim100$ RAVE
stars that overlap with a star observed in Gaia-ESO, half of these are
situated around the $\eta$ Chamaeleontis Cluster \citep{mamajek99},
and a third are in the vicinity of the Gamma Velorum cluster
\citep{jeffries14}.
%
%Figure~\ref{stelparam_ge} shows the comparison of stellar parameters between the
%RAVE and Gaia-ESO overlap stars.  
The overlap sample is small and new internal values are being analysed currently; 
still Table~\ref{tab:externalcomps} quantifies the results between these two surveys.

%\begin{figure}[htb]  
%\includegraphics[width=9cm]{stelparam_ge_dr5-eps-converted-to.pdf}
%\caption{A comparison between the stellar parameters of the stars presented here with those
%from Gaia-ESO.  There are 53, 11 and 28 stars in the top, middle and bottom panels,
%respectively.
%\label{stelparam_ge}}
%\end{figure}

\begin{table*}
\begin{scriptsize}
\begin{center}
\caption{Estimates of the external errors in the stellar parameters.  
} 
\label{tab:externalcomps}
\begin{tabular}{ p{3.85cm}p{0.95cm}p{1.35cm}p{0.95cm}p{0.95cm}p{0.95cm}}
stellar type & N & $\sigma (T_{\rm eff})$ & $\sigma (\log g)$ & $\sigma (\rm [M/H])$ & $\sigma (T_{\rm eff, IRFM})$ \\ 
\hline
dwarfs   ($\log g> 3.5$) \\
\hline
hot, all metallicities DR5 & 375 & 442 & 0.39 & 0.41 & 129 \\ %_2
hot, metal-poor DR5 & 38 & 253 & 0.48 & 0.95 & 258 \\ %_2
hot, metal-rich DR5 & 337 & 453 & 0.38 & 0.95 & 233 \\ %_2
cool, all metallicities DR5 & 332 & 250 & 0.75 & 0.41 & 187 \\ %_3
cool, metal-poor DR5 & 68 & 303 & 0.87 & 0.61 & 301 \\ %_3
cool, metal-rich DR5 & 264 & 233 & 0.72 & 0.29 & 146 \\ %_3
\hline
hot, all metallicities RAVE-on & 510 & 411 & 0.56 & 0.37 &  \\ %_2
hot, metal-poor RAVE-on & 95 & 498 & 0.94 & 0.55 &  \\ %_2
hot, metal-rich RAVE-on & 415 & 389 & 0.41 & 0.32 &  \\ %_2
cool, all metallicities RAVE-on & 267 & 291 & 0.62 & 0.24 &  \\ %_3
cool, metal-poor RAVE-on & 49 & 417 & 0.75 & 0.32 &  \\ %_3
cool, metal-rich RAVE-on & 218 & 255 & 0.57 & 0.20 &  \\ %_3
\hline
$\SNR > 40$ \\
\hline
hot, all metallicities DR5 & 260 & 210 & 0.29 & 0.16 &  \\ %_2
hot, metal-poor DR5 & 30 & 260 & 0.39 & 0.16 &  \\ %_2
hot, metal-rich DR5 & 230 & 201 & 0.28 & 0.15 &  \\ %_2
cool, all metallicities & 185 & 202 & 0.50 & 0.17 \\ %_3
cool, metal-poor & 48 & 256 & 0.70 & 0.21 \\ %_3
cool, metal-rich & 137 & 164 & 0.41 & 0.13 \\ %_3
\hline
hot, all metallicities RAVE-on & 314 & 273 & 0.34 & 0.21 &  \\ %_2
hot, metal-poor RAVE-on & 55 & 354 & 0.61 & 0.36 &  \\ %_2
hot, metal-rich RAVE-on & 259 & 253 & 0.24 & 0.16 &  \\ %_2
cool, all metallicities RAVE-on & 187 & 250 & 0.54 & 0.17 &  \\ %_3
cool, metal-poor RAVE-on & 35 & 303 & 0.65 & 0.21 &  \\ %_3
cool, metal-rich RAVE-on & 152 & 237 & 0.49 & 0.15 &  \\ %_3
\hline
\hline
Giants  ($\log g$ $<$ 3.5) \\
\hline
all, all metallicities DR5 & 1294 & 156 & 0.48 & 0.17 & 110 \\ %_4
hot DR5 & 28 & 240 & 0.45 & 0.30 & 261 \\
cool, metal-poor DR5 & 260 & 211 & 0.58 & 0.20 & 93 \\ %_4
cool, metal-rich DR5 & 1006 & 125 & 0.46 & 0.15& 96 \\ %_4
\hline
all, all metallicities RAVE-on & 1318 & 140 & 0.41 & 0.20 &  \\ %_4
hot RAVE-on & 5 & 270 & 0.62 & 0.27 &   \\
cool, metal-poor RAVE-on & 293 & 195 & 0.55 & 0.27 &  \\ %_4
cool, metal-rich RAVE-on & 1020 & 110 & 0.36 & 0.17 &  \\  %_4
\hline
SNR $>$ 40 \\
\hline
hot DR5 & 22 & 189 & 0.46 & 0.24 \\
cool, metal-poor DR5 & 225 & 210 & 0.58 & 0.20 \\ %_4
cool, metal-rich DR5 & 843 & 113 & 0.44 & 0.13 \\ %_4
\hline
hot RAVE-on & 3 & 120 & 0.28 & 0.23 \\
cool, metal-poor RAVE-on & 248 & 159 & 0.52 & 0.23 \\ %_4
cool, metal-rich RAVE-on & 810 & 88 & 0.33 & 0.15 \\ %_4
\hline
\hline
\hline
Giants (asteroseismically calibrated sample) & $\rm N_s$ & $\sigma(T_{\rm eff, IRFM}$) & $\sigma(\log g_s)$ & $\sigma(\rm [Fe/H]_c$) \\
\hline
all, all metallicities & 332 & 169 & 0.37 & 0.21 \\
hot & 11 & 640 & 0.39 & 0.28 \\
cool, metal-poor & 180 & 161 & 0.40 & 0.23 \\
cool, metal-rich & 835 & 107 & 0.29 & 0.15 \\ 
\hline
SNR $>$ 40 \\
\hline
hot & 5 & 471 & 0.42 & 0.15 \\
cool, metal-poor & 154 & 170 & 0.38 & 0.21 \\
cool, metal-rich & 701 & 95 & 0.28 & 0.12 \\ 
\hline
\end{tabular}
\end{center}
\end{scriptsize}
\end{table*}

\begin{table*}
\begin{scriptsize}
\begin{center}
\caption{RAVE External Comparisons By Survey} 
\label{tab:externalcomps_1}
\begin{tabular}{ p{2.0cm}p{2.5cm}p{2.5cm}p{2.5cm}}
 & ${\tt AlgoConv}\neq1$ & ${\tt AlgoConv}=0$,  $\SNR< 50$ & ${\tt
 AlgoConv}=0$,  $\SNR> 50$  \\ 
\hline
APOGEE & $T_{\rm eff}$:~$-$30$\pm$277 $\log g$:~$-$0.22$\pm$0.60 $\rm [Fe/H]$:~0.08$\pm$0.44 $\rm Num$:~711 $\log g_{sc}$:~0.03$\pm$0.29 $\rm Num_{sc}$:~317 & $T_{\rm eff}$:~4$\pm$342 $\log g$:~$-$0.35$\pm$0.70 $\rm [Fe/H]$:~0.05$\pm$0.52 $\rm Num$:~190 $\log g_{sc}$:~0.06$\pm$0.31 $\rm Num_{sc}$:~129 & $T_{\rm eff}$: $-$75$\pm$107 $\log g$: $-$0.05$\pm$0.37 $\rm [Fe/H]$:~0.16$\pm$0.14 $\rm Num$:~221 $\log g_{sc}$:~0.00$\pm$0.27 $\rm Num_{sc}$:~184 \  \\
\hline
GAIA-ESO & $T_{\rm eff}$:~243$\pm$477 $\log g$:~$-$0.12$\pm$0.89 $\rm [Fe/H]$:~0.25$\pm$0.93 $\rm Num$:~53 $\log g_{sc}$:~0.17$\pm$0.64 $\rm Num_{sc}$:~18 & $T_{\rm eff}$:~613$\pm$659 $\log g$:~$-$0.82$\pm$0.91 $\rm [Fe/H]$:~$-$0.10$\pm$0.30 $\rm Num$:~11 $\log g_{sc}$:~0.19$\pm$0.35 $\rm Num_{sc}$:~3 & $T_{\rm eff}$: 52$\pm$266 $\log g$: 0.08$\pm$0.46 $\rm [Fe/H]$:~0.13$\pm$0.21 $\rm Num$:~28 $\log g_{sc}$:~0.16$\pm$0.69 $\rm Num_{sc}$:~15 \\
\hline
Clusters & $T_{\rm eff}$:~38$\pm$309 $\log g$:~$-$0.12$\pm$0.63 $\rm [Fe/H]$:~$-$0.10$\pm$0.28 $\rm Num$:~75 $\log g_{sc}$:~$-$0.39$\pm$0.45 $\rm Num_{sc}$:~14 & $T_{\rm eff}$:~$-$62$\pm$422 $\log g$:~$-$0.42$\pm$1.13 $\rm [Fe/H]$:~$-$0.21$\pm$0.39 $\rm Num$:~15 $\log g_{sc}$:~$-$0.59$\pm$0.29 $\rm Num_{sc}$:~6 & $T_{\rm eff}$: 106$\pm$244 $\log g$: 0.13$\pm$0.29 $\rm [Fe/H]$:~0.01$\pm$0.16 $\rm Num$:~26 $\log g_{sc}$:~$-$0.17$\pm$0.50 $\rm Num_{sc}$:~7 \\ %%
\hline
Misc.~Field~Stars & $T_{\rm eff}$:~126$\pm$397 $\log g$:~$-$0.05$\pm$0.95 $\rm [Fe/H]$:~$-$0.09$\pm$0.40 $\rm Num$:~317 $\log g_{sc}$:~$-$0.25$\pm$0.90 $\rm Num_{sc}$:~51 & $T_{\rm eff}$:~251$\pm$517 $\log g$:~$-$0.33$\pm$1.17 $\rm [Fe/H]$:~$-$0.17$\pm$0.48 $\rm Num$:~57 $\log g_{sc}$:~$-$0.37$\pm$0.95 $\rm Num_{sc}$:~16 & $T_{\rm eff}$: 111$\pm$196 $\log g$: 0.15$\pm$0.51 $\rm [Fe/H]$:~0.01$\pm$0.18 $\rm Num$:~169 $\log g_{sc}$:~$-$0.18$\pm$0.90 $\rm Num_{sc}$:~33 \\
\hline
LAMOST & $T_{\rm eff}$: 30$\pm$325 $\log g$: 0.12$\pm$0.48 $\rm [Fe/H]$:~0.05$\pm$0.27 $\rm Num$:~2700 $\log g_{sc}$:~0.14$\pm$0.40 $\rm Num_{sc}$:~557 & $T_{\rm eff}$: $-$4$\pm$364 $\log g$: 0.08$\pm$0.49 $\rm [Fe/H]$:~0.00$\pm$0.27 $\rm Num$:~2026 $\log g_{sc}$:~0.24$\pm$0.45 $\rm Num_{sc}$:~224 & $T_{\rm eff}$: 58$\pm$208 $\log g$: 0.16$\pm$0.36 $\rm [Fe/H]$:~0.09$\pm$0.15 $\rm Num$:~987 $\log g_{sc}$:~0.06$\pm$0.33 $\rm Num_{sc}$:~313 \\ %% 5 flds 
\hline
GALAH & $T_{\rm eff}$: $-$36$\pm$274 $\log g$: 0.0$\pm$0.50 $\rm [Fe/H]$:~$-$0.02$\pm$0.33 $\rm Num$:~1700 $\log g_{sc}$:0.04$\pm$0.45 $\rm Num_{sc}$:~1255 & $T_{\rm eff}$: $-$43$\pm$376 $\log g$: $-$0.02$\pm$0.59 $\rm [Fe/H]$:~$-$0.07$\pm$0.45 $\rm Num$:~526 $\log g_{sc}$:~0.0$\pm$0.56 $\rm Num_{sc}$:~443 & $T_{\rm eff}$: $-$6$\pm$144 $\log g$: 0.06$\pm$0.35 $\rm [Fe/H]$:~0.04$\pm$0.13 $\rm Num$:~663 $\log g_{sc}$:~0.06$\pm$0.32 $\rm Num_{sc}$:~613 \\ %% 5 flds 
\hline
\end{tabular}
\end{center}
\end{scriptsize}
\end{table*}

\section{Elemental abundances}
\label{chemicalpipeline}

The elemental abundances for Aluminium, Magnesium, Nickel, Silicon, Titanium, and Iron are
determined for a number of RAVE stars using a dedicated 
chemical pipeline that relies on an equivalent width library encompassing
604 atomic and molecular lines in the RAVE wavelength range.  This chemical
pipeline was first introduced by \citet{boeche11} and then improved upon for the DR4 data release.

Briefly, equivalent widths are computed for a grid of stellar parameter values in the 
following ranges: $T_{\rm eff}$ from 4000 to 7000\,K, $\log g$ from 0.0 to 0.5~dex, 
$\rm [M/H]$ from $-$2.5 to +0.5~dex  and five levels of abundances from 
$-$0.4 to +0.4~dex relative to the metallicity, in steps of 0.2~dex,
using the solar abundances of \citet{grevesse98}.
Using the calibrated RAVE effective temperatures,
surface gravities and metallicities (see \S5), the pipeline searches for
the best-fitting model spectrum by minimizing the $\chi^{2}$ between the models
and the observations.  

The line list and specific aspects of the equivalent width library are given
in \citet{boeche11} and the full scheme to compute the
abundances is given in \S5 of \citet{kordopatis13}.
Abundances from the RAVE chemical abundance pipeline are only
provided for stars fulfilling the following criteria:
\begin{itemize}
\item  $T_{\rm eff}$ must be between 4000 and 7000\,K
\item  $\SNR >20$
\item Rotational velocity, $\rm V_{rot}<50\kms$.
\end{itemize}

The highest quality of abundances will be determined for the stars
that have the following additional constraints:
\begin{itemize}
\item  $\chi^{2}< 2000$, where $\chi^{2}$ quantifies the mismatch between
the observed spectrum and the  
best-matching model.
\item  ${\tt frac} > 0.7$, where {\tt frac} represents the fraction of the observed spectrum 
that satisfactorily matches the model.
\item  $c1$, $c2$ and $c3$ classification flags indicate that the spectrum is ``normal" \citep[see][for details
on the classification flags]{matijevic12}.
\item  {\tt AlgoConv} value indicates the stellar parameter pipeline
converged.  ${\tt AlgoConv}=0$ indicates the highest quality result.
\end{itemize}

The precision and accuracy of the resulting elemental abundances are
assesed in two ways.  First, uncertainties in the elemental abundances are
investigated from a sample of 1353 synthetic spectra.  The typical
dispersions are $\sigma\sim0.05\,$dex for $\SNR = 100$ spectra,
$\sigma\sim0.1\,$dex for $\SNR=40$ spectra and $\sigma\sim0.25\,$dex for $\SNR = 20$
spectra.  The exception is the element Fe, which has a smaller dispersion by
a factor of two, and the element Ti, which has a larger dispersion by a
factor of 1.5 - 2 \citep[see][for details]{boeche11, kordopatis13}.

The number of measured absorption lines for an element, which is also
provided in the DR5 data release, is, like SNR, a good indicator of the
reliability of the abundance.  The higher the number of measured lines, the
better the expected precision.  The relatively low uncertainty in the Fe
abundances reflects the large number of its measurable lines at all stellar
parameter values.  

A second assessment of the performance of the chemical pipeline is provided
by comparing the DR5 abundances in 98 dwarf stars with values given in
\citet{soubiran05} and in 203 giant stars with abundances in \citet{ruchti11}.  The
dwarfs in \citet{soubiran05} typically have  RAVE
 $\SNR>100$, and the giants in \citet{ruchti11}
have RAVE SNR in the  range 30 to 90.

\begin{figure}[htb]  
\includegraphics[width=9cm]{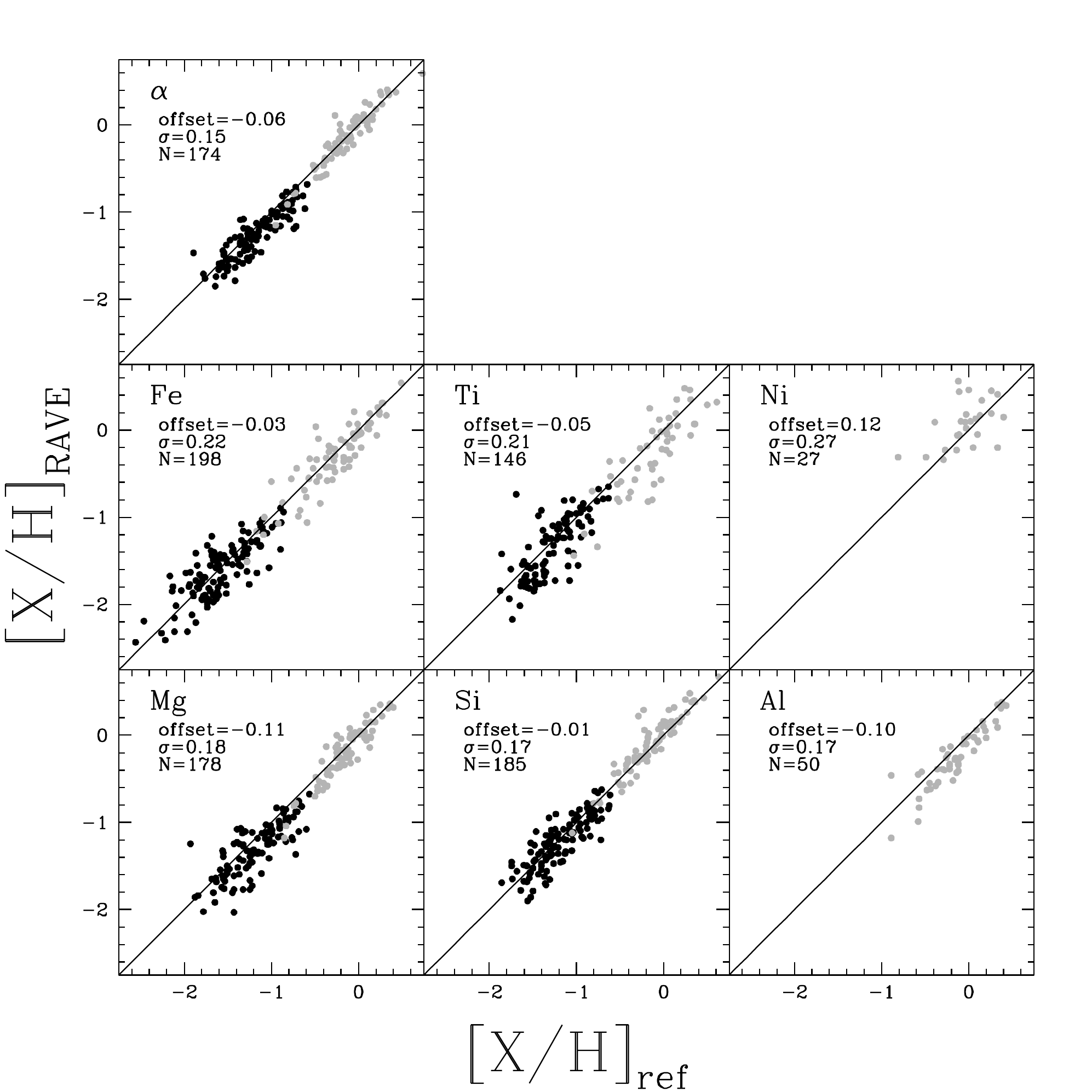}
\caption{Comparison between high-resolution elemental abundances from \citet{soubiran05} (grey) and
\citet{ruchti11} (black) compared to the derived elemental abundances from the RAVE chemical pipeline.  The 
input stellar parameters for the RAVE chemical pipeline are those presented here (see \S5).
\label{El_a}}
\end{figure}

\begin{figure}[htb]  
\includegraphics[width=9cm]{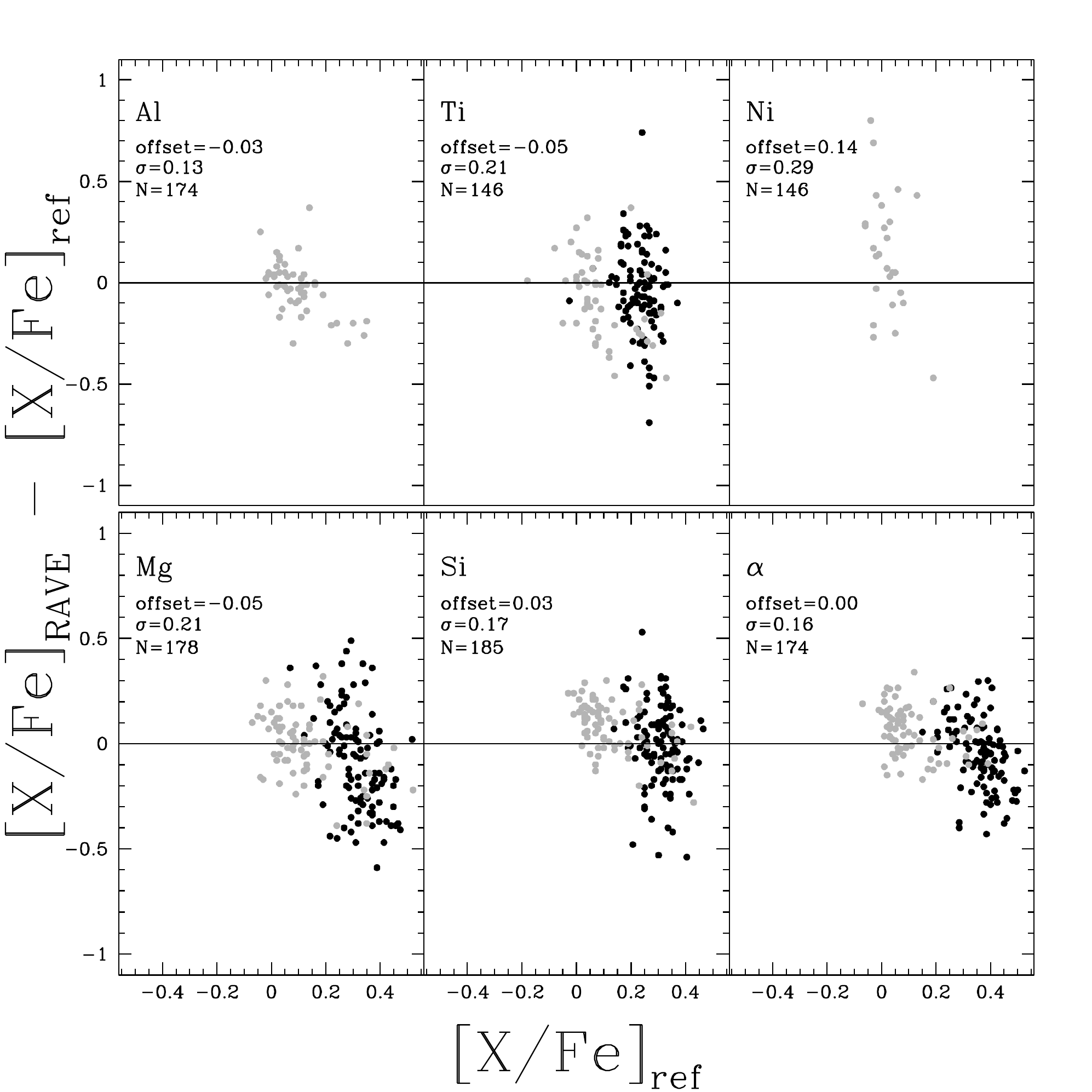}
\caption{A comparison between the literature relative elemental abundance
and residual abundances (RAVE-minus-literature).  The stellar parameters
and symbols used are as in Figure~\ref{El_a}.
\label{El_a1}}
\end{figure}

Figures~\ref{El_a} and~\ref{El_a1} show the results obtained for the six
elements from the RAVE chemical pipeline.  In general, there is a slight
improvement in the external comparisons from DR4, likely resulting from the
improved DR5 calibration for the stellar parameters.  The accuracy of the
RAVE abundances depends on many variables, which can be inter-dependent in a
non-linear way, making it non-trivial to provide one value to quantify the
accuracy of the RAVE elemental abundances.  We also have not taken into
account the errors in abundance measurements from high-resolution spectra.
Here is a summary of the expected accuracy of the DR5 abundances, element by
element.

\begin{itemize}

\item {\it magnesium}:  The uncertainty is $\sigma_{\rm Mg}\sim0.2\,$dex, 
slightly worse for stars with  $\SNR< 40$.
\item {\it aluminum}:  This is measured in RAVE spectra from only two isolated 
lines. Abundance errors are $\sigma_{\rm Al}\sim0.2\,$dex, and slightly worse 
for stars with  $\SNR<40$.
\item {\it silicon}:  This is one of the most reliably determined elements, with
$\sigma_{\rm Si}\sim0.2\,$dex, and slightly worse for stars with  $\SNR<40$.
\item {\it titanium}:  The estimates are best for high-SNR, cool giants
($\rm T_{eff} < 5500$\,K and $\log g < 3$).  We suggest rejecting Ti
abundances for dwarf stars.  Uncertainties for cool giants are
$\sigma_{\rm Ti}\sim0.2\,$dex, and slightly worse for stars with  $\SNR<40$.
\item {\it iron}:  A large number of measurable lines is available at all
stellar parameter values. 
The expected errors are $\sigma_{\rm Fe}\sim0.2\,$dex.
\item {\it nickel}: Ni estimates should be used for high SNR, cool stars only
($T_{\rm eff} < 5000$\,K).  In this regime, $\sigma_{\rm Ni}\sim0.25\,$dex,
but correlates with number of measured lines (i.e., with SNR).
\item $\alpha$-{\it enhancement}: This is the average of $\rm [Mg/Fe]$ and
$\rm [Si/Fe]$, and is a particularly useful measurement at low SNR.  The
expected uncertainty is $\sigma_\alpha \sim0.2\,$dex.

\end{itemize}

\begin{figure}[htb]  
\includegraphics[width=9cm]{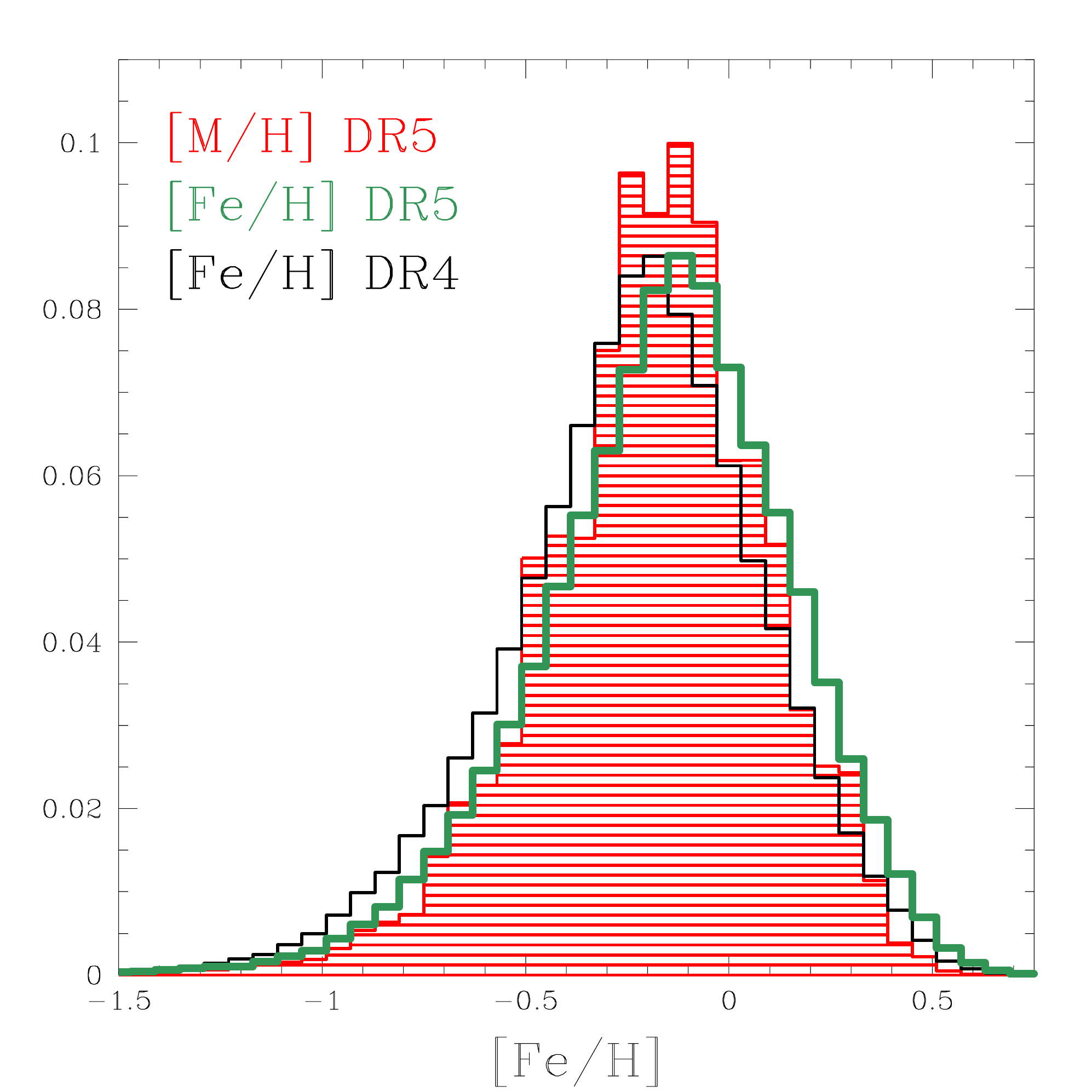}
\caption{A comparison between the \feh derived with the chemical pipeline to
the calibrated \mh values from the stellar parameter pipeline.
Also shown is the \feh distribution from DR4.
\label{Met_Met2}}
\end{figure}

The green histogram in Figure~\ref{Met_Met2} shows the distribution of \feh
from the chemical pipeline. This is similar to the black histogram of \feh
values in DR4 but  shifted to slightly larger \feh. The red histogram of \mh
values in DR5 is slightly narrower than either \feh histogram and peaks at
slightly lower values than the DR5 \feh histogram.

\section{Distances, Ages and Masses}\label{sec:D}

In DR4 we included for the first time distances derived by the Bayesian method 
developed by \cite{burnett10}.  This takes as its input the stellar parameters 
$T_{\rm eff}$, $\log g$ and $\mh$ determined from the RAVE spectra, and 
$J$, $H$ and $K_{\rm s}$ magnitudes from 2MASS.  This method was extended 
by \cite{binney14}, who included dust extinction in the modelling, and introduced an 
improvement in the description of the distance to the stars by providing multi-Gaussian 
fits to the full probability density function (pdf) in distance modulus. Previous data 
releases included distance estimates from different sources \citep{breddels10, zwitter10}, 
but the Bayesian pipeline has been shown to be more robust when dealing with 
atmospheric parameter values with large uncertainties, so it provided the 
recommended distance estimates for DR4, and the only estimates that we provide with DR5.

We provide distance estimates for all stars except those for which we do not believe 
we can find reliable distances, which include stars with the following DR5 characteristics:
\begin{itemize} 
\item ${\tt AlgoConv}=1$ or $\SNR<20$,
\item $T_{\rm eff}<4000$\,K and $\log g>3.5$ (i.e. cool dwarfs), and
\item $T_{\rm eff}>7400$\,K and $\rm [M/H]<-1.2$.
\end{itemize}

The distance pipeline applies the simple Bayesian statement
\[
P(\hbox{model}|\hbox{data})=\frac{P(\hbox{data}|\hbox{model})P(\hbox{model})}{P(\hbox{data})},
\]
where in our case ``data'' refers to the inputs described above for a single
star, and ``model'' comprises a star of specified initial mass ${\cal M}$,
age $\tau$, metallicity $\mh$, and location, observed through a specified
line-of-sight extinction. $P(\hbox{data}|\hbox{model})$ is determined
assuming uncorrelated Gaussian uncertainties on all inputs, and using
isochrones to find the values of the stellar parameters and absolute
magnitudes of the model star. The uncertainties of the stellar parameters are
assumed to be the quadratic sum of the quoted internal uncertainties and the
external uncertainties calculated from stars with $\SNR>40$
(Table~\ref{tab:externalcomps}).  $P(\hbox{model})$ is our prior, and
$P(\hbox{data})$ is a normalisation which we can safely ignore.

\begin{figure}[htb]  
%\centerline{\includegraphics[width=9cm]{DR4DR5Comp.pdf}} 
\centerline{\includegraphics[width=9cm]{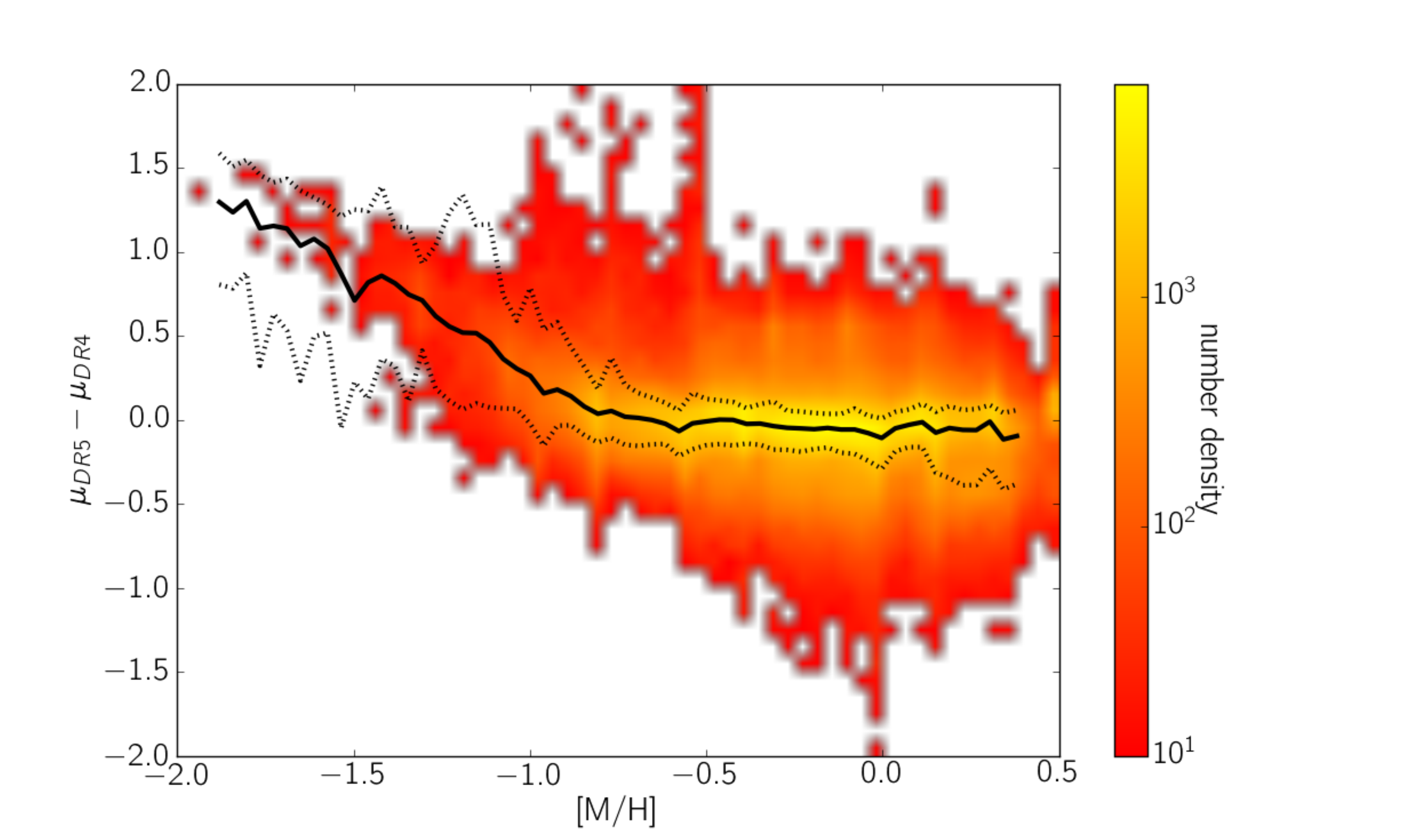}} 
\caption{Difference
between the derived distance modulus found in DR5 and DR4, as a function of
DR5 $\mh$. While there is some scatter at all metallicities, the clearest
trend is towards higher distances in DR5 at $\mh\lesssim-1$. This is due to
the absence of isochrones with $\mh<-0.9$ in the set used to derive distances
for DR4. The solid black line indicates the median in bins of 0.03 dex in $[M/H]$, 
and the dotted lines indicate the 1$\sigma$ equivalent range.
\label{fig:distancedifference}} 
\end{figure}

The method we use to derive the distances for DR5 is nearly the same as that
used by DR4, and we refer readers to \cite{binney14} for details. We apply
the same priors on stellar location, age, metallicity, and initial mass, and
on the line-of-sight extinction to the stars. These are all described in \S2
of \cite{binney14}.  The isochrone set that we use has been updated, to the
PARSEC v1.1 set \citep{bressan12}, which provide values for 2MASS $J$, $H$
and $K_{\rm s}$ magnitudes, so we
no longer need to obtain 2MASS magnitudes by transforming
Johnston-Cousins-Glass magnitudes, as we did when calculating the distances
for DR4. Whereas the isochrones used by \cite{binney14} went no lower in
metallicity than $Z=0.00220$ ($\mh=-0.914$), the new isochrones
extend to $Z=0.00010$ ($\mh=-2.207$) -- see
Table~\ref{tab:isochrones}.  The new isochrones have a clear impact on
distances to stars at lower metallicities
(Figure~\ref{fig:distancedifference}).  Experiments on a subset of stars
using isochrones more closely spaced in $Z$ found that the inclusion of more
isochrones has negligible impact on the derived properties of the stars.

\begin{table}
  \begin{center}
    \caption{Metallicities of the PARSEC v1.1 isochrones used, taking $Z_\odot = 0.0152$ and applying scaled solar composition, with  $Y=0.2485+1.78Z$.\label{table:Zs}}
    \begin{tabular}{rrr}
      \hline
      $Z$ & $Y$ & $\mh$ \\
      \hline
      0.00010 & 0.249 & -2.207 \\ 
0.00020 & 0.249 & -1.906 \\ 
0.00040 & 0.249 & -1.604 \\ 
0.00071 & 0.250 & -1.355 \\ 
0.00112 & 0.250 & -1.156 \\ 
0.00200 & 0.252 & -0.903 \\ 
0.00320 & 0.254 & -0.697 \\ 
0.00400 & 0.256 & -0.598 \\ 
0.00562 & 0.259 & -0.448 \\ 
0.00800 & 0.263 & -0.291 \\ 
0.01000 & 0.266 & -0.191 \\ 
0.01120 & 0.268 & -0.139 \\ 
0.01300 & 0.272 & -0.072 \\ 
0.01600 & 0.277 & 0.024 \\ 
0.02000 & 0.284 & 0.127 \\ 
0.02500 & 0.293 & 0.233 \\ 
0.03550 & 0.312 & 0.404 \\ 
0.04000 & 0.320 & 0.465 \\ 
0.04470 & 0.328 & 0.522 \\ 
0.05000 & 0.338 & 0.581 \\ 
0.06000 & 0.355 & 0.680 \\ 
      \hline
    \end{tabular}
  \end{center}
  \label{tab:isochrones}
\end{table}

The distance pipeline determines a full pdf, $P(\hbox{model}|\hbox{data})$, for all  the 
parameters used to describe the stars and their positions. We characterise this pdf in terms 
of expectation values and formal uncertainties for $\mh$, $\log_{10} (\tau)$, initial mass, 
and $\log_{10} (A_V)$ (marginalising over all other properties). 
For the distance we provide several characterisations of the pdf: expectation values and 
formal uncertainties for the distance itself ($s$), for the distance modulus ($\mu$) and 
for the parallax $\varpi$. As pointed out by \cite{binney14}, it is inevitable that the 
expectation values (denoted as e.g., $\left\langle s \right\rangle$) are such 
that $\left\langle s \right\rangle > s_{\left\langle \mu \right\rangle} > 1/\left\langle \varpi \right\rangle$ 
(where  $\log_{10}s_{\left\langle \mu \right\rangle} = 1+\left\langle \mu \right\rangle/5$ 
and $s$ is in ${\rm pc}$). In addition we provide multi-Gaussian fits to the pdfs in distance modulus. 

As shown in \cite{binney14}, the pdfs in distance are not always well
represented by an expectation value and uncertainty (which are conventionally
interpreted as the mean and dispersion of a Gaussian distribution).  A number
of the pdfs are double or even triple peaked (typically because it can not be
definitively determined whether the star is a dwarf or a giant), and
approximating this as single Gaussian is extremely misleading.  The
multi-Gaussian fits to the pdfs in $\mu$ provide a compact representation of
the pdf, and can be written as 
\begin{equation} 
%\[
\label{eq:defsfk}
P(\mu) = \sum_{k=1}^N {f_k\over \sqrt{2\pi\sigma_k^2}}
\exp\bigg(-{(\mu-\mu_k)^2\over2\sigma_k^2}\bigg), 
%\]
\end{equation} 
where the number of components $N$, the means $\mu_k$, weights $f_k$, and
dispersions $\sigma_k$ are determined by the pipeline.  DR5 gives these values 
as $number\_of\_Gaussians\_fit$ (for $N$), and for $k=1,2,3$ as 
$mean\_k$, $sig\_k$ and $frac\_k$ (corresponding to $\mu_k$, $\sigma_k$, 
and $f_k$ respectively). 

To determine whether a distance pdf is well represented by a given multi-Gaussian representation in $\mu$ we take bins in distance modulus of width $w_i = 0.2$, which contain a fraction $p_i$ of the total probability taken from the computed pdf and a fraction $P_i$ from the Gaussian representation, and compute the goodness-of-fit statistic 
%\[
\begin{equation}
\label{eq:defsF}
F = \sum_i \left(\frac{p_i}{w_i}-\frac{P_i}{w_i}\right)^2\tilde{\sigma} w_i,
\end{equation}
%\]
where the weighted dispersion
\[
\tilde{\sigma}^2 \equiv \sum_{k=1,N} f_k \sigma_k^2
\]
is a measure of the overall width of the pdf. Our strategy is to represent
the pdf with as few Gaussian components as possible, but if the value of $F$
is greater than a threshold value ($F_t=0.04$), or the dispersion associated
with the model differs by more than 20 per cent from that of the complete
pdf, then we conclude that the representation is in not adequate, and add
another Gaussian component to the representation (to a maximum of 3
components). For around 45 per cent of the stars, a single Gaussian component
proves adequate, while around 51 per cent are fitted with two Gaussians, and
only 4 per cent require a third component. The value of $F$ is provided in
the database as {\tt CHISQ\_Binney} and we also include a flag (denoted
{\tt FitFLAG\_Binney}) which is non-zero if the dispersion of the fitted model
differs by more than 20 per cent from that of the computed pdf. Typically the
problems flagged are rather minor \citep[as shown in Fig.~3 of][]{binney14}.

\begin{figure}[htb]  
%\centerline{\includegraphics[width=9cm]{CMdiagramDR5.pdf}}
\centerline{\includegraphics[width=9cm]{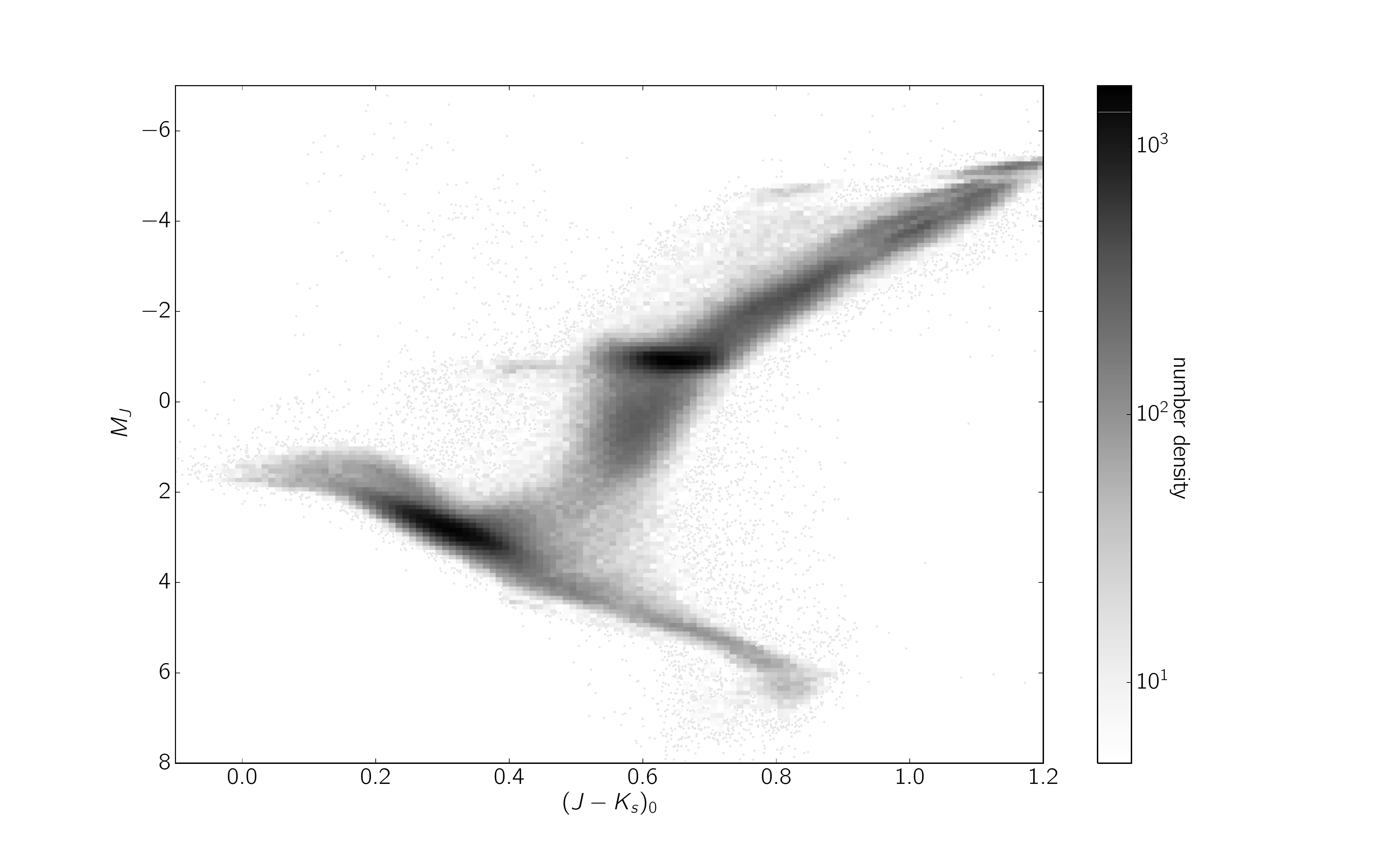}}
\caption{Absolute colour magnitude diagram, derived from the pipeline outputs, for all stars in the filtered distance catalogue. The values are found from the values in the catalogue as $M_J = J-\left\langle \mu \right\rangle-0.282\times\tilde{A_V}$ and $(J-K_s)_{\rm abs}=J-K_s-0.17\times\tilde{A_V}$ where $\log_{10}\tilde{A_V}=\left\langle \log_{10} A_V\right\rangle$
Shading indicates the number of stars in bins of width $0.01$ mag in $(J-K_s)_0$ and 0.1 mag in $M_J$. If there are less than 5 stars in a bin, they are represented as points.
\label{fig:CMdiagram}} 
\end{figure}

Using the derived distance moduli and extinctions, it is simple to plot an
absolute colour-magnitude diagram, from which we can check that the pipeline
produces broadly sensible results. It was inspection of this plot which led
us to filter out dwarfs with $T_{\rm eff}\leq4000\,$K and hot, metal poor
stars, because they fell in implausible regions of the diagram. We show this
plot, constructed from the filtered data, in Figure~\ref{fig:CMdiagram}.

To test the output from the pipeline, we compare the derived parallaxes (and uncertainties) 
with those found by {\it Hipparcos} \citep{vanLeeuwen07} for the ${\sim}5\,000$ stars common 
to the two catalogues. It is important to compare parallax with parallax, because, as noted 
before, $\left\langle \varpi \right\rangle>1/\left\langle s \right\rangle$, so this is the only fair test. 
We therefore consider the statistic $\Delta$, which we define as 
%\[ 
\begin{equation}
\label{eq:Delta}
\Delta = \frac{\left\langle \varpi_{\rm DR5} \right\rangle - \varpi_{\rm H}}
{ \sqrt{\sigma_{\varpi,{\rm DR5}}^2+\sigma_{\varpi,H}^2} },
\end{equation}
%\] 
where $\varpi_{\rm H}$ is the quoted {\it Hipparcos} parallax, and
$\sigma_{\varpi,H}$ the quoted uncertainty, while $\varpi_{\rm DR5}$ and
$\sigma_{\varpi,{\rm DR5}}$ are the same quantities from the distance
pipeline. Ideally, $\Delta$ would have a mean value of zero and a dispersion
of unity. 

\begin{figure}[htb]  
\centerline{\includegraphics[width=4.5cm]{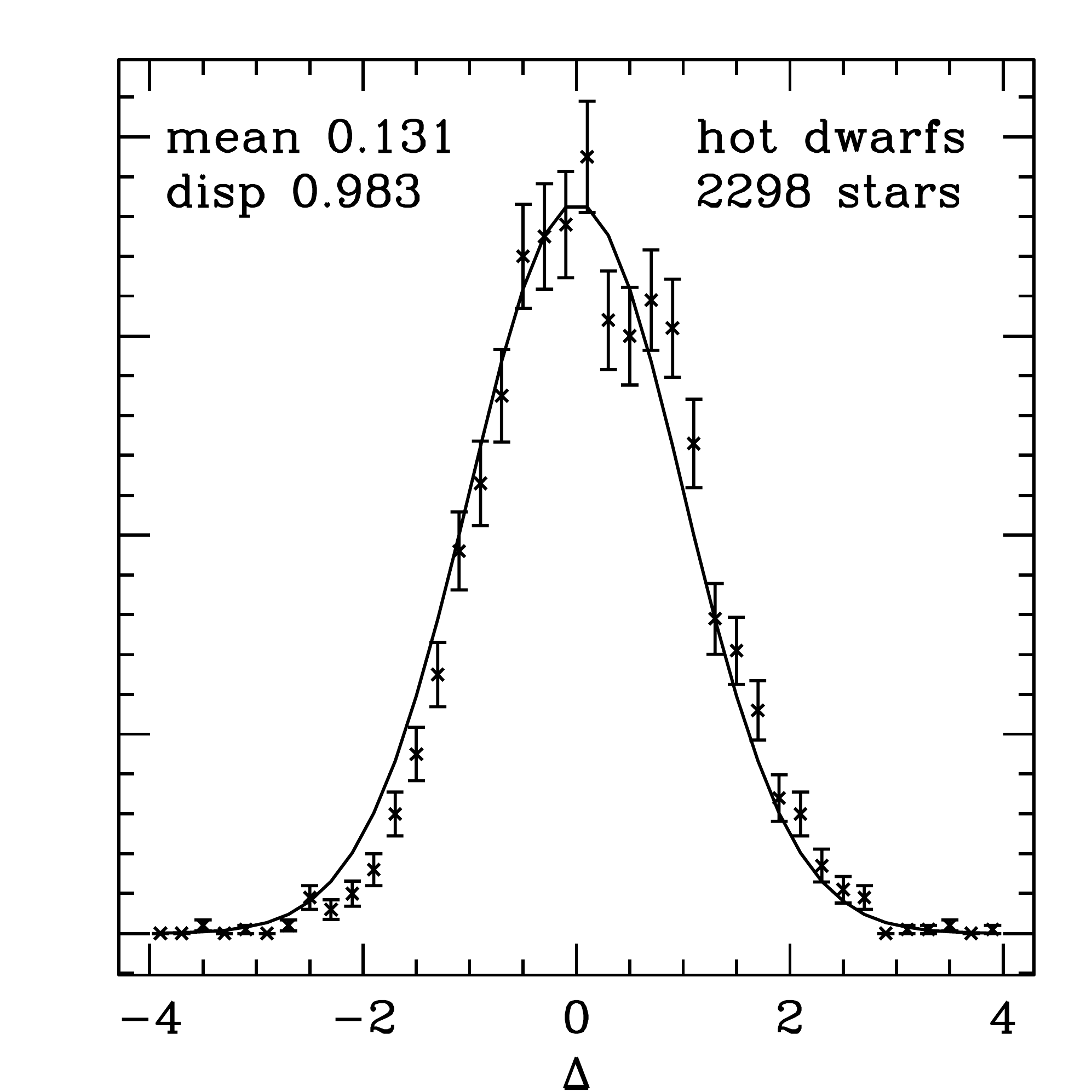}
	\includegraphics[width=4.5cm]{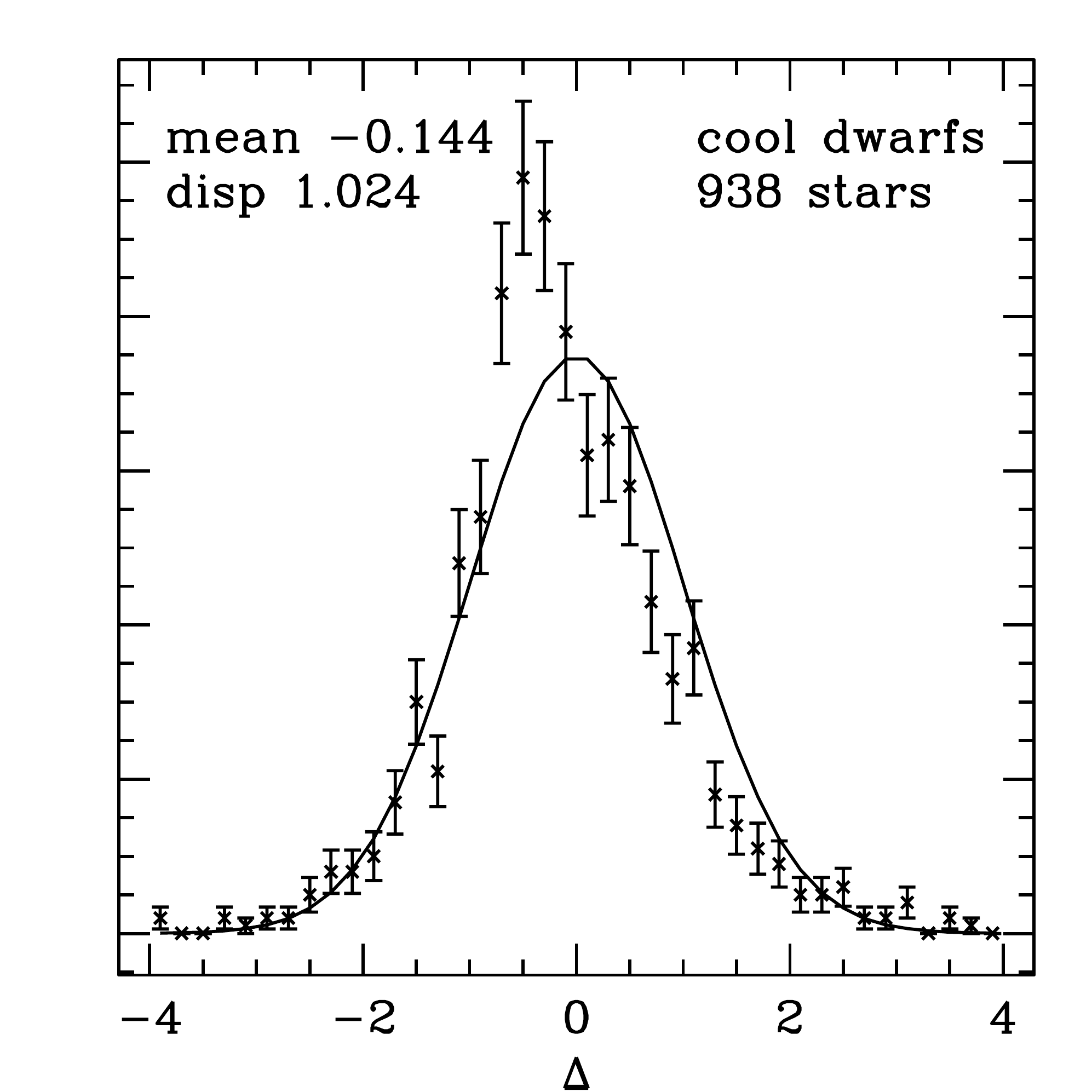}}\vspace{0.5mm}
	\centerline{\includegraphics[width=4.5cm]{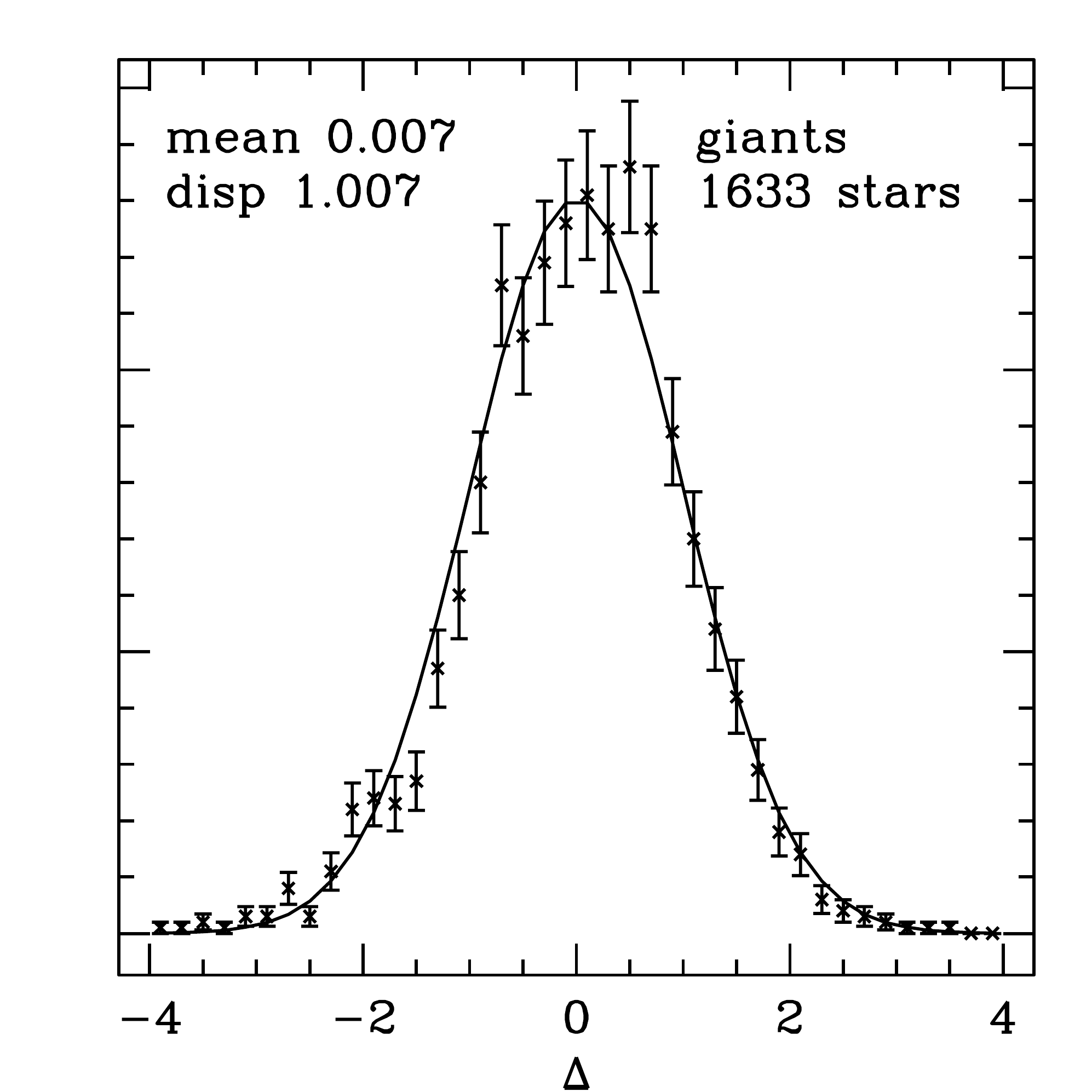}
	\includegraphics[width=4.5cm]{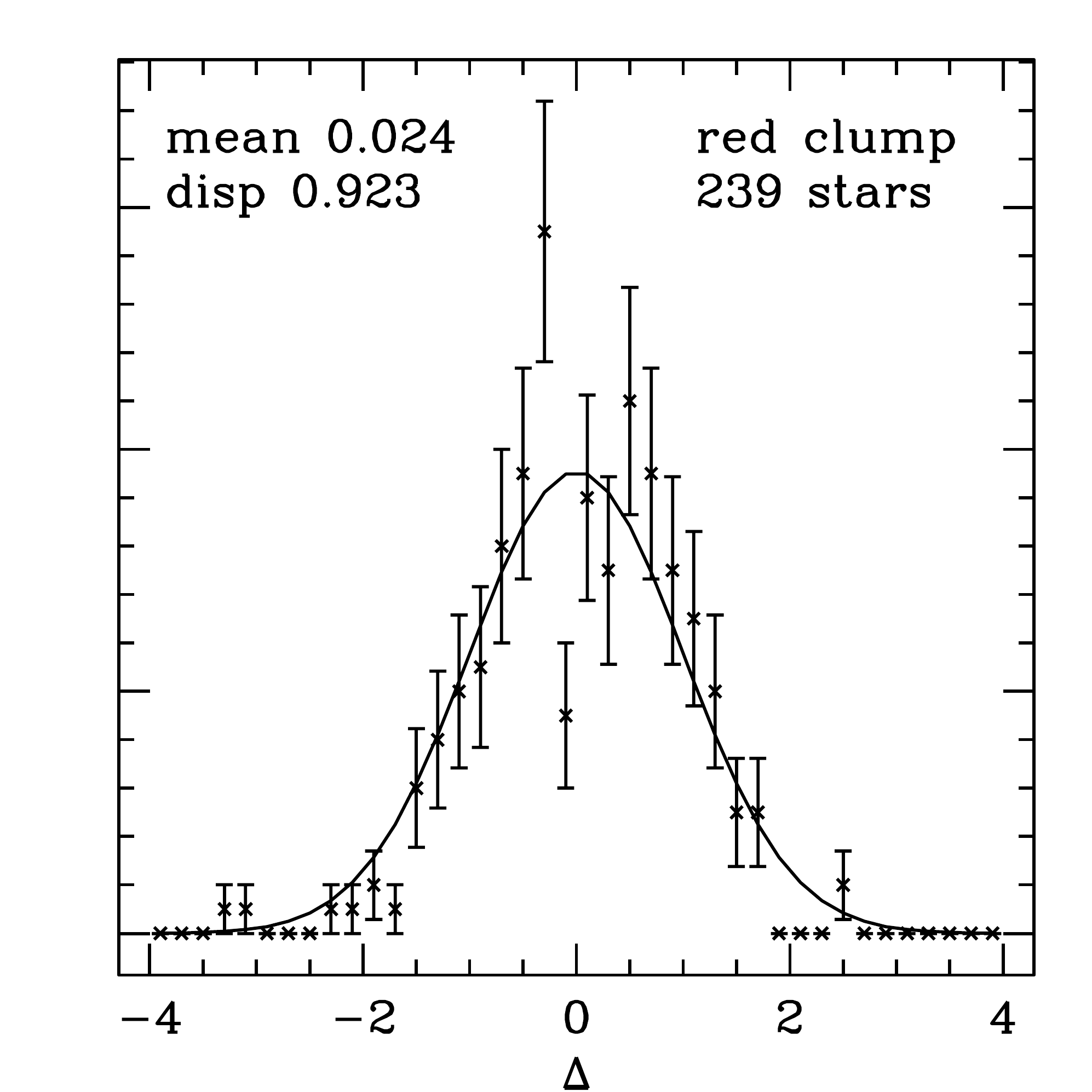}}
\caption{A comparison of the parallax estimates found by the DR5 pipeline and
those found by Hipparcos. The statistic $\Delta$ is defined in
equation~\ref{eq:Delta}, and ideally has a mean of zero and dispesion of
unity. The points are a histogram of $\Delta$, with error bars given by the
expected $\sqrt{N}$ Poisson noise in each bin. The solid line is a Gaussian
with the desired mean and dispersion. Stars are divided into `hot dwarfs'
($T_{\rm eff}>5500$K and $\log{g}>3.5$), `cool dwarfs' ($T_{\rm eff}<5500$K
and $\log{g}>3.5$), and `giants' ($\log{g}>3.5$), as labelled. The `red
clump' stars are a subset of the giants, with $1.7<\log g<2.4$ and
$0.55<{J}-{K}_{\rm s}<0.8$.  \label{fig:HippDist}} 
\end{figure}

In Figure~\ref{fig:HippDist} we plot a histogram of the values of $\Delta$
for these stars separated into giants ($\log g\leq3.5$), cool dwarfs ($\log
g>3.5$ and $T_{\rm eff}\leq5500$K), and hot dwarfs ($\log g>3.5$ and $T_{\rm
eff}>5500\,$K), as well as for the subset of giants that we associate with the
red clump ($1.7<\log g<2.4$ and $0.55<{J}-{K}_{\rm s}<0.8$). We have
`sigma clipped' the values, such that none of the (very few) stars with $
|\Delta |>4$ contribute to the statistics. The results are all pleasingly
close to having zero mean and dispersion of unity, especially the giants. We
tend to slightly overestimate the parallaxes of the hot dwarfs, and slightly
underestimate those of the cool dwarf (corresponding to \emph{under}estimated
distances to the hot dwarfs and \emph{over}estimated distances to the cool
dwarfs. This represents an improvement over the comparable figures for DR4,
except for a very slightly worse mean value for the cool dwarfs (and even for
these stars, there is an improvement in that the dispersion is now closer to
unity).

With the release of the TGAS data it becomes possible to construct a figure
like Figure~\ref{fig:HippDist} using the majority of RAVE stars. Thus much
more rigorous checks of our distance (parallax) estimates are now possible.
When that has been done and and systematics calibrated out, we will be able
to provide distances to all stars that are more accurate than those based on
either DR5 or TGAS alone, by feeding the TGAS data, including
parallaxes, into the distance pipeline. 

Where stars have been observed more than once by RAVE, we recommended using
the distance (and other properties) obtained from the spectrum with the
highest signal to noise ratio.  However, DR5 reports distances from each
spectrum.

\section{Infrared Flux Method Temperatures}\label{sec:IRFM}

The Infrared Flux Method (IRFM)
\citep{bs77,blackwell79} is one of the most accurate techniques to derive
stellar effective temperatures in an almost model independent way. The basic
idea is to measure for each star its bolometric flux and a monochromatic
infrared flux.  Their ratio is then compared to that obtained for a surface
at $T_{\rm eff}$, i.e., $\rm \sigma T_{eff}{^4}$ divided by the theoretical
monochromatic flux. The latter quantity is relatively easy to predict for
spectral types earlier than $\sim\rm{M}0$, because the near infrared region
is dominated by the continuum, and the monochromatic flux is proportional to
$T_{\rm eff}$ (Rayleigh-Jeans regime), so  dependencies on
other stellar parameters (such as $\rm [Fe/H]$ and $\log g$) and model
atmospheres are minimized \citep[as extensively tested in the literature,
  e.g.,][]{aamr96,casagrande06}. The 
method thus ultimately depends on a proper derivation of stellar fluxes,
from which $T_{\rm eff}$ can then be derived. Here we adopt an updated version of
the IRFM implementation described in \citet{casagrande06} and \citet{c10} which has been
validated against interferometric angular diameters \citep{c14b} and combines
APASS $BVg'r'i'$ together with 2MASS $JHK_s$ to recover bolometric and
infrared flux of each star. The flux outside photometric bands (i.e.~the
bolometric correction) is derived using a theoretical model flux at a given
$T_{\rm eff}$, $\rm [Fe/H]$, $\log g$. An iterative procedure in $T_{\rm eff}$
is adopted to cope with the mildly model dependent nature of the bolometric
correction and of the theoretical surface infrared monochromatic flux. For each
star, we interpolate over a grid of synthetic model fluxes, starting with an
initial estimate of the stellar effective temperature and fixing $\rm [Fe/H]$
and $\log g$ to the RAVE values, until convergence is reached within 1\,K in
effective temperature.

In a photometric method such as the IRFM, reddening can have a non-negligible
impact, and must be corrected for. For each target RAVE provides an estimate
of $E(B-V)$ from \cite{schlegel98}. These values however are integrated over
the line of sight, and in the literature there are several indications
suggesting that reddening from this map is overestimated, particularly in
regions of high extinction \citep[e.g.][]{ag99,sf11}. To mitigate this
effect, we recalibrate the \cite{schlegel98} map using the intrinsic colour
of clump stars, identified as number overdensities in colour distribution
(and thus independently of the RAVE spectroscopic parameters).  We take the
2MASS stellar catalogue, tessellate the sky with boxes of $10 \times 10$
degrees, and select stars in the magnitude range of RAVE.  Within each box we
can easily identify the overdensity due to clump stars, whose position in
$J-K_s$ colour is little affected by their age and metallicity. Thus, despite
the presence of metallicity and age gradients across the Galaxy
\citep[e.g.][]{bsp14,c16}, we can regard the average $J-K_s$ colour of clump
stars as a standard crayon. We take the sample of clump stars from
\cite{c14a}, for which reddening is well constrained, and use their median
unreddened $(J-K_s)_0$ against the median measured at each $n-$tessellation,
to derive a value of reddening at each location
$E(J-K)_n=(J-K_s)_n-(J-K_s)_0$.  We then compare these values of reddening
with the median ones obtained using the \cite{schlegel98} map over the same
tessellation. The difference between the reddening values we infer and those
from the \cite{schlegel98} map is well fitted as function of $\log(b)$ up to
$\simeq 40^{\circ}$ from the Galactic plane. We use this fit to rescale the
$E(B-V)$ from the \cite{schlegel98} map, thus correcting for its tendency to
overestimate reddening, while at the same time keeping its superior spatial
resolution ($\sim$ arcmin). For $|b| \gtrsim 40^{\circ}$ the extinction is low and well described.

\begin{figure}[htb]  
\includegraphics[width=9cm]{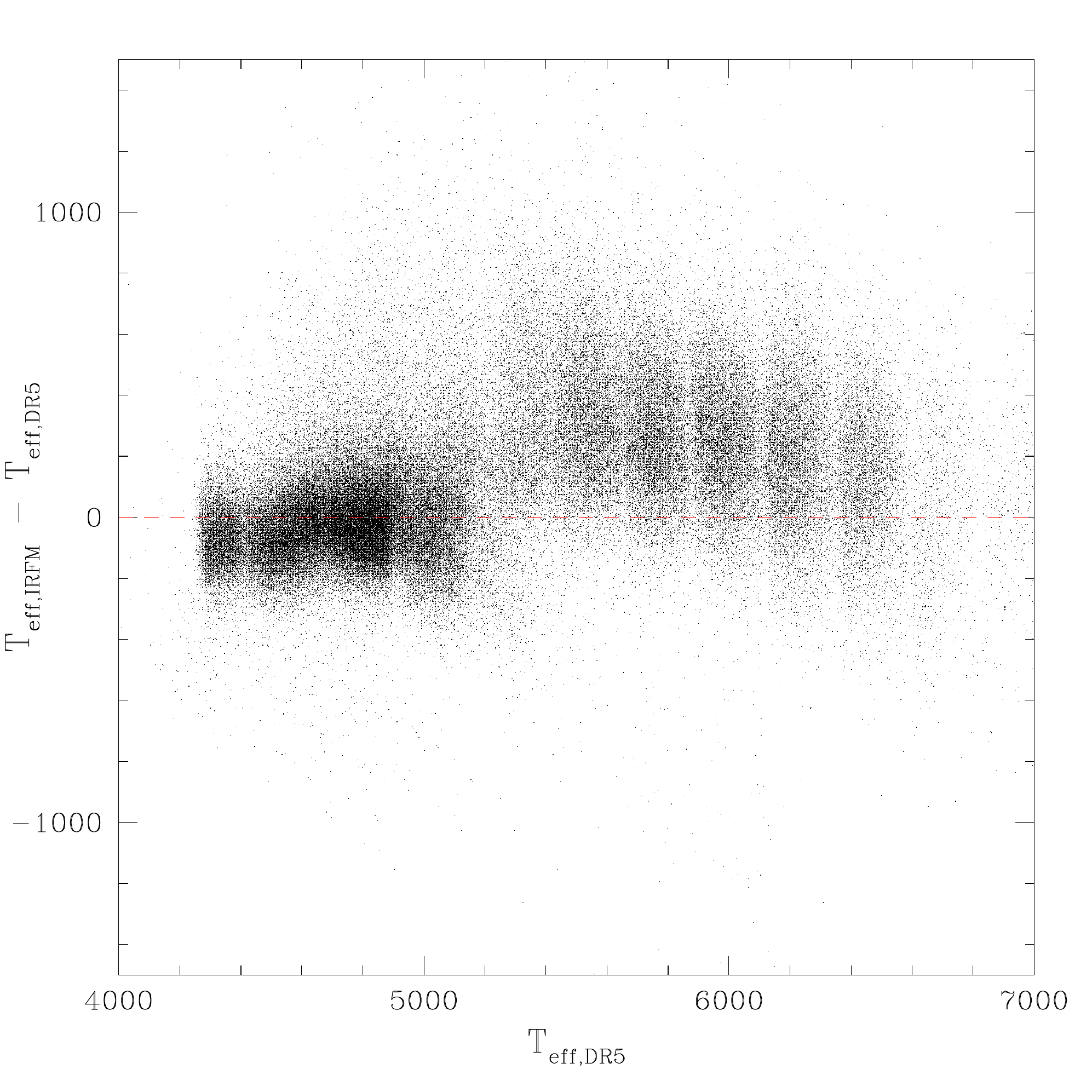}
\caption{A comparison between the temperatures derived from the IRFM with those
in DR5.  Only stars with  $\SNR> 50$ and ${\tt AlgoConv} = 0$ are shown.  
The giants, with $\rm T_{eff} <$ 5500, have temperatures that agree well with IRFM
temperatures, but there is a systematic offset to the main-sequence/turn-off stars.
The pixelisation, an artifact of the RAVE stellar parameter pipeline, is apparent as vertical bands. 
\label{irfmtemps}}
\end{figure}

Figure~\ref{irfmtemps} shows a comparison between the DR5 temperatures and
those from the IRFM, $T_{\rm eff,IRFM}$.  Stars with temperatures cooler than
$T_{\rm eff}\sim$5300\,K show a good agreement between $T_{\rm eff,IRFM}$ and
$T_{\rm eff,DR5}$, with a scatter of $\sim$150\,K, which is the typical
uncertainty of the RAVE temperatures.  Stars hotter than $T_{\rm eff}$ 5300\,K
have an offset in temperature, in the sense that $\rm T_{eff,IRFM}$ is
approximately 350\,K warmer than $T_{\rm eff,DR5}$ at 5500\,K.  As the
temperature increases, the temperature offset decreases to $\sim$100\,K at 7000\,K.
This offset is consistent to what is seen in comparison between RAVE and other 
datasets (see e.g., Table~\ref{tab:externalcomps} and Figures~\ref{field_teff1} 
and \ref{stelparam_lamo}) thus suggesting that the 
offset is unlikely to stem from the IRFM only.  From
Table~\ref{tab:externalcomps} it is evident that the IRFM temperatures for 
especially the cool dwarfs are in better agreement with high-resolution studies than
the spectroscopic DR5 temperatures.

Nevertheless, we remark that 
various reasons might be responsible for this trend:  first, the
rescaling of the \cite{schlegel98} map is based on clump stars, so it is not
surprising that best agreement is found for giant stars. Turn-off and main
sequence stars are on average closer than intrinsically brighter giants, so
despite of the rescaling, $E(B-V)$ will  on average still be overestimated
implying hotter effective temperatures in the IRFM. Also, at the hottest
$T_{\rm eff}$ the contribution of optical photometry becomes increasingly
important so does proper control over the standardization, and absolute
calibration of the APASS photometry. 

\section{Asteroseismically Calibrated Red Giant Catalog}\label{sec:AC}
\label{aseis}

Asteroseismic data provide a very accurate way to determine surface 
gravities of red giant stars \citep[e.g.,][]{stello08, mosser10, bedding11}.
When solar-like pulsations in red giants can be detected, the pulsation frequencies, 
such as the average large frequency separation, $\langle \Delta \nu \rangle$, and the frequency of
maximum oscillation power, $\nu_{\rm max}$, can be used to obtain the density and 
surface gravity of the star.  Exquisite datasets with which to search
for oscillations have arisen in the space-based missions {\it CoRoT} and {\it
Kepler},
and it has already been shown that their long dataset in time gives the frequency 
resolution needed to extract accurate estimates of the basic parameters 
of individual modes covering several radial orders, such as frequencies, 
frequency splittings, amplitudes, and damping rates.

Pulsations in red giants have significantly longer periods and larger
amplitudes than solar-type stars, so 
oscillations may be detected in fainter (more numerous) targets observed
with long cadence.  Further, the seismic $\log g$ values are almost fully independent 
of the input physics in the stellar evolution models that are used \citep[e.g.,][]{gai11}.
This makes the use of red giants with asteroseismic $\log g$ values ideal
to check and calibrate surface gravities that are obtained spectroscopically.

V16 present 72 RAVE stars with solar-like oscillations
detected by the K2 mission.  
The finite length and cadence of the observations of a K2 field means
there is a limit in our ability to extract properties from solar-like oscillations,
and hence for how well $\langle \Delta \nu \rangle$, and $\nu_{\rm max}$ 
can be obtained \citep[e.g.,][]{davies16}.  This means that the asteroseismic 
calibration based on the K2 stars is limited to roughly the range of 2.1 $< \log g <$ 3.35~dex. 
For the colour interval $0.50<J-K_s<0.85$,
which was shown to be appropriate for selecting red giant stars in the Kepler
field, the spectroscopic gravities present in the RAVE
catalogue are calibrated gainst the seismic gravities.  This calibration is a function only
of RAVE $\log g_{\rm p}$ and does not depend on photometric colour, metallicity or
SNR.  Whereas the \citet{schlegel98} reddening maps indicate that the
$(J-K)$ reddening in the K2 field is negligible, RAVE observes many reddened stars.
Therefore, the dereddened colour range is kept unchanged, and DR5 includes 
$\log g$ calibrated according to V16 only when
the dereddened colour $(J-K_s)_0$ lies in the interval $(0.50,0.85)$. 

There are 207\,050 RAVE stars that fall within $0.50<(J-K_s)_0<0.85$;
200\,524 of these have a RAVE $\log g$, enabling the application of 
an asteroseismic calibration.  Because of the RAVE $\log g$
uncertainties, misclassifications of red giants can occur, i.e., red
giants can have gravities that indicate they are dwarfs or supergiant stars.
Therefore each asteroseismically calibrated RAVE star has a flag, {\tt
Flag050}, indicating if the seismically calibrated $\log g$, $\log g_{sc}$, and
the DR5 $\log g_{\rm p}$ are within 0.5 dex of each other.  The flag {\tt Flag\_M} specifies if
all 20 classification flags of \citet{matijevic12} point to the star being
``normal", which likely means the star is indeed a typical red giant.
Therefore, stars with both {\tt Flag050}=1 and {\tt Flag\_M}=1 point to
an especially desirable sample of asteroseismically calibrated
giants.   

\begin{figure}
\includegraphics[width=\linewidth]{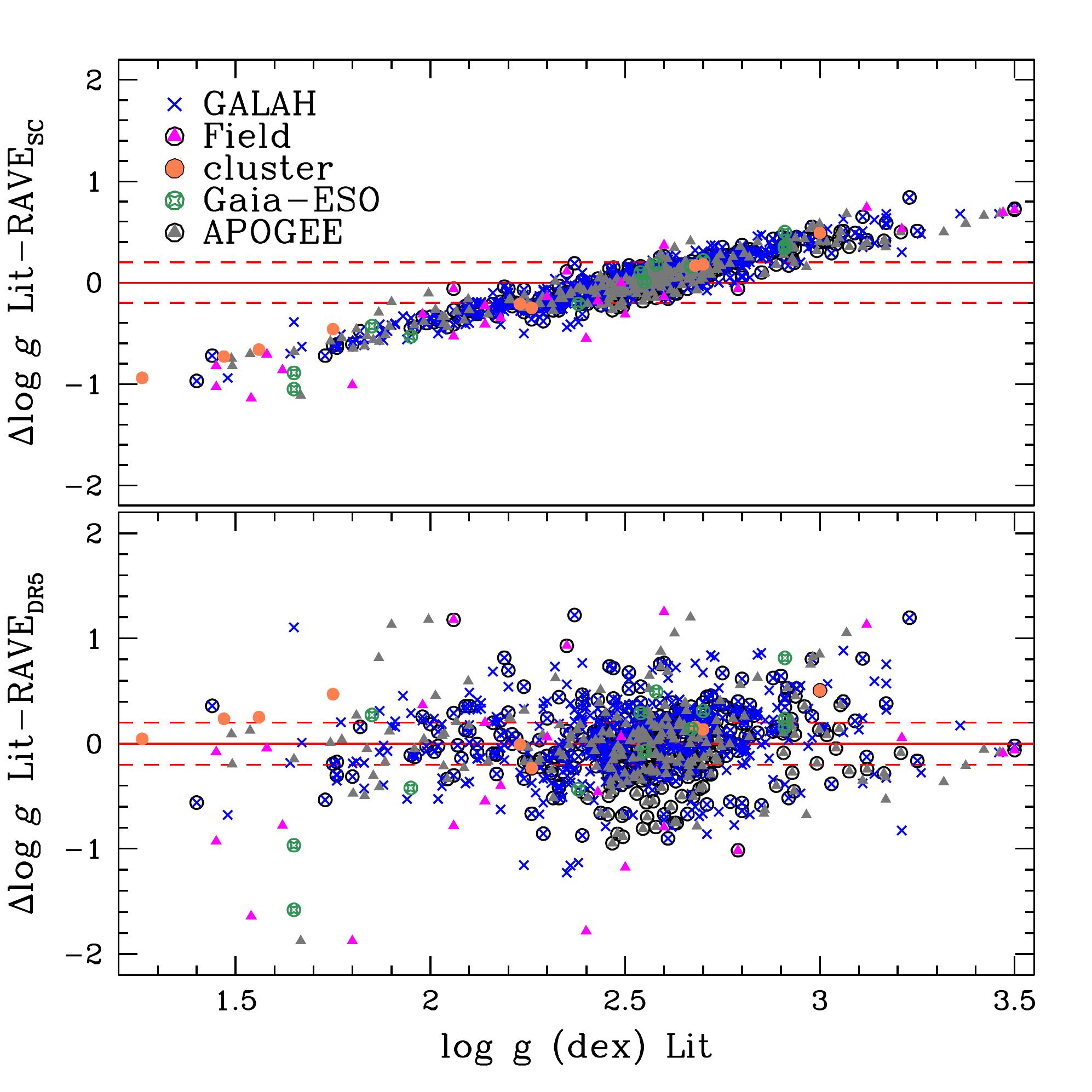}\\
\caption{ {\it Top:}  The difference in the asteroseismically calibrated gravities, $\log g_{sc}$ 
and that from various sources in the literature as a function of literature $\log g$.
Only stars with SNR $>$ 40, {\tt Flag\_M}=1 and {\tt Algo\_Conv}=0 are shown.  
The black open circles designate those stars with {\tt Flag\_050}=1, which in general
are the stars at the extremes of the calibration. 
{\it Bottom:}  The same stars as in the top panel, but the gravities in the main DR5 catalog 
are used.
\label{stelparam_dr5_lg}}
\end{figure}

Figure~\ref{stelparam_dr5_lg} shows $\log g_{sc}$ compared to the
gravities from the RAVE stars observed by the 
APOGEE, GALAH and Gaia-ESO surveys, as well
as the RAVE cluster and external stars (from \S\ref{sec:EV}).
The scatter about these 906 stars with SNR $>$ 40, {\tt Flag\_M}=1 and {\tt Algo\_Conv}=0 
is $\sigma \log g_{sc}$ $=$ 0.35~dex.  This is a 12\% smaller scatter than when 
using the RAVE DR5 $\log g$ from the main catalog.
When additionally imposing the {\tt Flag\_050}=1 criterion, the $\sigma \log g_{sc}$ $=$ 0.26~dex,
which is a 25\% smaller scatter than when using the RAVE DR5 $\log g$.  

Tables~\ref{tab:externalcomps} and \ref{tab:externalcomps_1} summarise 
how $\log g_{sc}$ compares with external results.  The {\tt Flag\_M}=1 criterion
is implemented in these comparisons.

\begin{figure}[htb]  
\includegraphics[width=\linewidth]{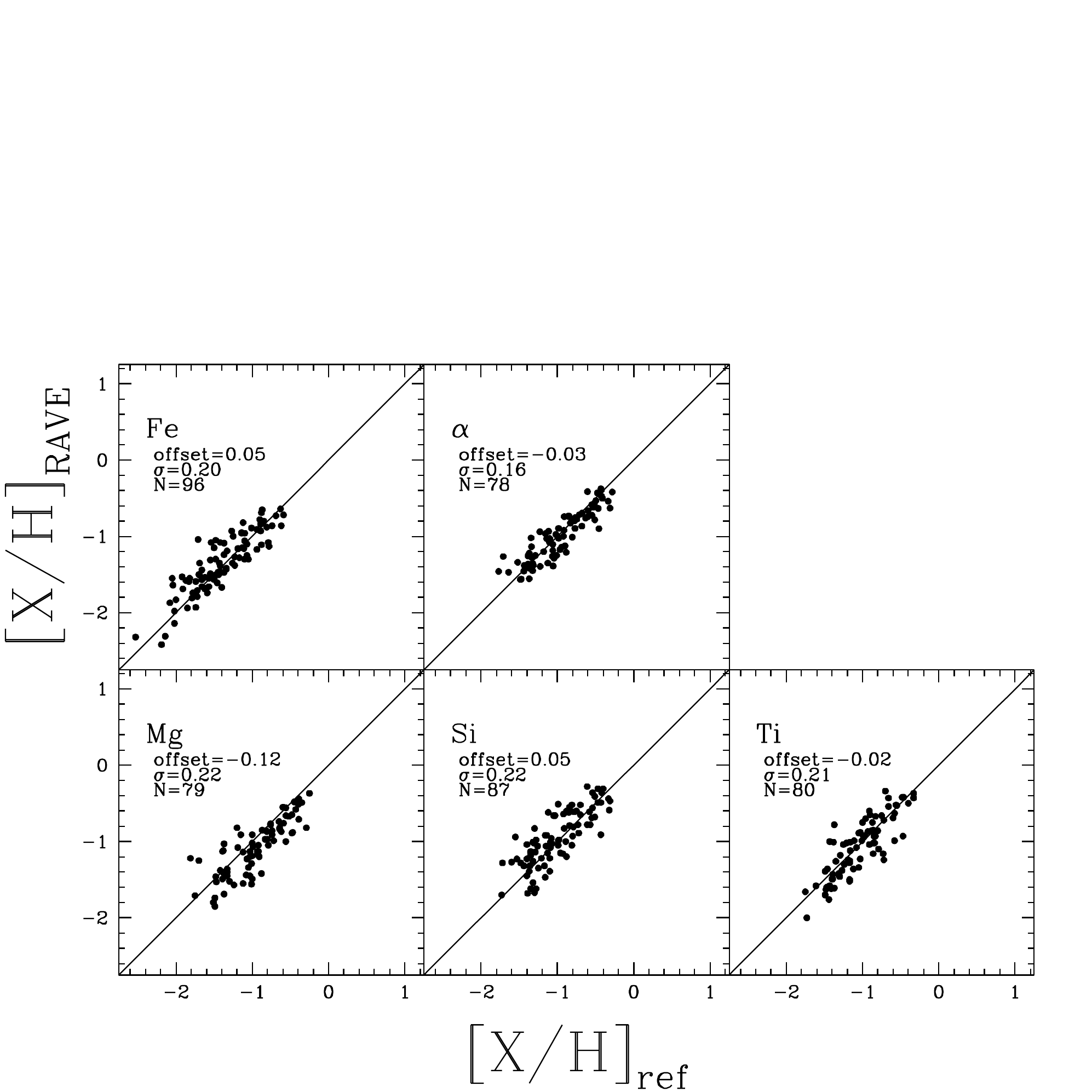}\\
\includegraphics[width=\linewidth]{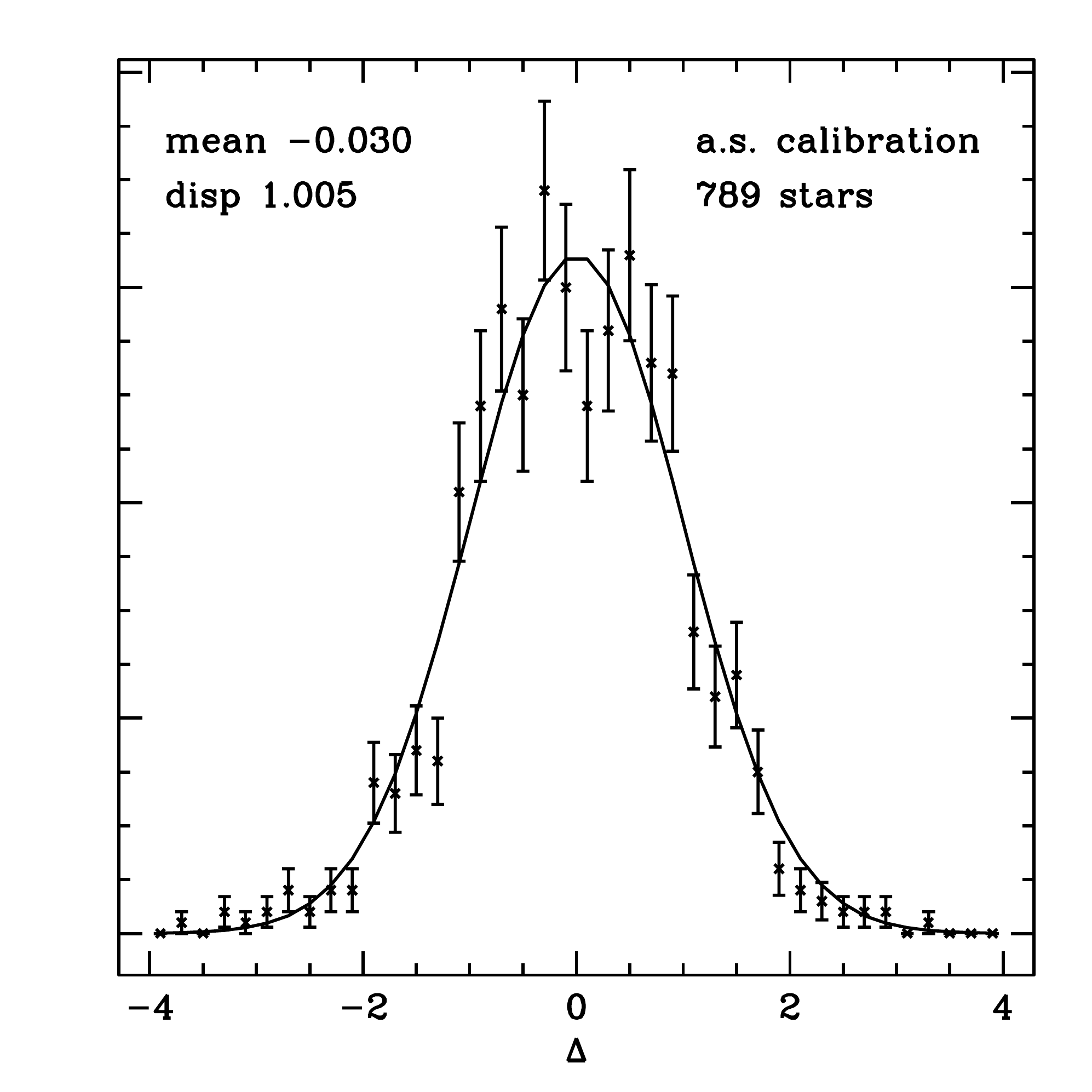}
\caption{A comparison of the elemental abundances from the 
RAVE chemical pipeline (top) and parallax estimates found from the DR5 distance 
pipeline but using $\log~g_{sc}$ and IRFM temperatures as an input.
Only stars with {\tt Flag\_M}=1 are considered.  
\label{LgCal_hipp_2flags}}
\end{figure}

Combining the $\log g_{sc}$ with the temperatures from the IRFM, the RAVE 
chemistry (\S\ref{chemicalpipeline}) and distance pipeline(\S\ref{sec:D}) are re-run.
Neither the uncertainty in chemical pipeline nor the uncertainty in distance 
changes when using the more accurate
$\log g_{sc}$ and IRFM temperatures as an input, as seen
in Figure~\ref{LgCal_hipp_2flags}. %(see Figure~\ref{fig:HippDist}).
The seismically calibrated giants are presented in a separate table, along with
the elemental abundances and distances derived.

\section{Use of different RAVE stellar parameters}\label{sec:BP}

\subsection{ DR5 main catalog vs RAVE-on }
While our official DR parameters are constantly under improvement, other approaches 
to determine parameters from RAVE spectra have become public. One example is the result
from C16, who present the RAVE-on catalog by the data-driven approach {\it The Cannon}.  
In short, this method is based on training the data on a set for which more information is 
known by independent means (i.e. spectra of the stars at other wavelength domains, 
asteroseismic observations, etc).  The disadvantage however is that the performance of 
the results relies fully on the training set. 
For example, as seen in C16, if the training set does not contain 
metal-poor stars, the derived metallicities from survey stars will lack a metal-poor population 
as well.  
The RAVE training sample used in C16 was inhomogeneous,
using RAVE overlap stars from APOGEE, \citet{fulbright10} and \citet{ruchti11} for the giants and 
RAVE overlap stars from LAMOST and the fourth RAVE data release for the main-sequence stars.
Unlike for the giants, the training sample for the main-sequence stars did not have known elemental
abundances, so no elemental abundances could be derived for main-sequenceto be 
non-trivial stars.

The main RAVE DR5 catalog, on the other hand, is based on stellar physics -- the use
of a grid of synthetic spectra over a large parameter space is utilised to derive stellar 
parameters.  Therefore for each star there is a physical justification ensuring the 
coherence of the obtained stellar parameters.  This leads to cases in which no feasible 
match to a theoretical spectra can be made, and so unlike in {\it The Cannon}, there
are instances in which the algorithm does not converge.  Also, stellar parameters
are obtained along the gridlines of the synthetic spectra, leading to pixelation of the values,
different visually to the smooth interpolation of {\it The Cannon}.

\begin{figure}
\includegraphics[width=\linewidth]{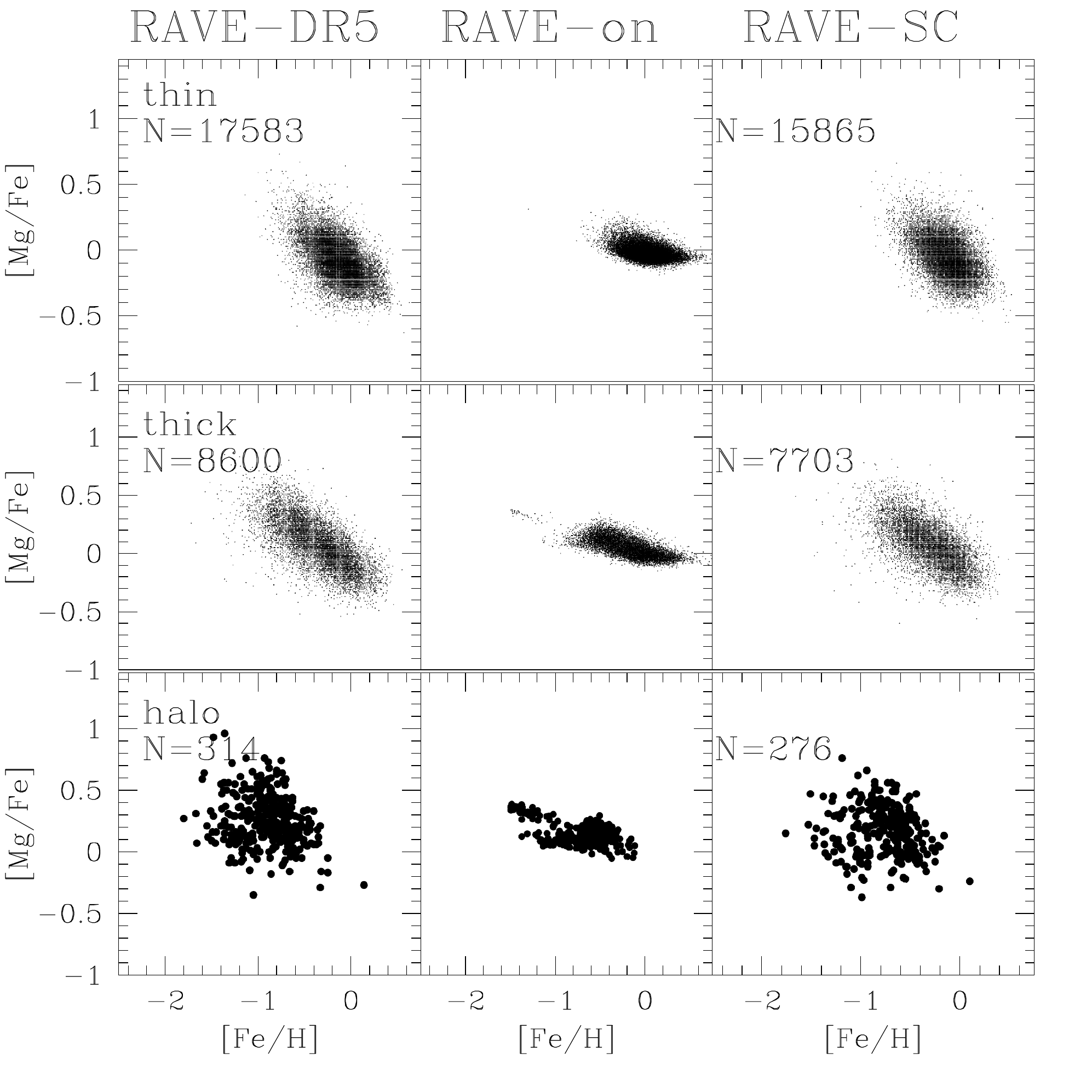}
\caption{Abundance ratio $\rm [Mg/Fe]$ versus the metallicity $\rm [Fe/H]$ for the thin disk 
component (top), the thick disk component (middle panel), and the halo component (bottom panel)
for parameters from RAVE-on, RAVE DR5 and the seismically calibrated RAVE stars.  
\label{refreport}}
\end{figure}

Figure~\ref{refreport} shows the metallicities and Mg elemental abundances of thin disk,
thick disk and halo RAVE stars for the RAVE DR5 and RAVE-on stars.  The maximum 
distance above the plane ($z_{max}$), rotational velocity and eccentricity were 
used to separate between these components
as described by \citet{boeche13b}.  These parameters were
computed by integrating the orbits of the RAVE stars 
using {\tt galpy} \citep{bovy15}, where the input parameters were the radial velocities and 
distances presented here, as well as the TGAS proper motions.  We opted 
to not use the TGAS parallaxes to determine distances, as this is 
non-trivial \citep{astraatmadja16, bailerjones15}.

Figure~\ref{refreport} illustrates the narrower chemical sequences of RAVE-on, due in part 
to smaller formal uncertainties in  $\rm [Mg/Fe]$ and $\rm [Fe/H]$, and the smooth 
interpolation of the stellar parameters (i.e., no pixelisation).  It can also be seen that
RAVE DR5 has a larger sample of stars with elemental abundances, and a more physical
distribution for stars with $\rm [Fe/H] < -$1~dex.  This is due to the difficulty of obtaining
main-sequence stars needed to train {\it The Cannon} (C16).

Table~\ref{tab:externalcomps} quantifies the agreement from external stars of the $T_{\rm eff}$, 
$\log g$ and $\rm [Fe/H]$ presented in RAVE-on and in RAVE DR5.  C16 performed external
validation of the RAVE-on stellar parameters on cool stars (F,G, and K stars), and here we extend this.
There is no significant difference in the precision when comparing the RAVE-on and RAVE DR5
stellar parameters to those from high-resolution stars.  RAVE-on lacks metal-poor stars in 
the training sample, leading to a worse agreement for stars with $\rm [Fe/H]$ metallicities 
more metal-poor than $-$1~dex.  It also is on a different metallicity scale than RAVE DR5, 
on average 0.15~dex more metal-poor than RAVE DR5.  There are more RAVE stars with derived
stellar parameters, $T_{\rm eff}$, $\log g$ and $\rm [Fe/H]$, in RAVE-on, and more stars with
elemental abundances in RAVE DR5.

\subsection{ DR5 main catalog $T_{\rm eff}$ vs IRFM $T_{\rm eff}$}
The IRFM temperatures and those from the main DR5 are similar, as shown in 
Figure~\ref{irfmtemps}, and as discussed in \S\ref{sec:IRFM}.  However, 
there is better agreement between RAVE stars observed from high-resolution 
studies and $T_{\rm eff,IRFM}$ (see Table~\ref{tab:externalcomps}).
Moreover, $T_{\rm eff,IRFM}$ is available for 95\% of the RAVE stars, 
and is independent of SNR.  Temperatures from the IRFM are critical for the RAVE
stars that were released in DR1, because during the first year of
RAVE operations, no blocking filter was used to isolate the spectral range
required and as a result, the spectra collected were contaminated by the
second order.  Hence, although the determination of radial velocities is
still straight-forward, stellar parameters cannot be reliably determined from
the spectra.  IRFM temperatures are further especially valuable for stars 
with temperatures cooler than 4000~K and for stars hotter than 8000~K, 
as the main DR5 catalog is only able to determine temperatures for stars 
within 4000 - 8000~K.

\subsection{ DR5 main catalog $\log g$ vs $\log g_{sc}$ }
For RAVE stars with colors between $0.50<(J-K_s)_0<0.85$, a direct asteroseismic 
calibration can be carried out, as described by V16.
This calibration uses the raw DR5 $\log g$ as a starting point, and
therefore any problems in the derivation of the raw DR5 $\log g$
is also carried over to the $\log g_{sc}$.  
Figure~\ref{stelparam_dr5_lg} shows how $\log g_{sc}$ compares to
$\log g_{DR5}$ for external stars observed with high resolution.
The $\log g_{sc}$ agrees with external estimates $\sim$12\% better than
$\log g_{DR5}$.  However, we note the linear relation between $\log g_{sc}$ and
gravities from the literature, suggesting minor biases are present in $\log g_{sc}$, in a sense
that $\log g$ values less than 2.3~dex are underestimated and 
$\log g$ values greater than 2.8~dex are overestimated.  This can be
minimised by selecting stars with  {\tt Flag050}=1.  There is no
correlation between literature $\log g$ and $\log g_{DR5}$.

\section{Differences between DR4 and DR5}\label{sec:diff}

RAVE DR5 differs from DR4 in a number of ways, as listed below. 
\begin{itemize}

\item  The DR5 RAVE sample is larger than DR4 by $\sim$30\,000 stars.  This is due in part
to the inclusion of the 2013 data, but mainly due to the improvement of the DR5
reduction pipeline, which now processes data on a fibre-by-fibre basis instead of
a field-by-field basis.  

\item The DR1 data are now ready to be ingested through the same reduction
pipeline, improving the homogeneity of the DR5 radial velocities compared
to those in DR4.

\item The error spectra now available for all RAVE stars have yielded more accurate
uncertainties on the RAVE radial velocities and stellar parameters,
especially for low-SNR and hot stars.  We plan to extend the error spectra analysis 
to the chemical elements in a future release.

\item A new $T_{\rm eff}$, $\log g$ and \mh~calibration has been applied,
increasing the accuracy of the stellar parameters by up to 15\%.  This
calibration is employed mainly because there are now RAVE stars with $\log g$
values determined asteroseismically (V16).  The metal-rich tail
of the RAVE stars has also been re-investigated, by increasing the number of
calibration stars in the super-solar metallicity regime.  Hence the updated
DR5 stellar parameters mainly improve the gravities of the giants and the
super-solar $\rm [M/H]$ stars.  Figure~\ref{stellparamdr4dr5} shows how the
atmospheric parameters in DR5 differ from those in DR4.

\begin{figure}
\includegraphics[width=\linewidth]{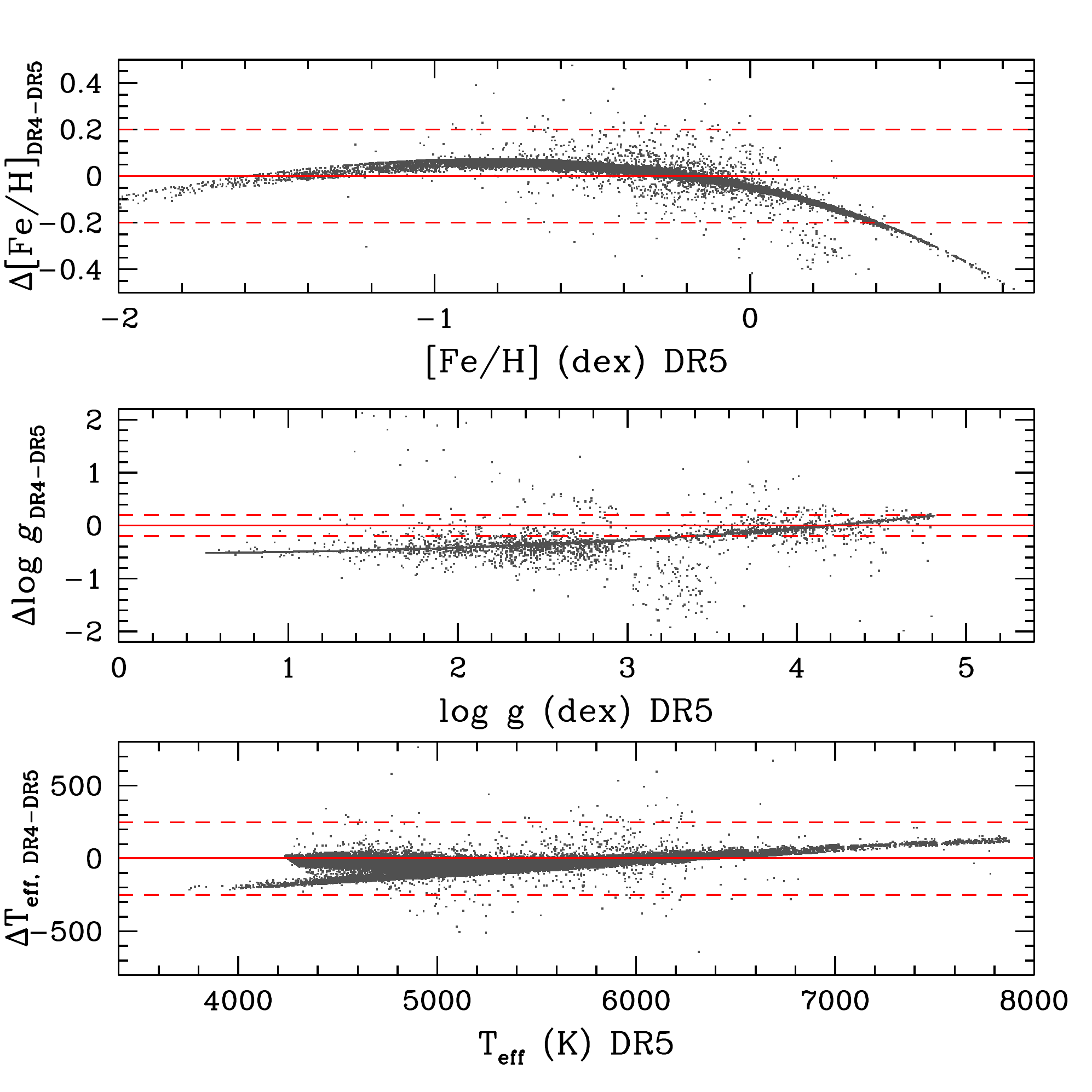}\\
\caption{The difference in the stellar parameters $T_{\rm eff}$ (bottom), 
$\log g$ (middle) and $\rm [M/H]$ (top) between RAVE DR4 and DR5.
Only stars with  $\SNR>40$ and ${\tt AlgoConv} = 0$ are shown.
\label{stellparamdr4dr5}}
\end{figure}

\item A sample of RAVE giants is provided for which the V16
asteroseismic calibration can be applied.  These $\log~g$ parameters are the
most accurate, but can only be applied to stars that fall within
$0.50<(J-K_S)_0<0.85$.

\item Although the chemical pipeline is the same as the one employed in DR4, the
stellar parameters fed into this pipeline are better calibrated, and hence
the resulting elemental abundances are slightly changed.  The $\rm [Fe/H]$ and
$\rm [X/Fe]$ abundances are shifted by $\sim0.1\,$dex to be more metal-rich
than in DR4.

\item The distance pipeline has been improved, especially for the metal-poor
stars.  In DR5, we list individual distances per spectrum and not per star,
for stars that have been observed more than once (indicated by the
{\tt Rep\_Flag}), we recommend use of the distance from the spectrum with the highest
SNR.

\item For the first time, photometry from APASS and WISE can be matched with
RAVE stars. This development opens new ways to do science with the database.
For example, Figure~\ref{wisecmd} shows the RAVE giants in a 2MASS-WISE
colour-colour plot.  The most metal-poor giants observed by RAVE ($\rm [Fe/H] <
-$2~dex) are over-plotted in red.  These metal-poor stars have been identified by
projecting all RAVE spectra on a low-dimensional manifold using the
t-Distributed Stochastic Neighbor Embedding (t-SNE) and then re-analysing the
metallicity, via the CaT lines, of all RAVE stars in the manifold that is
mostly populated by very metal-poor stars (Matijevi\u{c} et~al. 2016, in
prep.).  It is evident that they occupy a distinct WISE colour range.  The
comprehensive RAVE dataset may be used as a test bed to define cuts in colour
space to select metal-poor candidates, which can then be applied to fainter
samples than RAVE probed or regions RAVE has not surveyed
\citep[e.g.,][]{schlaufman14}.

\begin{figure}
\includegraphics[width=\linewidth]{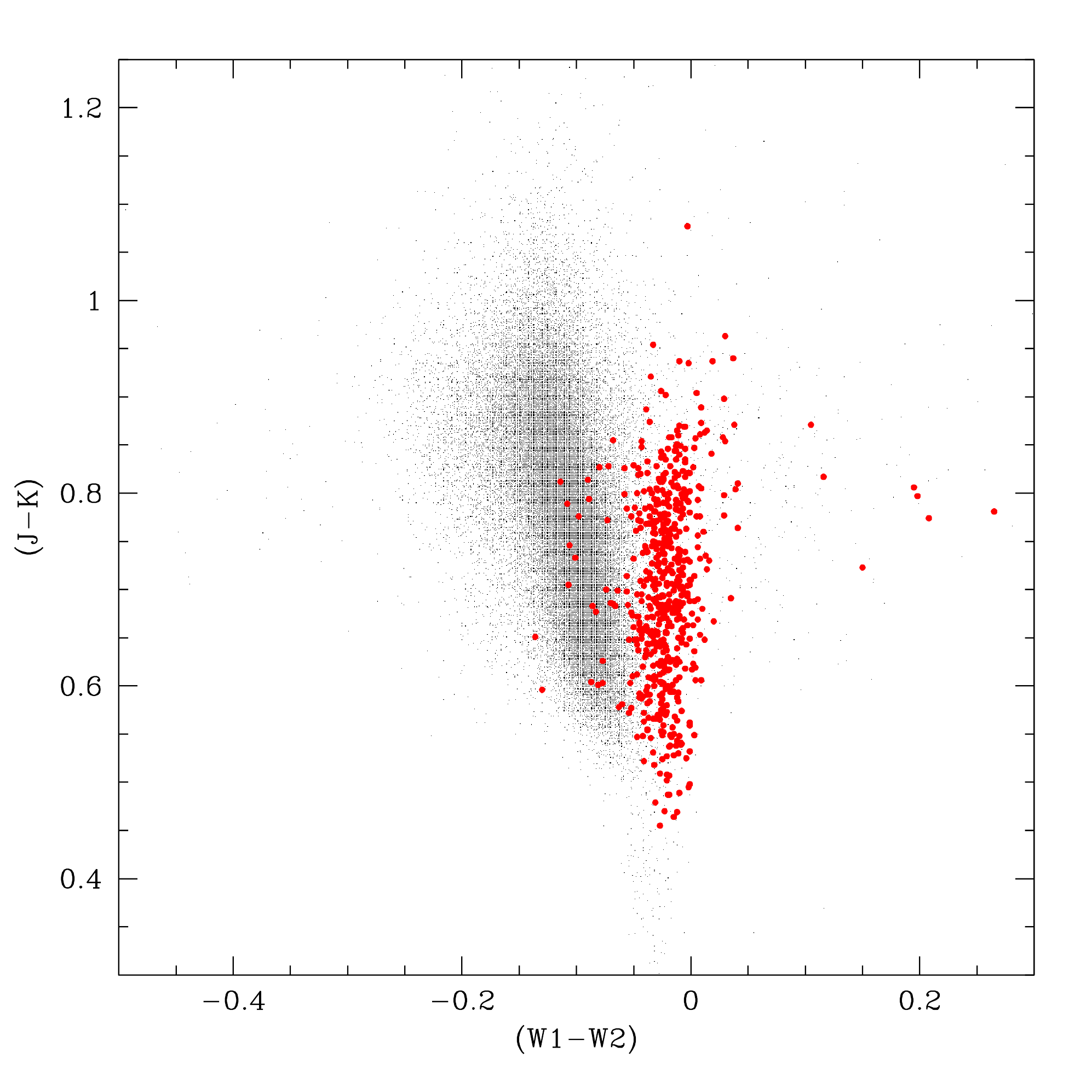}\\
\caption{Colour-colour diagram of the RAVE giants, with the most metal-poor RAVE stars
over-plotted in red.  RAVE can be used to refine criteria and quantify likelihood to photometically
select metal-poor stars for spectroscopic follow-up.
\label{wisecmd}}
\end{figure}
 
\item The inclusion of APASS photometry also allows for the determination of IRFM temperatures, which
are provided for more than 95\% of the RAVE sample.

\end{itemize}

\section{Conclusions}\label{sec:conclude}

The RAVE DR5 presents radial velocities for 457\,589 individual
stars in the brightness range $9< I< 12\,$mag, obtained from spectra with
a resolution of 7\,500 covering the CaT regime.  This catalog can be accessed
by \doi{10.17876/rave/dr.5/001} and in the supplemental online data.  The typical SNR of a RAVE
star is 40 and the typical uncertainty in radial velocity is $<2\kms$.
Stellar parameters are derived from the DR4 stellar parameter pipeline, based
on the algorithms of MATISSE and DEGAS, but an updated calibration improves
the accuracy of the DR5 stellar parameters by up to 15\%.  This pipeline is valid
for stars with temperatures between 4000~K and 8000~K.  The uncertainties
in $T_{\rm eff}$, $\log g$ and \mh~are approximately 250\,K, 0.4\,dex and
0.2\,dex, respectively, but vary with stellar population and SNR.  
The best stellar parameters have {\tt Algo\_Conv}=0, SNR $>$ 40, and {\tt c1}=n,
{\tt c2}=n and {\tt c3}=n.  An error spectrum has been computed for each 
observed spectrum, and is then used to assess the
uncertainties in the radial velocities and stellar parameters.

Temperatures from the Infrared Flux Method are derived for $>95$\% of all
RAVE stars, and for a sub-sample of stars that can be calibrated
asteroseismically ($\sim45$\% of the RAVE sample), the asteroseismically
calibrated $\log g$ is provided.  
The RAVE stars in the asteroseismically calibrated sample are 
given in \doi{10.17876/rave/dr.5/002} and described in Table~8 of the Appendix.
As in \citet{matijevic12}, 
binarity and morphological flags are given for each spectrum.  Photometric
information and proper motions are compiled for each star.

The abundances of Al, Si, Ti, Fe, Mg and Ni are provided for approximately
2/3 of the RAVE stars.  These are generally good to $\sim0.2\,$dex, but their
accuracy varies with SNR and, for some elements, also of the stellar
population.  Distances, ages, masses and the interstellar extinctions are
computed using the methods presented in \citet{binney14}, but upgraded,
especially for the more metal-poor stars.

%For 80\% of the stars in the RAVE volume, space velocities can be derived to
%better than $20\kms$ by combining DR5 distances, radial velocities and the
%UCAC4 proper motions listed in DR5.  
The astrometry and parallaxes from the first Gaia data 
combined with the RAVE DR5 radial velocities ensure that $10\kms$ uncertainties in 
space velocities for 70\% of the RAVE-TGAS stars can be derived.
Further, because Gaia astrometry provides completely new constraints on
distances and tangential velocities, we can now use the RAVE pipelines to
derive yet more accurate stellar parameters and distances for the TGAS stars,
and even improve the parameters and distances of RAVE stars that are not in
TGAS.  The RAVE stars that have TGAS counterparts are 
provided in \doi{10.17876/rave/dr.5/004}.

\acknowledgments
Funding for RAVE has been provided by: the Australian Astronomical Observatory; the 
Leibniz-Institut fuer Astrophysik Potsdam (AIP); the Australian National University; the 
Australian Research Council; the French National Research Agency; the German Research 
Foundation (SPP 1177 and SFB 881); the European Research Council (ERC-StG 240271 Galactica); 
the Istituto Nazionale di Astrofisica at Padova; The Johns Hopkins University; the National Science 
Foundation of the USA (AST-0908326); the W. M. Keck foundation; the Macquarie University; the 
Netherlands Research School for Astronomy; the Natural Sciences and Engineering Research 
Council of Canada; the Slovenian Research Agency; the Swiss National Science Foundation; the 
Science \& Technology Facilities Council of the UK; Opticon; Strasbourg Observatory; and the 
Universities of Groningen, Heidelberg and Sydney.
Based on data products from observations made with ESO Telescopes at the La Silla Paranal
Observatory under programme ID 188.B-3002.

\appendix
\section{Appendix Bookkeeping}

%\\
In total, there are 2505 RAVE DR4 stars that are not in this data release.  These fall into five categories:
\\
\begin{enumerate}
\item  Doubled field -- identical field was published twice under a different name

20060123\_0456m20 is doubled with 20060126\_0456m20

20060123\_0456m20 is removed

\item    Renamed fields -- fields that were renamed

20060627\_0003m13 is renamed 20060629\_0003m13 

20070207\_0734m34 is renamed  20070918\_0734m34

\item    Incorrect FITS headers -- coordinates in header do not appear to be correct, so the proper stars that were observed can not be identified; these fields were removed

20050814\_2314m31 

20060629\_0003m13 

\item    Poor quality fields that were released in DR4 %-- not sure why these are poor, check with Harry \& Gal
20110705\_2028m00b 

20091201\_0206m84 %(Alessandro did fix this field, but it does not process through SPARV now)

\item    DR4 stars with SNR $<$ 10, spectra of too poor quality to process

\end{enumerate}

We are left with 296 DR4 stars with SNR $>$ 10 that were not able to be processed with SPARV.  

\section{Appendix material}
The contents of the individual columns of the main DR5 catalog are specified in 
Table~\ref{tab:DR5}, and the contents of the individual columns of the asteroseismically calibrated
red giant catalog are specified in Table~\ref{tab:DR5SC}. The catalog is accessible online at
\url{http://www.rave-survey.org} and via the CDS VizieR service.

\begin{table}
\caption{Main DR5 Catalogue description} 

\label{tab:DR5}
\begin{center}
\begin{tabular}{llcclll}
\hline 
Col  &  Format   &   Units &   NULL & Label      &               Explanations \\           \hline
%--------------------------------------------------------------------------------------------------------------------------
1       & char     & -          & N     & RAVE\_OBS\_ID                 & Target designation                                                     \\
2       & char     & -          & N     & HEALPix                      & Hierarchical Equal-Area iso-Latitude Pixelisation value (Note 1)      \\
3       & char     & -          & N     & RAVEID                      & RAVE target designation                                                \\
4       & double       & deg        & N     & RAdeg                       & Right ascension                                                        \\
5       & double       & deg        & N     & DEdeg                       & Declination                                                            \\
6       & double       & deg        & N     & Glon                        & Galactic longitude                                                     \\
7       & double       & deg        & N     & Glat                        & Galactic latitude                                                      \\
8       & float        & km/s       & N     & HRV                         & Heliocentric radial velocity                                           \\
9       & float        & km/s       & N     & eHRV                        & HRV error                                                              \\
10       & float        & km/s       & N     & StdDev\_HRV            & Standard deviation in HRV from 10 resampled spectra      \\
11      & float        & km/s       & N     & MAD\_HRV            & Median absolute deviation in HRV from 10 resampled spectra        \\
12 	  & float        &               & Y     & STN\_SPARV			& Signal-to-noise ratio calculated by SPARV (Note 2)  \\
13      & float        & -          & Y     & SNR\_K                       & Signal to Noise value (Note 2)                                         \\
14      & float        & K          & Y     & Teff\_K                      & Effective temperature (Note 2)                                         \\
15      & float        & K          & Y     & Teff\_N\_K                      & Calibrated effective temperature (Note 2)                                         \\
16      & float        & K          & Y     & eTeff\_K                     & Error Effective temperature (Note 2)                                   \\
17      & float        & K       & N     & MAD\_Teff\_K            & Median absolute deviation in Teff\_K from 10 resampled spectra        \\
18       & float        & K      & N     & StdDev\_Teff\_K            & Standard deviation in Teff\_K from 10 resampled spectra      \\
19      & float        & dex        & Y     & logg\_K                      & Log gravity (Note 2)                                                   \\
20      & float        & dex        & Y     & logg\_N\_K                      & Calibrated log gravity (Note 2)                                                   \\
21      & float        & dex        & Y     & elogg\_K                     & Error Log gravity (Note 2)                                             \\
22      & float        & dex       & N     & MAD\_logg\_K            & Median absolute deviation in logg\_K from 10 resampled spectra        \\
23       & float        & dex     & N     & StdDev\_logg\_K            & Standard deviation in logg\_K from 10 resampled spectra      \\
24      & float        & dex        & Y     & Met\_K                       & Metallicity [m/H](Note 2)                                              \\
25      & float        & dex        & Y     & Met\_N\_K                     & Calibrated Metallicity [m/H](Note 2)                                              \\
26      & float        & dex        & Y     & eMet\_K                      & ErrorMetallicity [m/H] (Note 2)                                        \\
27      & float        & dex       & N     & MAD\_Met\_K            & Median absolute deviation in Met\_K from 10 resampled spectra        \\
28       & float        & dex      & N     & StdDev\_Met\_K            & Standard deviation in Met\_K from 10 resampled spectra      \\
29      & float        & -          & Y     & CHISQ\_K                    & $\chi$2 of the Stellar Parameter Pipeline (Note 2) \\
30      & float        & -          & Y     & Algo\_Conv\_K                 & Quality Flag for Stellar Parameter pipeline [0..4] (Note 2, Note 4)    \\
31      & float        & K          & Y     & Teff\_IR                      & Temperature from infrared flux method                                          \\
32      & float        & K          & Y     & eTeff\_IR                      & Internal error on Teff\_IR              \\
33      & char        & -          & N     & IR\_direct                     & infrared flux method flag (Note 5)                                   \\
34      & float        & dex        & Y     & Mg                          & Abundance of Mg [Mg/H]                                                 \\
35      & int          & -          & Y     & Mg\_N                        & Number of used spectral lines for calculation of abundance             \\
36      & float        & dex        & Y     & Al                          & Abundance of Al [Al/H]                                                 \\
37      & int          & -          & Y     & Al\_N                        & Number of used spectral lines for calculation of abundance             \\
38      & float        & dex        & Y     & Si                          & Abundance of Si [Si/H]                                                 \\
39      & int          & -          & Y     & Si\_N                        & Number of used spectral lines for calculation of abundance             \\
40      & float        & dex        & Y     & Ti                          & Abundance of Ti [Ti/H]                                                 \\
41      & int          & -          & Y     & Ti\_N                        & Number of used spectral lines for calculation of abundance             \\
42      & float        & dex        & Y     & Fe                          & Abundance of Fe [Fe/H]                                                 \\
43      & int          & -          & Y     & Fe\_N                        & Number of used spectral lines for calculation of abundance             \\
44      & float        & dex        & Y     & Ni                          & Abundance of Ni [Ni/H]                                                 \\
45      & int          & -          & Y     & Ni\_N                        & Number of used spectral lines for calculation of abundance             \\
46      & float          & dex         & Y     & Alpha\_c                     & Alpha-enhancement from chemical pipeline  (Note 2)          \\
47      & float        & -        & Y     & CHISQ\_c                        & $\chi$2 of the chemical pipeline (Note 2)                                           \\
48      & float          & -          & Y     & frac\_c                        & Fraction of spectrum used for calculation of abundances (Note 2)      \\
49      & float          & mag         & Y     & AV\_Schlegel                 & Total Extinction in V-band from \citet{schlegel98} \\	
50     & float        & kpc        & Y     & distance                  & Spectrophotometric Distance  \citep{binney14} \\
51     & float        & kpc        & Y     & edistance                      & Error on Distance  \citep{binney14}                                         \\
52     & float        & mag         & Y     & log\_Av                          & Log Av Extinction  \citep{binney14}                                       \\
53     & float        & mag     & Y     & elog\_Av                        & Error on log\_Av  \citep{binney14}                                    \\
54    & float        & mas        & Y     & parallax                    & Spectrophotometric parallax \citep{binney14}                                          \\
55     & float        & mas        & Y     & eparallax                  & Error on parallax  \citep{binney14}                              \\
56     & float        & mag          & Y     & DistanceModulus\_Binney      & Distance modulus  \citep{binney14}                           \\
57     & float        & mag       & Y     & eDistanceModulus\_Binney     & Distance modulus  \citep{binney14}                           \\
58     & int        & -       & Y     & Fit\_Flag\_Binney     & See final paragraph \S3 of  \citet{binney14}          \\
59     & float        & -          & Y     & FitQuality\_Binney      & given by symbol "F" in eq 15 of  \citet{binney14}                    \\
60     & int 	& -       & Y     & N\_Gauss\_fit		&     Number of components required for multi-Gaussian distance modulus fit      \\
61     & int 	& -       & Y     & Gauss\_mean\_1	&     Property of multi-Gaussian distance modulus fit, see Section~\ref{sec:D}, eq.~\ref{eq:defsfk}     \\
62     & float 	& -       & Y     & Gauss\_sigma\_1	&    Property of multi-Gaussian distance modulus fit, see Section~\ref{sec:D}, eq.~\ref{eq:defsfk}      \\
63     & float 	& -       & Y     & Gauss\_frac\_1	&   Property of multi-Gaussian distance modulus fit, see Section~\ref{sec:D}, eq.~\ref{eq:defsfk}       \\
64     & float 	& -       & Y     & Gauss\_mean\_2	&    Property of multi-Gaussian distance modulus fit, see Section~\ref{sec:D}, eq.~\ref{eq:defsfk}      \\
65     & float 	& -       & Y     & Gauss\_sigma\_2	&  Property of multi-Gaussian distance modulus fit, see Section~\ref{sec:D}, eq.~\ref{eq:defsfk}        \\
66     & float 	& -       & Y     & Gauss\_frac\_2	&   Property of multi-Gaussian distance modulus fit, see Section~\ref{sec:D}, eq.~\ref{eq:defsfk}       \\
67     & float 	& -       & Y     & Gauss\_mean\_3	&    Property of multi-Gaussian distance modulus fit, see Section~\ref{sec:D}, eq.~\ref{eq:defsfk}      \\
\end{tabular}
\end{center}
\end{table}

\setcounter{table}{6}
\begin{table}
\caption{Catalogue description (continued)} 
\begin{center}
\begin{tabular}{llcclll}
\hline 
Col  &  Format   &   Units &   NULL & Label      &               Explanations \\           \hline
68     & float 	& -       & Y     & Gauss\_sigma\_3	&  Property of multi-Gaussian distance modulus fit, see Section~\ref{sec:D}, eq.~\ref{eq:defsfk}        \\
69     & float 	& -       & Y     & Gauss\_frac\_3	&    Property of multi-Gaussian distance modulus fit, see Section~\ref{sec:D}, eq.~\ref{eq:defsfk}      \\
% 71     & float        & -       & Y     & QFlag\_Binney     &not sure         \\
70     & char      & -          & Y     & c1                          & n.th minimum distance (Note 6)                                         \\
71     & char      & -          & Y     & c2                          & n.th minimum distance (Note 6)                                         \\
72     & char      & -          & Y     & c3                          & n.th minimum distance (Note 6)                                         \\
73     & char      & -          & Y     & c4                          & n.th minimum distance (Note 6)                                         \\
74     & char      & -          & Y     & c5                          & n.th minimum distance (Note 6)                                         \\
75     & char      & -          & Y     & c6                          & n.th minimum distance (Note 6)                                         \\
76     & char      & -          & Y     & c7                          & n.th minimum distance (Note 6)                                         \\
77     & char      & -          & Y     & c8                          & n.th minimum distance (Note 6)                                         \\
78     & char      & -          & Y     & c9                          & n.th minimum distance (Note 6)                                         \\
79     & char      & -          & Y     & c10                         & n.th minimum distance (Note 6)                                         \\
80     & char      & -          & Y     & c11                         & n.th minimum distance (Note 6)                                         \\
81     & char      & -          & Y     & c12                         & n.th minimum distance (Note 6)                                         \\
82     & char      & -          & Y     & c13                         & n.th minimum distance (Note 6)                                         \\
83     & char      & -          & Y     & c14                         & n.th minimum distance (Note 6)                                         \\
84     & char      & -          & Y     & c15                         & n.th minimum distance (Note 6)                                         \\
85     & char      & -          & Y     & c16                         & n.th minimum distance (Note 6)                                         \\
86     & char      & -          & Y     & c17                         & n.th minimum distance (Note 6)                                         \\
87     & char      & -          & Y     & c18                         & n.th minimum distance (Note 6)                                         \\
88     & char      & -          & Y     & c19                         & n.th minimum distance (Note 6)                                         \\
89     & char      & -          & Y     & c20                         & n.th minimum distance (Note 6)                                         \\
90 	& int    	& -	    & N       & Rep\_Flag		     & 0: single observation, 1: more than one observation \\
91 	& int	    	& -	    & N       & CluStar\_Flag		     & 0: not a targeted observation, 1: targeted observation \\
92 	& int	    	& -	    & N       & FootPrint\_Flag		     & 0: outside RAVE selection function footprint, 1: inside footprint \\
93 	& char 	& -		& Y       & ID\_TGAS\_source		     & TGAS Target designation  \\
94 	& char 	& -		& Y       & MatchFlag\_TGAS		     &  Crossmatch quality flag (Note 7)    \\
95      & float      & deg        & Y     & RA\_TGAS       & TGAS Right ascension (J2015)                                     \\
96      & float        & deg     & Y     & DE\_TGAS                & TGAS Declination (J201                                      \\
97      & float        & mas/yr     & Y     & pmRA\_TGAS                & Proper motion RA from TGAS -- $\dot{\alpha }\mathrm{cos}(\delta )$
                                \\
98      & float        & mas/y     & Y     & pmRA\_error\_TGAS                 & standard error of proper motion in RA from TGAS                                    \\
99      & float        & mas/yr     & Y     & pmDE\_TGAS                & Proper motion in DE from TGAS -- $\dot{\delta }$               \\
100      & float        & mas/yr     & Y     & pmDE\_error\_TGAS                 & standard error of proper motion in DE from TGAS                                            \\
101      & float        & mas     & Y     & parallax\_TGAS                & parallax from TGAS                       \\
102      & float        & mas     & Y     & parallax\_error\_TGAS                & standard error of parallax from TGAS                       \\
103      & float        & mag     & Y     & phot\_g\_mean\_mag\_TGAS              & $G$-band mean magnitude  from TGAS                                           \\
104      & float        & e-/s    & Y     & phot\_g\_mean\_flux\_TGAS              & $G$-band mean flux from TGAS                                         \\
105      & float        & e-/s    & Y     & phot\_g\_mean\_flux\_error\_TGAS              & Error on $G$-band mean flux from TGAS                                           \\
106 	& char 	& -	& Y       & ID\_Hipparcos		     & Hipparcos Target designation  \\
107      & char     & -          & Y     & ID\_TYCHO2                   & TYCHO2 Target designation                                              \\
108     & float        & arcsec     & Y     & Dist\_TYCHO2                 & Center distance to target catalog                                      \\
109      & char      & -          & Y     & MatchFlag\_TYCHO2       & Crossmatch quality flag (Note 6)                                       \\
110      & float        & mag     & Y     & BTmag\_TYCHO2                 & $B_T$ magnitude from TYCHO2                                           \\
111      & float        & mag     & Y     & eBTmag\_TYCHO2                & error on $B_T$ mag from TYCHO2                                  \\
112      & float        & mag     & Y     & VTmag\_TYCHO2                 & $V_T$ magnitude from TYCHO2                                       \\
113      & float        & mag     & Y     & eVTmag\_TYCHO2                & error $V_T$ magnitude from TYCHO2                       \\
114      & float        & mas/yr     & Y     & pmRA\_TYCHO2                 & Proper motion RA from TYCHO2                                           \\
115      & float        & mas/yr     & Y     & epmRA\_TYCHO2                & error Proper motion RA from TYCHO2                                  \\
116      & float        & mas/yr     & Y     & pmDE\_TYCHO2                 & Proper motion DE from TYCHO2                                           \\
117     & float        & mas/yr     & Y     & epmDE\_TYCHO2                & error Proper motion DE from TYCHO2                                  \\
118      & char     & -          & Y     & ID\_UCAC4                    & UCAC4 Target designation                                               \\
119      & float        & arcsec     & Y     & Dist\_UCAC4                  & Center distance to target catalog                                      \\
120      & char      & -          & Y     & MatchFlag\_UCAC4        & Crossmatch quality flag (Note 7)                                       \\
121     & float        & mas/yr     & Y     & pmRA\_UCAC4                  & Proper motion RA from UCAC4                                            \\
122      & float        & mas/yr     & Y     & epmRA\_UCAC4                 & error Proper motion RA from UCAC4                                      \\
123      & float        & mas/yr     & Y     & pmDE\_UCAC4                  & Proper motion DE from UCAC4                                            \\
124     & float        & mas/yr     & Y     & epmDE\_UCAC4                 & error Proper motion DE from UCAC4                                      \\
125     & char     & -          & Y     & ID\_PPMXL                    & PPMXL Target designation                                               \\
126      & float        & arcsec     & Y     & Dist\_PPMXL                  & Center distance to target catalog                                      \\
127      & char      & -          & Y     & MatchFlag\_PPMXL        & Crossmatch quality flag (Note 7)                                       \\
128      & float        & mas/yr     & Y     & pmRA\_PPMXL                  & Proper motion RA from PPMXL                                            \\
129      & float        & mas/yr     & Y     & epmRA\_PPMXL                 & error Proper motion RA from PPMXL                                      \\
130      & float        & mas/yr     & Y     & pmDE\_PPMXL                  & Proper motion DE from PPMXL                                            \\
131      & float        & mas/yr     & Y     & epmDE\_PPMXL                 & error Proper motion DE from PPMXL                                      \\
132      & char     & -          & Y     & ID\_2MASS                    & 2MASS Target designation                                               \\
133      & float        & arcsec     & Y     & Dist\_2MASS                  & Center distance to target catalog                                      \\

   \end{tabular}
   \end{center}
\end{table}

\setcounter{table}{6}
\begin{table}
\caption{Catalogue description (continued)} 
\begin{center}
\begin{tabular}{llcclll}
\hline 
Col  &  Format   &   Units &   NULL & Label      &               Explanations \\           \hline
134      & char      & -          & Y     & MatchFlag\_2MASS        & Crossmatch quality flag (Note 7)                                       \\
135      & double       & mag        & Y     & Jmag\_2MASS                  & $J$  magnitude                                                             \\
136      & double       & mag        & Y     & eJmag\_2MASS                 & Error $J$  magnitude                                                       \\
137      & double       & mag        & Y     & Hmag\_2MASS                  & $H$ magnitude                                                             \\
138      & double       & mag        & Y     & eHmag\_2MASS                 & Error $H$  magnitude                                                       \\
139      & double       & mag        & Y     & Kmag\_2MASS                  & $K$  magnitude                                                             \\
140      & double       & mag        & Y     & eKmag\_2MASS                 & Error $K$  magnitude                                                       \\
141      & char     & -          & Y     & ID\_ALLWISE                    & WISE Target designation                                               \\
142      & double       & arcsec     & Y     & Dist\_ALLWISE                  & Centre distance to target catalog                                      \\
143      & char      & -          & Y     & MatchFlag\_ALLWISE        & Crossmatch quality flag (Note 7)                                       \\
144      & double       & mag        & Y     & W1mag\_ALLWISE                  & $W1$ magnitude                                                             \\
145      & double       & mag        & Y     & eW1mag\_ALLWISE                 & Error $W1$  magnitude                                     \\
146      & double       & mag        & Y     & W2mag\_ALLWISE                  & $W2$ magnitude                                                             \\
147      & double       & mag        & Y     & eW2mag\_ALLWISE                 & Error $W2$  magnitude                                            \\
148      & double       & mag        & Y     & W3mag\_ALLWISE                  & $W3$ magnitude                                                             \\
149      & double       & mag        & Y     & eW3mag\_ALLWISE                 & Error $W3$ magnitude                                        \\
150      & double       & mag        & Y     & W4mag\_ALLWISE                  & $W4$ magnitude                                                             \\
151      & double       & mag        & Y     & eW4mag\_ALLWISE                 & Error $W4$ magnitude                                             \\
152      & char    & -        & Y     & cc\_flags\_ALLWISE                  & prioritized artifacts affecting the source in each band\\
153      & int       & -        & Y     & ext\_flg\_ALLWISE                 & probability source morphology is not consistent with single PSF\\
154      & char    & -        & Y     & var\_flg\_ALLWISE                  &      probability that flux varied in any band greater than amount expected from unc.s                                                        \\
155      & char    & mag        & Y     & ph\_qual\_ALLWISE                 & photometric quality of each band (A=highest, U=upper limit)\\
156     & double & arcsec     & Y     & Dist\_APASSDR9                 & Centre distance to target catalog                                      \\
157     & char     & -          & Y     & MatchFlag\_APASSDR9       & Crossmatch quality flag (Note 7)                                       \\
158     & double       & mag        & Y     & Bmag\_APASSDR9                & $B$ magnitude                                                            \\
159     & double       & mag        & Y     & eBmag\_APASSDR9                & error $B$ magnitude \\
160     & double       & mag        & Y     & Vmag\_APASSDR9                & $V$ magnitude                                                            \\
161     & double       & mag        & Y     & eVmag\_APASSDR9                & error $V$ magnitude                                                  \\
162     & double       & mag        & Y     & gpmag\_APASSDR9                 & $g'$  magnitude                                                             \\
163     & double       & mag        & Y     & egpmag\_APASSDR9                 & error $g'$  magnitude                                                     \\
164     & double       & mag        & Y     & rpmag\_APASSDR9                 & $r'$  magnitude                                                             \\
165     & double       & mag        & Y     & erpmag\_APASSDR9                 & error $r'$ magnitude                                                  \\
166     & double       & mag        & Y     & ipmag\_APASSDR9                 & $i'$  magnitude                                                             \\
167     & double       & mag        & Y     & eipmag\_APASSDR9                 & error $i'$ magnitude                                              \\
168      & char     & -          & Y     & ID\_DENIS                    & DENIS Target designation                                               \\
169      & double       & arcsec     & Y     & Dist\_DENIS                  & Centre distance to target catalog                                      \\
170      & char      & -          & Y     & MatchFlag\_DENIS        & Crossmatch quality flag (Note 7)                                       \\
171      & double       & mag        & Y     & Imag\_DENIS                  & I  magnitude                                                             \\
172      & double       & mag        & Y     & eImag\_DENIS                 & Error I  magnitude                                                       \\
173      & double       & mag        & Y     & Jmag\_DENIS                  & J  magnitude                                                             \\
174      & double       & mag        & Y     & eJmag\_DENIS                 & Error J  magnitude                                                       \\
175      & double       & mag        & Y     & Kmag\_DENIS                  & K  magnitude                                                             \\
176      & double       & mag        & Y     & eKmag\_DENIS                 & Error K  magnitude                                                       \\
177      & char     & -          & Y     & ID\_USNOB1                   & USNOB1 Target designation                                              \\
178     & double       & arcsec     & Y     & Dist\_USNOB1                 & Centre distance to target catalog                                      \\
179     & char      & -          & Y     & MatchFlag\_USNOB1       & Crossmatch quality flag (Note 7)                                       \\
180     & double       & mag        & Y     & B1mag\_USNOB1                & B1  magnitude                                                            \\
181     & double       & mag        & Y     & R1mag\_USNOB1                & R1  magnitude                                                            \\
182     & double       & mag        & Y     & B2mag\_USNOB1                & B2  magnitude                                                            \\
183     & double       & mag        & Y     & R2mag\_USNOB1                & R2  magnitude                                                            \\
184     & double       & mag        & Y     & Imag\_USNOB1                 & I  magnitude                                                             \\
185      & int        & mas/yr     & Y     & pmRA\_USNOB1                  & Proper motion RA from USNOB1                                            \\
186      & int        & mas/yr     & Y     & epmRA\_USNOB1                & error Proper motion RA from USNOB1                                      \\
187      & int        & mas/yr     & Y     & pmDE\_USNOB1                  & Proper motion DE from USNOB1                                            \\
188      & int        & mas/yr     & Y     & epmDE\_USNOB1                 & error Proper motion DE from USNOB1                                      \\
189     & int     & -          & N     & Obsdate                     & Observation date yyyymmdd                                              \\
190      & char     & -          & N     & FieldName                   & Name of RAVE field (RA/DE) \\
191      & int          & -          & N     & FiberNumber                 & Number of optical fiber [1,150]                                        \\
192      & int          & -          & N     & PlateNumber                 & Number of field plate [1..3]                                           \\
193     & double      & day   & N & MJD\_OBS  & Modfied Julian Date \\
194     & char       &   -      & N & LST\_start & exposure start in Local Sidereal Time \\
195     & char       &   -      & N & LST\_end & exposure end  in Local Sidereal Time \\
196     & char       & - 	& N & UTC\_start & exposure start in Coordinated Universal Time \\
197     & char       & - 	& N & UTC\_end & exposure end in Coordinated Universal Time \\

\hline

   \end{tabular}
   \end{center}
\end{table}

\setcounter{table}{6}
\begin{table}
\caption{Catalogue description (continued)} 
\begin{center}
\begin{tabular}{llcclll}
\hline 
\hline
%\end{tabular}
\footnotetext{{\bf Notes.}
(1): HEALPix values were computed using the resolution parameter $N_{\rm side} = 4096$
(resolution index of 12) and the NESTED numbering scheme. Any lower-resolution index HEALPix value
can be computed from the given one by dividing it by $4^(12 - n)$, where $n < 12$
is the desired resolution index.
(2): Originating from:
   \_K indicates values from Stellar Parameter Pipeline, 
   \_N\_K indicates a calibrated value,
   \_c indicates values from Chemical Pipeline,
   \_SPARV indicates values of Radial Velocity Pipeline (used in DR3 also).
(3): Flag value of the form FGSH, F being for the entire plate, G for the 50 
fibres group to which the fibre belongs. S flags the zero-point correction used: C 
for cubic and S for a constant shift. If H is set to * the fibre is close to a 15 
fibre gap. For F and G the values can be A, B, C, D, or E
    $A =$~dispersion around correction lower than 1km/s
    $B =$~dispersion between 1 and 2km/s
    $C =$~dispersion between 2 and 3km/s
    $D =$~dispersion larger than 3km/s
    $E =$~less than 15 fibres available for the fit.
(4): Flag of Stellar Parameter Pipeline     
   $0 =$~Pipeline converged. 
   $1 =$~no convergence.
   $2 =$~MATISSE oscillates between two values and the mean has been performed. 
   $3 =$~results of MATISSE at the boundaries or outside the grid and the DEGAS value has been adopted
   $4 =$~the metal-poor giants with SNR$<$20 have been re-run by degas with a scale factor (ie, internal parameter of DEGAS) of 0.40
(5): Cross-identification flag as follows:
   $IRFM$~Temperature derived from infrared flux method
   $CTRL$~Temperature computed via color-$T_{\rm eff}$ relations
   $NO$~No temperature derivation possible
(6): Morphological Flag 
   n.th minimum distance to base spectrum given by one of 
   the types {\it a,b,c,d,e,g,h,n,o,p,t,u,w} \citep[see][]{matijevic12}.
(7): Cross-identification flag as follows:
   $A = 1$~association within 2 arcsec.
   $B = 2$~associations within 2 arcsec.
   $C =$~More than 2 associations within 2 arcsec.
   $D =$~Nearest neighbour more than 2 arcsec. away.
   $X =$~No association found (within 10 arcsec limit ).
     }
   \end{tabular}
   \end{center}
\end{table}

The contents of the individual columns of the asteroseismically calibrated giant sample 
in the Fifth Data Release catalog are specified in Table~\ref{tab:DR5SC}. The catalog is accessible online at
\url{http://www.rave-survey.org} and via the CDS VizieR service.

\begin{table}
\caption{Asteroseismically Calibrated Red Giant Catalog description} 

\label{tab:DR5SC}
\begin{center}
\begin{tabular}{llcclll}
\hline 
Col  &  Format   &   Units &   NULL & Label      &               Explanations \\           \hline
%--------------------------------------------------------------------------------------------------------------------------
1       & char(32)     & -          & N     & RAVE\_OBS\_ID                 & Target designation                                                     \\
2 	& float	    & dex	& Y       & logg\_SC		     & Log gravity calibrated asteroseismically (V16) \\
3	& float	    & dex	& Y       & elogg\_SC		     & error on logg\_MV (V16) \\
4 	& int	    & dex	& Y       & Flag050		     & Difference between logg\_MV and logg\_K is less than 0.5 dex. 1= true 0=false \\
5 	& int	    & dex	& Y       & Flag075		     & Difference between logg\_MV and logg\_K is less than 0.75 dex. 1= true 0=false \\
6 	& int	    & dex	& Y       & Flag\_M     & Normal star, meaning c1 - c20 are all ``n". 1= true 0=false \\
7      & float        & K          & Y     & Teff\_IR                      & Temperature from infrared flux method                                          \\
8      & float        & dex        & Y     & Mg                          & Abundance of Mg [Mg/H]                                                 \\
9      & int          & -          & Y     & Mg\_N                        & Number of used spectral lines for calculation of abundance             \\
10      & float        & dex        & Y     & Al                          & Abundance of Al [Al/H]                                                 \\
11      & int          & -          & Y     & Al\_N                        & Number of used spectral lines for calculation of abundance             \\
12      & float        & dex        & Y     & Si                          & Abundance of Si [Si/H]                                                 \\
13      & int          & -          & Y     & Si\_N                        & Number of used spectral lines for calculation of abundance             \\
14      & float        & dex        & Y     & Ti                          & Abundance of Ti [Ti/H]                                                 \\
15      & int          & -          & Y     & Ti\_N                        & Number of used spectral lines for calculation of abundance             \\
16      & float        & dex        & Y     & Fe                          & Abundance of Fe [Fe/H]                                                 \\
17      & int          & -          & Y     & Fe\_N                        & Number of used spectral lines for calculation of abundance             \\
18      & float        & dex        & Y     & Ni                          & Abundance of Ni [Ni/H]                                                 \\
19      & int          & -          & Y     & Ni\_N                        & Number of used spectral lines for calculation of abundance             \\
20      & float          & dex         & Y     & Alpha\_c                     & Alpha-enhancement from chemical pipeline         \\
21      & float        & -        & Y     & CHISQ\_c                        & $\chi$2 of the chemical pipeline                                  \\
22      & float          & -          & Y     & frac\_c                        & Fraction of spectrum used for calculation of abundances      \\
23      & float          & mag         & Y     & AV\_Schlegel                 & Total Extinction in V-band from \citet{schlegel98} \\	
24     & float        & kpc        & Y     & distance                  & Spectrophotometric Distance  \citep{binney14} \\
25     & float        & kpc        & Y     & e\_distance                      & Error on Distance  \citep{binney14}                                         \\
26     & float        & mag         & Y     & log\_Av                          & Log Av Extinction  \citep{binney14}                                       \\
27     & float        & mag     & Y     & elog\_Av                        & Error on log\_Av  \citep{binney14}                                    \\
28    & float        & mas        & Y     & parallax                    & Spectrophotometric parallax \citep{binney14}                                          \\
29     & float        & mas        & Y     & e\_parallax                  & Error on parallax  \citep{binney14}                              \\
30     & float        & mag          & Y     & DistanceModulus\_Binney      & Distance modulus  \citep{binney14}                           \\
31     & float        & mag       & Y     & eDistanceModulus\_Binney     & Distance modulus  \citep{binney14}                           \\
32     & float        & -       & Y     & Fit\_Flag\_Binney     & See final paragraph \S3 of  \citet{binney14}          \\
33     & float        & -          & Y     & FitQuality\_Binney      & given by symbol "F" in eq 15 of  \citet{binney14}                    \\
34     & float 	& -       & Y     & N\_Gauss\_fit		&   Number components required for multi-Gaussian distance modulus fit        \\
35     & int 	& -       & Y     & Gauss\_mean\_1	&     Property of multi-Gaussian distance modulus fit, see Section~\ref{sec:D}, eq.~\ref{eq:defsfk}     \\
36     & float 	& -       & Y     & Gauss\_sigma\_1	&    Property of multi-Gaussian distance modulus fit, see Section~\ref{sec:D}, eq.~\ref{eq:defsfk}      \\
37     & float 	& -       & Y     & Gauss\_frac\_1	&   Property of multi-Gaussian distance modulus fit, see Section~\ref{sec:D}, eq.~\ref{eq:defsfk}       \\
38     & float 	& -       & Y     & Gauss\_mean\_2	&    Property of multi-Gaussian distance modulus fit, see Section~\ref{sec:D}, eq.~\ref{eq:defsfk}      \\
39     & float 	& -       & Y     & Gauss\_sigma\_2	&  Property of multi-Gaussian distance modulus fit, see Section~\ref{sec:D}, eq.~\ref{eq:defsfk}        \\
40     & float 	& -       & Y     & Gauss\_frac\_2	&   Property of multi-Gaussian distance modulus fit, see Section~\ref{sec:D}, eq.~\ref{eq:defsfk}       \\
41     & float 	& -       & Y     & Gauss\_mean\_3	&    Property of multi-Gaussian distance modulus fit, see Section~\ref{sec:D}, eq.~\ref{eq:defsfk}      \\
42     & float 	& -       & Y     & Gauss\_sigma\_3	&  Property of multi-Gaussian distance modulus fit, see Section~\ref{sec:D}, eq.~\ref{eq:defsfk}        \\
43     & float 	& -       & Y     & Gauss\_frac\_3	&    Property of multi-Gaussian distance modulus fit, see Section~\ref{sec:D}, eq.~\ref{eq:defsfk}      \\
   \end{tabular}
   \end{center}
\end{table}

\clearpage

\clearpage

\end{document}